%% file: arxiv-main.tex
\documentclass[aps,prl,reprint,twocolumn,superscriptaddress]{revtex4-2}

\usepackage{graphicx,amsmath,amssymb}
\usepackage[colorlinks]{hyperref} 
\usepackage{xcolor}
\newcommand{\gw}{\mathrm{gw}}

\newcommand{\FG}{\mathrm{FG}}
\newcommand{\BG}{\mathrm{BG}}
\newcommand{\joint}{\mathrm{joint}}

\newcommand{\vp}{\vec{\phi}}

\begin{document}

\title{The Stochastic Siren: \\ Astrophysical Gravitational-Wave Background Measurements of the Hubble Constant}

\input{authors}

\begin{abstract}
We report the first measurement of the Hubble constant $H_0$ using the stochastic gravitational-wave background arising from binary black hole mergers. This astrophysical background is sensitive to the expansion history of the Universe and thus can be used for cosmological parameter inference independently of not only electromagnetic methods, but also gravitational-wave standard siren approaches. We describe the background's cosmological dependence and show how it can be used as a ``stochastic siren'' to measure $H_0$. By analyzing existing resolved binary black hole mergers and the current non-detection of the background, we find that $H_0$ can be measured more accurately relative to using resolved mergers alone. We also note that the stochastic siren may serve a unique role in the Hubble tension in that the lower bound of the $H_0$ measurement would progressively increase with continued non-detection of the background.

\end{abstract}

\maketitle

\noindent \textit{Introduction.} While it has been known for over two decades that the Universe is undergoing an accelerating expansion~\citep{riess1998observational}, there are significant discrepancies in the measured expansion rate. In particular, the early-~\citep{aghanim2020planck} and late-universe~\citep{kamionkowskiHubbleTensionEarly2022,di2021realm} measurements of the present-day expansion rate (the Hubble constant, $H_0$) are in $>5\sigma$ conflict~\citep{kamionkowskiHubbleTensionEarly2022}, now named the Hubble tension. 
Late-universe probes (e.g.,~type Ia supernovae and the cosmic distance ladder~\citep{riessLargeMagellanicCloud2019a,Riess:2021jrx,broutPantheonAnalysisCosmological2022,freedmanMeasurementsHubbleConstant2021}, gravitationally-lensed systems~\citep{suyu2010dissecting,wongH0LiCOWXIII242020}, and cosmic chronometers~\citep{jimenezLocalDistantUniverse2019}) suggest higher values of $H_0$ ($\sim~72\mbox{--}74~\rm km~s^{-1}~Mpc^{-1}$), while early-universe probes (e.g.,~the cosmic microwave background~\citep{aghanim2020planck,balkenholMeasurementCMBTemperature2023}, 
baryon acoustic oscillations~\citep{alamCompletedSDSSIVExtended2021,collaborationDESI2024VI2024}, and the inverse distance ladder~\citep{macaulayFirstCosmologicalResults2019}) suggest lower values ($\sim67\mbox{--}68~\rm km~s^{-1}~Mpc^{-1}$).  

More recently, observations of gravitational waves (GWs) from the network of LIGO~\citep{aasi2015advancedligo}, Virgo~\citep{acernese2014advancedvirgo}, and KAGRA ~\citep{somiya2012kagra} detectors offer unique methods to study the Hubble tension. 
The GW amplitude of a merger provides a direct measurement of its luminosity distance, without the need for a distance ladder, so only a measurement of the merger's redshift is required to compute $H_0$. 
This so-called ``standard siren'' approach~\citep{schutzDeterminingHubbleConstant1986,holzUsingGravitationalWave2005} can be applied via ``bright'' sirens, wherein an electromagnetic counterpart of the merger is identified to obtain the redshift~\citep{dalalShortGRBBinary2006,nissanke2013determining,vitale2018nsbhH0,ligo2017gw170817H0,chen2022:190521H0,gupta2023using}, or via ``dark'' sirens whose host galaxy redshift can be inferred statistically using galaxy catalogs~\citep{schutzDeterminingHubbleConstant1986,macleod2008precision,delpozzo2012inference,nishizawa2017measurement,chenTwoCentHubble2018,nair2018measuring,fishbach2019gw170817H0,DES:2019ccw,gray2020cosmological,yu2020hunting,palmese2020statistical,borhanian2020dark,finke2021cosmology,abbott2023cceh,gray2023joint,bom2024dark}. 
Other GW methods obtain the redshift by correlating galaxy distributions with merger events~\citep{oguri2016measuring,bera2020incompleteness,mukherjee2020probing,mukherjee2021accurate,mukherjee2024cross} or using the tidal deformability of neutron stars~\citep{messenger2012measuring,chatterjee2021cosmology}. 

Furthermore, features in the mass spectrum of mergers have been utilized as ``spectral sirens'' to break the mass-redshift degeneracy~\citep{chernoff1993gravitational,taylor2012cosmology1,taylor2012cosmology2,farr2019future,you2021standard,ezquiaga2021jumping,ezquiaga2022spectral, Mali:2024wpq}, providing a measurement of $H_0$ using only GWs. The most recent spectral siren analysis from the LIGO-Virgo-KAGRA (LVK) collaboration~\citep{abbott2023cceh} measures $H_0 = 46^{+49}_{-26}\,\, \rm km~s^{-1}~Mpc^{-1}$ (maximum a-posteriori probability and $68.3\%$ highest density interval) using 42 binary black hole (BBH) candidates from the third GW transient catalog (GWTC-3).

Beyond detecting individual mergers, the LVK collaboration searches for the GW background (GWB) arising from unresolved signals~\citep{LIGOScientific:2016fpe,abbottUpperLimitsIsotropic2021,KAGRA:2021gwtc3population}. 
The GWB is thought to be dominated by distant compact binary mergers that cannot be individually detected~\citep{rosado2011agwb1,zhu2011agwb2,marassi2011agwb3,abbottUpperLimitsIsotropic2021}, but it may also contain contributions from other astrophysical sources and/or early-universe phenomena~\citep{starobinskii1979earlygwb1,kibble1976stringsearlygwb2,easther2006earlygwb3,caprini2018cosmological}. 
While there has been a recent possible detection of a GWB produced from supermassive BBHs using pulsar timing arrays~\citep{agazie2023nanograv,antoniadis2023pta1,reardon2023pta2,xu2023pta3}, the stochastic astrophysical GWB composed of stellar-mass mergers has not yet been detected~\citep{abbottUpperLimitsIsotropic2021}. 
However, given that this GWB is composed of individual mergers, its non-detection can already be used to probe the BBH population~\citep{callisterShoutsMurmursCombining2020,abbottUpperLimitsIsotropic2021,turbang2024metallicity}.

A generic GWB is characterized by $\Omega_\gw$, the Universe's current energy density in GWs.
For an astrophysical background, this energy density depends on both the total number of binary mergers integrated over cosmic time and the available volume over which they are distributed, the latter of which depends on the cosmic expansion.
Thus, the overall strength of the astrophysical GWB is affected by the Universe's expansion history, suggesting that it can be used to measure $H_0$.

In this Letter, we propose a novel method to measure $H_0$ by combining the search for the GWB with population inference of individual mergers, an idea we name the ``stochastic siren.''
The conceptual basis of the method is as follows: a smaller value of $H_0$ implies larger comoving volumes in the Universe, leading to more compact object mergers for a fixed merger rate and hence a larger $\Omega_\gw$, meaning that a non-detection of the GWB can exclude lower values of $H_0$. 
Thus, by considering the GWB during cosmological inference, $H_0$ could be measured more accurately.
We demonstrate this by analyzing BBH data from the LVK's first three observing runs, showing that even the current non-detection of the GWB improves measurements of $H_0$ relative to the spectral siren alone.

We review the dependence of the GWB energy spectrum on $H_0$, construct a statistical method based on the likelihood of observing individual mergers with the GWB, and present the results of applying our method to the first three LVK observing runs~\cite{abbott2023gwtc3,abbottUpperLimitsIsotropic2021}. We conclude by discussing possible applications and extensions of the stochastic siren approach. 
Throughout, we refer to the resolved BBHs as ``FG'' (foreground) signals and unresolved BBHs as ``BG'' (background) signals. We use ``GWB'' to refer to the stochastic astrophysical background arising from stellar-mass BBH mergers, unless otherwise noted. 
We assume a flat cosmology ($\Omega_k=0$) \citep{aghanim2020planck} but allow other cosmological energy densities and $H_0$ to vary. We provide Supplemental Material~\citep{Cousins2025prl_suppmat} to explicitly show the cosmological dependence of the GWB, elaborate our full population parameter results, and provide forecasts for GWB detection in various cosmologies.

\noindent \textit{GWB dependence on $H_0$.} The dimensionless GW energy density from astrophysical sources is given by \citep{allenDetectingStochasticBackground1999,ferrariStochasticBackgroundGravitational1999,regimbau2008astrophysical,regimbauAstrophysicalGravitationalWave2011,romanoDetectionMethodsStochastic2017}
\begin{equation}\label{eq:Omega_GWB_def_01}
    \Omega_\gw =
    \frac{f}{\rho_c c^3} F(\vp,f) =
    \frac{f}{\rho_c c^3} \int p(\vp) \int \mathcal{F}(\vp, f) \frac{d \dot{N}^\text{o}}{d z}d\vp d z,
\end{equation}
for speed of light $c$, observer-frame GW frequency $f$, critical mass density of the Universe $\rho_c \equiv 3H_0^2/8\pi G$ (for gravitational constant $G$), and the sources' redshift-integrated flux $F$.
The integrated flux can be written---as above---in terms of the sources' fluence $\mathcal{F}$ (flux times time) and rate $\frac{d\dot{N}^\text{o}}{dz}$ (sources per observer-frame time per redshift interval), each of which is a function of source properties $\vp$ with probabilities $p(\vp)$.

For compact object mergers, Eq.~\eqref{eq:Omega_GWB_def_01} reduces to~\citep{ferrariStochasticBackgroundGravitational1999,regimbau2008astrophysical,regimbauAstrophysicalGravitationalWave2011}
\begin{align}\label{eq:Omega_GWB_def_02}
    \Omega_\gw 
    &= \frac{8\pi G }{ 3 c^2 H_0^3} f
    \int \frac{ \mathcal{R}(z)}{(1+z) E(z)} \biggr \langle \frac{dE_\gw}{d f_s} \biggr \rangle \biggr |_{f_s} d z \, ,
\end{align}
for rate density $\mathcal{R}(z)$ (mergers per comoving volume per source-frame time) and $E(z) \equiv H(z)/H_0$ (for Hubble parameter $H(z)$). The bracketed quantity in Eq.~\eqref{eq:Omega_GWB_def_02} is the average energy spectrum of a single binary evaluated at the source frame frequency $f_s = f (1+z)$:
\begin{equation}\label{eq:dEdf}
\biggr \langle \frac{dE_\gw}{d f_s} \biggr \rangle \biggr |_{f_s} =
\int d \vp \, p(\vp) \frac{d E_\gw (\vp, f_s)}{df_s}. 
\end{equation} 

Note that there is direct dependence on $H_0$ in Eq.~\eqref{eq:Omega_GWB_def_02}. This arises from Eq.~\eqref{eq:Omega_GWB_def_01} since $\mathcal{F} \sim H_0^{-2}$ (flux scales inversely with comoving area), canceling with $\rho_c^{-1} \sim H_0^2$, leaving $d\dot{N}^\text{o} \sim dV_C \sim H_0^{-3}$ for comoving volume element $dV_C$.
We provide a detailed derivation of Eq.~\eqref{eq:Omega_GWB_def_02} in~\cite{Cousins2025prl_suppmat}, but qualitatively outline the cosmological dependence here as follows.
The value of $H_0$ is inversely-proportional to the comoving distance between points in space at a given redshift (see \cite{hogg1999distance} and standard texts, e.g.,~\cite{dodelson2003modern}). 
A higher value of $H_0$ indicates a smaller comoving distance, so fewer mergers would be encompassed in any given comoving volume of space under a fixed merger rate density. Therefore, a universe with a higher value of $H_0$ would possess fewer mergers contributing to $\Omega_\gw$, leading to a smaller, harder-to-detect GWB signal. The inverse is also true, meaning that ruling out higher values of $\Omega_\gw$ would correspondingly rule out lower values of $H_0$. 
Thus, the GWB can inform measurements of $H_0$ as a ``stochastic siren'' even if it has not yet been detected.

\noindent \textit{Likelihood construction.} GW merger observations follow an inhomogeneous Poisson distribution characterized by the total observer-frame merger rate across the Universe, $R$, and the model hyperparameters, $\vec\lambda$, that affect $R$.
Taking the fraction of these mergers that are observable as $P_{\rm det}(\vec{\lambda})$, the likelihood of observing a catalog of foreground GW events is given by~\citep{taylorminingGWcatalogs2018, mandelextractingdistributionparameters2018, Vitaleinferringpopulation2022}
\begin{equation}
    \mathcal{L}_\FG (\{d\} | \vec{\lambda}) \propto N^{N_{\rm obs}} e^{-N P_{\rm det}(\vec{\lambda})} \prod_{i=1}^{N_{\mathrm{obs}}} \mathcal{L}(d_i | \vec{\lambda}), 
\end{equation}
where $N = R \, T$ is the total number of mergers occurring during observing time $T$, and the per-event likelihood terms in the product are generally calculated by explicitly marginalizing over all possible signals assuming Gaussian detector noise.
We emphasize that we use a formulation of the foreground likelihood with explicit rate dependence because the rate is also explicitly required in the background likelihood.
Note that while $P_{\text{det}}$ can introduce inconsistencies in population inference~\citep{Farr2019selection,essick2024ensuring}, this can be mitigated by evaluating $P_{\text{det}}$ empirically via a synthetic injection catalog analysis, as is performed by standard analysis software such as \texttt{ICAROGW}~\cite{mastrogiovanniICAROGWPythonPackage2023}.

Searches for an astrophysical GWB typically assume that the background is stationary, Gaussian, unpolarized, and isotropic, although see~\cite{drasco2003DetectionMethods,smith2018OptimalSearch} for alternative approaches.
Such a GWB can be searched for by cross-correlating the strain measured by different pairs of detectors within the LVK network. 
Following~\cite{allenDetectingStochasticBackground1999} and~\cite{romanoDetectionMethodsStochastic2017}, for a baseline comprising the detectors $I$ and $J$, we define the following cross-correlation statistic, 
\begin{equation}\label{eq:C_f}
\hat{C}(f) = \frac{1}{\Delta t} \frac{20\pi^2}{3H_0^2} f^3 \tilde{s}_I(f) \tilde{s}_J^*(f), 
\end{equation}
where $\tilde{s}_I(f)$ and $\tilde{s}_J(f)$ are respectively the short-time Fourier transform of the strain measured by the detectors $I$ and $J$, and $\Delta t$ is the segment time length. 
The background likelihood ($\mathcal{L}_\BG$) of observing this cross-correlation statistic for the background given some $\vec{\lambda}$ is well approximated by a Gaussian~\citep{mandic2012spectralparameterGWB, callisterPolarizationBasedTestsGravity2017, callisterShoutsMurmursCombining2020}:
\begin{equation}\label{eq:likelihood_GWB}
        \mathcal{L}_\BG(\hat{C} | \vec{\lambda})
        \propto \exp \left[ -\frac{1}{2} \left(\hat{C} - \gamma \Omega_{\gw}(\vec{\lambda}) | \hat{C} - \gamma \Omega_{\gw}(\vec{\lambda}) \right)\right]\,,
\end{equation}
where $\gamma(f)$ is the overlap reduction function that gives the sensitivity of a particular pair of detectors to an isotropic background based on their relative geometry~\citep{ChristensenmeasuringGWB1992, FlanagansensitivityligoGWB1993, allenDetectingStochasticBackground1999}. 
The inner product in this likelihood is defined as 
 \begin{equation} (A|B) = 2T \left(\frac{3H_0^2}{10\pi^2} \right)^2 \int_0^\infty df \frac{\tilde{A}(f) \tilde{B}^*(f)}{f^6 P_I(f) P_J(f)}, 
\label{eq:innerProd}
\end{equation}
with $P_I(f)$ the one-sided noise power spectral density of detector $I$.
Note that this likelihood depends on $H_0$ through the inner product, the cross-correlation statistic, and the energy density of the GWB.

Following~\cite{callisterShoutsMurmursCombining2020}, we assume the foreground and background data are independent.
In practice, there is some cross-contamination due to the observed mergers contributing to the ``background'' signal unless they are explicitly removed. However, this effect is small for the GWTC-3 dataset we consider: in~\cite{Cousins2025prl_suppmat}, we calculate that the individually-resolved BBHs contribute to $\Omega_\gw$ at $\sim 0.3\%$ of the GWB upper limits. They hence can be excluded, allowing the joint likelihood to be written as
\begin{equation}
    \mathcal{L}_\joint = \mathcal{L}_\FG \mathcal{L}_\BG.\label{eq:Ljoint}
\end{equation}

\noindent \textit{Population models.} To apply the stochastic siren method to GW data, the source energy spectrum in Eq.~\eqref{eq:dEdf} and the rate density $\mathcal{R}(z)$ in Eq.~\eqref{eq:Omega_GWB_def_02} must be computed to obtain $\Omega_\gw$. We address this using parametrized models for the source redshift and component mass distributions involving 12 hyperparameters (described further in~\cite{Cousins2025prl_suppmat}). For Eq.~\eqref{eq:dEdf}, we exclusively consider BBHs since they form the majority of detected compact binaries and their population properties are better-understood than other merger types~\citep{KAGRA:2021gwtc3population}. Thus, we can approximate the merger energy spectra as that of an inspiralling circular binary in the Newtonian limit with source-frame masses $m_1$ and $m_2$ (as in, e.g., \cite{thorne1987gravitational}):
\begin{equation}
    \frac{dE_\gw}{df_s} = \frac{(\pi G)^{2/3} }{3 f_s^{1/3}} \frac{m_1 m_2}{\,\,\,\,\,(m_1+m_2)^{1/3}}. 
\end{equation} 

To model $\mathcal{R}(z)$, we follow~\cite{callisterShoutsMurmursCombining2020} and fit the redshift distribution of merging binaries with a functional form of the Madau-Dickinson star formation rate~\citep{madau2014cosmic}. This four-parameter model possesses a smooth transition between two power-law distributions with a peak at redshift $z_p$ and a local merger rate density $R_0 = \mathcal{R}(z=0)$.
As motivated by GWTC-3 observations~\citep{theligoscientificcollaborationBinaryBlackHole2019}, we model the BBH mass distribution using the {\tt PowerLaw+Peak} model~\citep{talbotMeasuringBlackHoleMass2018,talbot2024gwpopulation}, an eight-parameter function consisting of a truncated power-law component and a Gaussian component for the primary mass and a power-law distribution of mass ratios.
Both the primary mass and mass ratio distributions are smoothly tapered at low masses.

We neglect correlations between the mass and redshift distributions. Some correlation between these parameters is expected due to metallicity variation of progenitor stars~\citep{belczynski2010effect} and variation in the delay time of BBH formation and merger~\citep{kushnir2016gw150914,gallegos2021binary,van2022redshift}, which may influence population parameter inference~\citep{rinaldi2024evidence,torniamenti2024hierarchical} and thus may affect resulting $H_0$ constraints. However, recent work suggests that such a correlation is unlikely to be relevant or required for BBH data as of GWTC-3~\citep{LallemanBBHdistributionredshift2025}.

Beyond the 12 population parameters described above, we include three cosmological parameters. As in~\cite{abbott2023cceh}, we consider $H_0$, the present-day matter energy density $\Omega_{m,0}$, and one parameter for the dark energy equation of state $w(z) = w_0$ (which enters $\Omega_\gw$ via $E(z)$). This yields a total of 15 parameters for our full inference. 

\noindent \textit{Posterior distribution construction.} To measure $H_0$, each likelihood must be computed given available data.
For $\mathcal{L}_\FG$, we must evaluate the likelihood of observed BBHs given $\vec\lambda$. For $\mathcal{L}_\BG$, we must compute $\Omega_\gw$ given $\vec\lambda$ and determine the associated likelihood of that value of $\Omega_\gw$.

To evaluate $\mathcal{L}_\FG$, we use the \texttt{ICAROGW}~\citep{mastrogiovanniICAROGWPythonPackage2023} hyperparameter posteriors from the GWTC-3 population analysis~\citep{KAGRA:2021gwtc3population, abbott2023cceh}, which considered BBH mergers with signal-to-noise ratios (SNR) greater than $11$.
We generate a 15-dimensional kernel density estimator (KDE) using these posteriors, providing a probability function of a given $\vec\lambda$.

For $\mathcal{L}_\BG$, we consider a power-law form of $\Omega_\gw$ with spectral index $\alpha=2/3$, corresponding to a compact binary GWB~\citep{regimbauAstrophysicalGravitationalWave2011}, with a reference frequency of $25\,\rm Hz$ as in GWTC-3 GWB searches~\citep{KAGRA:2021gwtc3population,abbottUpperLimitsIsotropic2021}. 
We note that $\mathcal{L}_\BG$ can be evaluated using the existing posteriors of $\Omega_\gw$ obtained from previous GWB searches~\citep{abbottUpperLimitsIsotropic2021}, if we prescribe a uniform prior on $\Omega_\gw$. 
According to Bayes' theorem:
\begin{equation}
p_\BG(\Omega_\gw| \hat{C}) \propto \pi_\BG(\Omega_\gw) \mathcal{L}_\BG(\hat{C}|\Omega_\gw),
\label{eq:bayes-omegaGW1}
\end{equation}
where $p_\BG$ and $\pi_\BG$ are the posterior and prior distributions for $\Omega_\gw$, respectively. If we consider a uniform prior, then we can invert Eq.~\eqref{eq:bayes-omegaGW1} for the likelihood:
\begin{equation}
\mathcal{L}_\BG(\hat{C}|\Omega_\gw) \propto p_\BG(\Omega_\gw| \hat{C}).
\label{eq:bayes-omegaGW2}
\end{equation}
We obtain $p_\BG$ using the posterior of $\Omega_{\alpha = 2/3}$ constructed from the data obtained during the first three LVK observing runs~\citep{abbottUpperLimitsIsotropic2021}, under a uniform $\pi_\BG$, allowing us to compute $\mathcal{L}_\BG$. We use the $\Omega_{\alpha = 2/3}$ posterior samples to generate a 1-dimensional KDE to estimate $\mathcal{L}_\BG$ as a function of $\Omega_\gw$ within the range covered by the prior of $\Omega_{\alpha = 2/3}$. 
Then, since $\Omega_\gw$ is just a function of $\vec\lambda$, we can evaluate $\mathcal{L}_\BG$ using population and cosmological parameters.

These two individual likelihoods compose the joint likelihood of Eq.~\eqref{eq:Ljoint} as a function of 15 (hyper)parameters. We perform nested sampling~\citep{skilling2004nested,skilling2006nested} over these parameters to compute $\mathcal{L}_\joint$ using the \texttt{dynesty} sampler~\citep{Speagle:2019ivv}, as implemented in the \texttt{bilby} package \citep{Ashton:2018jfp}.

\begin{figure*}[tp!]
  \includegraphics[width=0.9\textwidth]{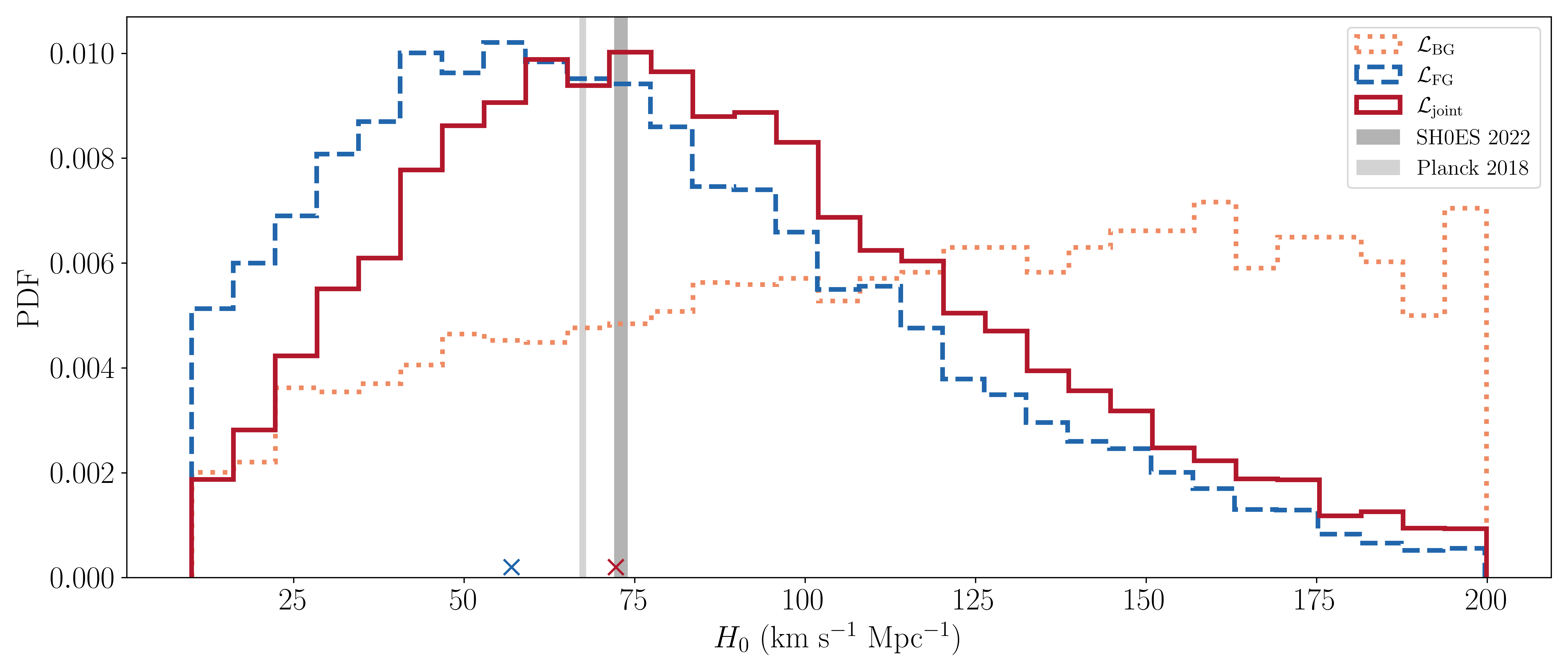}
  \caption{Posterior distributions of the Hubble constant obtained with the spectral siren ($\mathcal{L}_\FG$, dashed blue), the GWB search ($\mathcal{L}_\BG$, dotted orange), and the stochastic siren joint measurement ($\mathcal{L}_\joint$, red), each respectively marginalized over their 14 other parameters. Since no GWB has been detected, $\mathcal{L}_\BG$ shows less support for lower values of $H_0$. Combining this with the foreground shifts the posterior, as indicated by the maximum a-posteriori probabilities for $\mathcal{L}_\FG$ and $\mathcal{L}_\joint$ shown by $\times$ markers (no marker is shown for $\mathcal{L}_\BG$ since the GWB non-detection provides only a lower bound on $H_0$). The joint measurement shifts slightly closer to the Hubble tension region vertically demarcated by the {\em Planck} \citep{aghanim2020planck} and SH0ES \citep{Riess:2021jrx} values.
  }
  \label{fig:H0-all}
\end{figure*}

\noindent \textit{Analysis of GW observing runs.} We present the results of applying the stochastic siren method to GWTC-3 data in Fig.~\ref{fig:H0-all}. In addition to the full $\mathcal{L}_\joint$ result (solid red), we compare with the results obtained when considering only the foreground ($\mathcal{L}_\FG$, dashed blue) and background ($\mathcal{L}_\BG$, dotted orange) populations.
Since a uniform prior is prescribed for the population-model hyperparameters and cosmological parameters, all likelihood distributions are proportional to the corresponding posterior probability distribution, which is a function of $H_0$. 
We show the $H_0$ values from {\em Planck}\/~\citep{aghanim2020planck} and SH0ES~\citep{Riess:2021jrx} as vertical light- and dark-gray regions, respectively.

Observe that the measurement accuracy of $H_0$ from $\mathcal{L}_\joint $ is improved over that from $\mathcal{L}_\FG $. 
Specifically, the $H_0$ posterior obtained via $\mathcal{L}_\joint $ peaks at a value that is closer to both the {\em Planck}\/ and SH0ES values, which indicates improved consistency between GW and electromagnetic measurements of $H_0$. 
This can be seen when comparing the maximum a-posteriori probabilities of each method: $72^{+44}_{-37}\, \rm km~s^{-1}~Mpc^{-1}$ for $\mathcal{L}_\joint$ versus $57^{+43}_{-35} \, \rm km~s^{-1}~Mpc^{-1}$ for $\mathcal{L}_\FG$ or $46^{+49}_{-26} \, \rm km~s^{-1}~Mpc^{-1}$ from the spectral siren~\cite{abbott2023cceh}. Each posterior's $68.3\%$ highest density interval is quoted as its uncertainty.
The full results for all 15 (hyper)parameters are contained in~\cite{Cousins2025prl_suppmat}.

\noindent \textit{Discussion.} We introduced the ``stochastic siren'' measurement of the Hubble constant and applied it by combining the resolved binary black hole (BBH) population with the search for the gravitational-wave background (GWB). 
We have shown that the GWB's cosmological dependence allows it to aid in the measurement of $H_0$, as evidenced in our analysis of the LVK's first three observing runs. We found increased accuracy versus the spectral siren method, even though the GWB has yet to be detected.
This improvement is characterized by a shift of the maximum a-posteriori probability from lower $H_0$ values to values more consistent with electromagnetic measurements.
Note that this is a conservative result since we considered only BBHs, while the GWTC-3 GWB search results~\cite{abbottUpperLimitsIsotropic2021} are for all merger types and thus favor values of $\Omega_\gw$ higher than what BBHs alone could produce. Lower values of $\Omega_\gw$ are hence less probable, increasing the posterior support for higher values of $H_0$.

This could be addressed in future extensions of our current stochastic siren methodology. Other mergers classes can contribute to the GWB comparably to BBHs~\citep{abbottUpperLimitsIsotropic2021,KAGRA:2021gwtc3population}, which would enhance the stochastic siren method at the cost of additional population modeling and observations. We also considered only a single set of BBH population models in this work, but a variety of population models could be used. Aside from compact object mergers, other GW sources could produce observable GWBs~\citep{Ferrari:1998ut, Chowdhury:2024fdr}, which could likewise be included.
Finally, as in \cite{ferraiuolo2025inferring}, the proper integration of our method with existing population, stochastic, and cosmology analysis systems in the LVK \citep[e.g.,][]{Fischbach:stochmon,mastrogiovanniICAROGWPythonPackage2023,gray2023joint,renzini2023pygwb,talbot2024gwpopulation,renzini2024popstock} would help improve cosmological constraints as LVK observing runs continue.

As current observing runs progress, a continued non-detection of the GWB should raise the lower bound of $H_0$, thereby progressively improving the precision of the measurement.
This may allow the stochastic siren to serve a unique role in the Hubble tension by probing the lower early-universe measurement of $H_0$ in advance of the higher late-universe measurements.
Moreover, the LVK detectors are expected to observe the GWB in coming years \citep{LIGOScientific:2016fpe, LIGOScientific:2017zlf, renziniStochasticGravitationalWaveBackgrounds2022}, at which point the stochastic siren would yield even stronger improvements on the measurement of $H_0$. Our current projections~\citep{Cousins2025prl_suppmat} suggest that the GWB could be detected with an SNR of eight in less than a year using the 
anticipated A\# detector upgrades. 
In these projections, we also consider a variety of different cosmological and population parameters, showing what regions of parameter space will be probed by the detection or non-detection of the GWB.
This highlights that the GWB is sensitive to the full expansion history of the Universe and thus depends on other cosmological parameters such as $\Omega_m$ and $w_0$. While we show in ~\citep{Cousins2025prl_suppmat} that the current GWB non-detection cannot constrain these parameters, the stochastic siren could in principle aid in these measurements.

The future utility of combining resolved and unresolved BBHs for population and cosmological inference should be further explored through mock data analyses. A first step in this direction is made by~\cite{ferraiuolo2025inferring}, wherein this method is applied to simulations of two years of observations during a future LVK O5 observing run, further improving $H_0$ measurement accuracy, albeit with a different mass model and a fixed value of $w_0$. The full effect of the stochastic siren could be further explored via a mock data analysis that includes the detection of the GWB, such as during the post-O5 A\# detector upgrades~\cite{fritschel2022asharp}.

While upcoming LVK observing runs and upgrades are expected to enable the detection of the GWB, next-generation ground-based detectors are expected to come online soon after~\citep{punturoEinsteinTelescopeThirdgeneration2010,reitze2019cosmic,maggioreScienceCaseEinstein2020,evansHorizonStudyCosmic2021,Evans:2023euw,Branchesi:2023mws}.
However, the GWB observed by these future detectors is expected to be non-Gaussian and non-stationary \citep{Buscicchio:2022raf}, which would require modifications to the current approach of GWB analysis. Moreover, the assumption of independent likelihoods used to construct the joint likelihood (Eq.~\eqref{eq:Ljoint}) may also need to be revised given the greater number of resolved mergers.

Thus, there are various avenues to extend the stochastic siren approach, but it can already inform current investigations of the Hubble tension as an entirely novel cosmological probe.
It is one of the very few methods that can measure $H_0$ entirely independently of electromagnetic observations, allowing it to complement both GW and electromagnetic $H_0$ measurements. 
This will culminate with the eventual detection of the GWB and its resulting cosmological constraints, framing the stochastic siren as a promising tool for cosmology.

\begin{acknowledgements}
\noindent \textit{Acknowledgements.}~We thank the authors of \cite{ferraiuolo2025inferring} for constructive coordination and feedback during internal LVK review of this work. We also thank the anonymous referees for their helpful feedback. B.C.~acknowledges that this material is based upon work supported by the NSF Graduate Research Fellowship Program under Grant No. DGE 21-46756. 
K.~S.~would like to acknowledge the NSF Graduate Research Fellowship Program under Grant No. DGE–1746047 and the NSF under award PHY-2207650. 
A.K.W.C and N.Y. acknowledge the support from the Simons Foundation through Award No. 896696, the NSF through award PHY-2207650 and NASA through Grant No. 80NSSC22K0806. 
C.~T.~and T.~C.~are supported by the Eric and Wendy Schmidt AI in Science Postdoctoral Fellowship, a Schmidt Sciences program.
D.E.H is supported by NSF grant  PHY-2110507, and by the Kavli Institute for Cosmological Physics through an endowment from the Kavli Foundation and its founder Fred Kavli.
This work made use of the Illinois Campus Cluster, a computing resource that is operated by the Illinois Campus Cluster Program (ICCP) in conjunction with the National Center for Supercomputing Applications (NCSA) and which is supported by funds from the University of Illinois at Urbana-Champaign.
This material is based upon work supported by National Science Foundation (NSF) LIGO Laboratory, which is a major facility fully funded by the NSF.
This work made use of \texttt{astropy}~\citep{astropy2013,astropy2018,astropy2022}, \texttt{bilby}\citep{Ashton:2018jfp},
\texttt{h5py}~\citep{collette2013python}, \texttt{matplotlib}~\citep{hunter2007matplotlib}, \texttt{numpy}~\citep{harris2020numpy}, and \texttt{scipy}~\citep{virtanen2020scipy}.
The Supplemental Material~\citep{Cousins2025prl_suppmat} includes additional citations not referenced in the main text~\citep{renzini2024projections,meacher2015mock,perkins2021improved,linderExploringExpansionHistory2003,zhuGravitationalWaveBackground2013,christensenParameterEstimationGravitational2022,perigoisStartrackPredictionsStochastic2021,ligoGWBimplications:2017,flanagan1998measuring,wu2012accessibility,VoyagerAdhikariCryogenicSiliconInterferometer2020,ade2016planck}.
\end{acknowledgements}

% \bibliography{bibliography}
% \nocite{*}

% switch to one-column: jolcadtips.blogspot.com/2016/01/make-one-column-appendix-in-two-column.html
\pagebreak
\onecolumngrid

% manually set the appendix title
\renewcommand\appendixname{{\large \textbf{Supplemental Material for ``The Stochastic Siren: Astrophysical Gravitational-Wave Background Measurements of the Hubble Constant''}}}
\begin{center}
\appendixname
\end{center}

\section{Cosmology and the GWB}\label{app:cosmo}

\subsection{Definitions}\label{subapp:cosmo-defn}
We first make cosmological definitions with the Hubble parameter $H(z)$ and present-day cosmological density parameters for radiation, matter, curvature, and dark energy ($\Omega_{r,0},\, \Omega_{m,0},\, \Omega_{k,0},\, \mathrm{and}\, \Omega_{DE,0}$ respectively, with dark energy's energy density $\rho_{DE}$) for a generic wCDM model:
\begin{equation}
H(z) = H_0 \left[(1+z)^4 \Omega_{r,0} +  (1+z)^3 \Omega_{m,0} + (1+z)^2 \Omega_k 
+ \Omega_{DE,0} \frac{\rho_{DE}(z)}{\rho_{DE,0}}
\right]^{1/2},
\end{equation}
which allows us to define $E(z)$ as:
\begin{equation}
E(z) \equiv \frac{H(z)}{H_0} = \left[ (1+z)^4 \Omega_{r,0} + (1+z)^3 \Omega_{m,0} + (1+z)^2 \Omega_{k,0}
+ \Omega_{DE,0} \frac{\rho_{DE}(z)}{\rho_{DE,0}}\right]^{1/2}.\label{eq:Ez1}
\end{equation}
In our analysis, we consider only a flat universe with negligible radiation, so $\Omega_{k,0} = \Omega_{r,0} = 0$, then $\Omega_{DE,0} = 1-\Omega_{m,0}$. Eq.~\eqref{eq:Ez1} then becomes
\begin{equation}
    E(z) \equiv \frac{H(z)}{H_0} = \left[ (1+z)^3 \Omega_{m,0} + (1-\Omega_{m,0}) \frac{\rho_{DE}(z)}{\rho_{DE,0}}\right]^{1/2}.\label{eq:Ez2}
\end{equation}

We next define the requisite distance measures:
\begin{itemize}
    \item Hubble distance: \begin{equation}D_H \equiv c/H_0\end{equation}\label{eq:DH}
    \item line-of-sight comoving distance: \begin{equation}D_C = \frac{c}{H_0} \int_0^z{\frac{dz}{E(z)}} = D_H \int_0^z{\frac{dz}{E(z)}}\end{equation}
    \item transverse comoving distance:
    \begin{equation}
    D_{\mathrm{M}}=\left\{
    \begin{array}{l}
        {{D_{\mathrm{H}}\ \frac{1}{\sqrt{\Omega_{k}}}\mathrm{~sinh}\left[\sqrt{\Omega_{k}}\,D_{\mathrm{C}}/D_{\mathrm{H}}\right]}} {{\mathrm{~for~}\Omega_{k} > 0}}\\ D_C {{\mathrm{~for~}\Omega_k=0}}\\ {{D_{\mathrm{H}}\ \frac{1}{\sqrt{|\Omega_{k}|}}\,\mathrm{~sin}\left[\sqrt{|\Omega_{k}|}\,D_{\mathrm{C}}/D_{\mathrm{H}}\right]}} {{\mathrm{~for~}\Omega_{k} < 0}}
    \end{array}\right.
    \end{equation}
    \item angular diameter distance:
    \begin{equation} D_{\mathrm{{A}}}={\frac{D_{\mathrm{{M}}}}{1+z}} = \frac{D_C}{1+z} \,\, \mathrm{when} \,\, \Omega_k=0 \end{equation}
    \item luminosity distance:
    \begin{equation}\label{eq:DL} D_L = (1+z) D_M = (1+z) D_C  \,\, \mathrm{when} \,\, \Omega_k=0 \end{equation}
   
\end{itemize}

\subsection{E(z) for parametrized dark energy}\label{subapp:Ez-w}

Using the dark energy equation-of-state parameter $w(z)$, we follow~\cite{linderExploringExpansionHistory2003} and write the dark energy evolution term in Eq.~\eqref{eq:Ez2} as
\begin{equation}
\frac{\rho_{DE}(z)}{\rho_{DE,0}} =
\exp{\left[
3 \int_0^{\ln(1+z)} d\ln(1+z') [1+w(z')]
\right]}
.\label{eq:Ez-w0_1}
\end{equation}
We next modify this term using a parametrization of $w(z)$. In general, $w(z)$ can be solved for under a specific scalar field theory, or is more generically parametrized. In this work, however, we assume a constant equation of state parameter ($w(z) = w_0$, as in~\cite{abbott2023cceh}). Then, the integral in the exponential can be solved analytically: let $y \equiv \ln(1+z)$, so the integral becomes
\begin{equation}
\left[
\int_0^y dy (1+w_0)
\right]
= y(1+w_0)
= \ln(1+z) (1+w_0),
\end{equation}
which allows the exponential to be rewritten as
\begin{equation}
 \exp[3 \ln(1+z) (1+w_0)]
= (1+z)^{3(1+w_0)}.
\end{equation}
Equation~\eqref{eq:Ez2} is then simply
\begin{equation}
E(z, w_0, \Omega_{m,0}) = 
\left[ 
(1+z)^3 \Omega_{m,0} + (1-\Omega_{m,0})
(1+z)^{3(1+w_0)}
\right]^{1/2}.\label{eq:Ez-w0_2}
\end{equation}

\subsection{Derivation of the GWB energy density}\label{subapp:omegaGW}
We derive $\Omega_{\mathrm{gw}}$ as arising from astrophysical mergers, generally following the procedure of earlier work~\citep{ferrariStochasticBackgroundGravitational1999,regimbau2008astrophysical,regimbauAstrophysicalGravitationalWave2011,zhu2011agwb2,wu2012accessibility} while avoiding the occasional error of incorrect cosmological distances and/or frames that can result in incorrect factors of $(1+z)$ as noted in~\cite{zhuGravitationalWaveBackground2013}. Note that for consistency with current LVK convention, we express frequencies as $f$ instead of the occasionally-used $\nu$; we hence express the fluence here as $\mathcal{F}$ instead of $f$.

The dimensionless energy density $\Omega_{\mathrm{gw}}$ for an astrophysical GWB is typically expressed as:
\begin{equation}
\Omega_{\mathrm{gw}}={\frac{f}{\rho_c c^3}} F(\vp,f)\label{eq:omegaGW0}
\end{equation}
where $ f = f_s/(1+z)$ is the observer-frame GW frequency in terms of the source-frame frequency $f_s$ and $F$ is the integrated flux of the astrophysical sources:
\begin{equation}
F(\vp,f)=\int p(\vp) \int \mathcal{F}(\vp, f) \frac{d \dot{N}^\text{o}(\vp,\ z)}{d z}d\vp d z,\label{eq:int-flux1}
\end{equation}
for source parameters $\vp$ and associated probability distribution $p(\vp)$, source fluence $\mathcal{F}(\vp, f)$ (flux times time), and the total number of events per unit observer time per redshift interval ${d \dot{N}^\text{o}(\vp,z)}/{d z}$.

We first consider the fluence, which is the energy per unit area $A$ per unit frequency. We take its definition (in source frame and frequency) as Eq.~(2.45) of~\cite{flanagan1998measuring}:
\begin{equation}
\mathcal{F}(\vp,f) \equiv \frac{dE}{dAdf} = 
\frac{(1+z)^2}{D_L^{2}}\frac{d E_{\mathrm{gw}} (\vp,f_s)}{d\Omega d f_s} 
= \frac{(1+z)^2}{(1+z)^2 D_C^{2}}\frac{d E_{\mathrm{gw}} (\vp,f_s)}{d\Omega d f_s}
= \frac{1}{ D_C^{2}}\frac{d E_{\mathrm{gw}} (\vp,f_s)}{d\Omega d f_s}\label{eq:fluence}
\end{equation}
where we have used the luminosity distance as defined in Eq.~\eqref{eq:DL} and introduced the solid angle differential $d\Omega$. Note here that the emitted gravitational spectral energy is ${dE_{\mathrm{gw}}}(\vp, f)/{d f_s}$, as used in Eq.~(3) of the main text.

We next consider the total number of events per unit \textit{observer} time, per redshift interval,
\begin{equation}
\frac{d \dot{N}^\text{o}(\vp,\ z)}{d z} = \mathcal{R}^\text{o}(\vp,z)\frac{d V}{d z}(z),\label{eq:dNdot0}
\end{equation}
for observer-frame comoving event rate density $\mathcal{R}^\text{o}$ and comoving volume element
\begin{equation}
d V_{\mathrm{C}}
= D_{\mathrm{H}}\,\frac{(1+z)^{2}\,D_{\mathrm{A}}^{2}}{E(z)}\,d\Omega\,d z
= \frac{c}{H_0} \frac{D_C^2}{E(z)} \, d \Omega d z.\label{eq:dVc}
\end{equation}
Combining Equations \eqref{eq:dNdot0} and \eqref{eq:dVc} yields
\begin{equation}
\frac{d \dot{N}^\text{o}(\vp,\ z)}{d z} = \mathcal{R}^\text{o}(\vp,z) \frac{c}{H_0} \frac{D_C^2}{E(z)} \, d \Omega.\label{eq:dNdot1}
\end{equation}

To match current literature, we next convert to \textit{source} time. This requires redshifting to the source-frame comoving rate density $\mathcal{R} = (1+z)\mathcal{R}^\text{o}$ (since time redshifts as $t_s = t/(1+z)$), so we rewrite Eq.~\eqref{eq:dNdot1} as
\begin{equation}
\frac{d \dot{N}^\text{o}(\vp,\ z)}{d z} =
\frac{\mathcal{R}(\vp,z)}{(1+z)} \frac{c}{H_0} \frac{D_C^2}{E(z)} \, d \Omega.\label{eq:dNdot2}
\end{equation}

We then assemble the integrated flux by inserting Eq.~\eqref{eq:fluence} and Eq.~\eqref{eq:dNdot2} into Eq.~\eqref{eq:int-flux1}:
    \begin{align}
    F(\vp,f)
    &= \int p(\vp) d\vp \int \frac{1}{ D_C^{2}}\frac{d E_{\mathrm{gw}} (\vp,f_s)}{d\Omega d f_s} \frac{\mathcal{R}(\vp,z)}{(1+z)} \frac{c}{H_0} \frac{D_C^2}{E(z)} \, d \Omega \nonumber \\
    &= \frac{c}{H_0} \int \frac{\mathcal{R}(z)}{ (1+z) E(z)} \int p(\vp) \frac{d E_{\mathrm{gw}} (\vp, f_s)}{d f_s} d\vp dz \nonumber \\
    &= \frac{c}{H_0} \int \frac{\mathcal{R}(z)}{ (1+z) E(z)} \biggr \langle \frac{dE_\gw}{d f_s} \biggr \rangle \biggr |_{f_s} d z \, ,\label{eq:int-flux2}
    \end{align}
where we obtain the second line by assuming no correlation between source parameters and redshift ($\mathcal{R}(z,\vp)=\mathcal{R}(z)$) as elaborated in the main text. We obtain the third line by defining 
\begin{equation}
\biggr \langle \frac{dE_\gw}{d f_s} \biggr \rangle \biggr |_{f_s} \equiv \int p(\vp) \frac{d E_{\mathrm{gw}} (\vp,f_s)}{d f_s} d\phi\,.
\end{equation}

Finally, by inserting Eq.~\eqref{eq:int-flux2} into Eq.~\eqref{eq:omegaGW0}, we find that the GW energy density is
\begin{equation}
\boxed{
    \Omega_{\mathrm{gw}} 
    = \frac{f}{ \rho_{c} c^2 H_0} \int \frac{\mathcal{R}(z)}{(1+z) E(z)} \biggr \langle \frac{dE_\gw}{d f_s} \biggr \rangle \biggr |_{f_s} d z\label{omegagw_notheta}
    },
\end{equation}
which yields Eq.~(2) in the main text when using the definition of the critical mass density $\rho_c$.

\section{Parameter Inference Methods and Results}\label{app:results}

\subsection{Contribution of resolved mergers to $\Omega_\gw$}
In order for an analysis using a joint likelihood factorized into $\mathcal{L}_\BG$ and $\mathcal{L}_\FG$ to be valid, the magnitude of the resolved foreground's contributions to $\Omega_\gw$ must be negligible relative to the total $\Omega_\gw$. To quantify this contribution, we discretely computed the $\Omega_\gw$ contribution of the 42 resolved BBHs considered in our work following the formalism outlined in~\cite{meacher2015mock,renzini2024projections}.
Specifically, we write the integrated flux in Eq.~\eqref{eq:omegaGW0} as
\begin{equation}\label{eq:omegaGW-flux}
F(f)=T_{\mathrm{obs}}^{-1}\frac{\pi c^{3}}{2G}f^{2}
\sum_{k=1}^{N}(|\tilde{h}_{+,k}|^{2} + |\tilde{h}_{\times,k}|^{2})
\end{equation}
where $T_{\rm{obs}}$ is the observing time and $\tilde{h}^2_{+,k},\,\tilde{h}^2_{\times,k}$ are respectively the Fourier transforms of the plus and cross strain polarizations for the $k^{\rm{th}}$ merger, out of $N$ total mergers.
In the circular Newtonian regime (containing the GW frequency of interest, $25$ Hz), the Fourier-transformed strain polarizations are
\begin{align}\label{eq:h-plus}
    &|\tilde{h}_{+}(f)| = h_{z}{\frac{(1+\cos^{2}\iota)}{2}}f^{-7/6}\\
    &|\tilde{h}_{\times}(f)| = h_{z}\cos\iota f^{-7/6}.\label{eq:h-cross}
\end{align}
Here, the amplitude of the signal is
\begin{equation}\label{eq:h-amp}
h_{z}=\sqrt{\frac{5}{24}}\frac{[G \mathcal{M}(1+z)]^{5/6}}{\pi^{2/3}c^{3/2}D_{\mathrm{L}}(z)},
\end{equation}
for source-frame chirp mass $\mathcal{M}$.

Inserting Eq.~\eqref{eq:omegaGW-flux}-\eqref{eq:h-amp} into Eq.~\eqref{eq:omegaGW0} yields 
\begin{equation}\label{eq:omega-discrete}
    \Omega_\gw(f, \mathcal{M}, z, \iota) =
    \frac{5 \pi^{2/3} G^{5/3}}{18 c^3 H_0^2} f^{2/3}
    \sum_{k=1}^N \left\{ \frac{ \left[ \mathcal{M}_k (1+z_k)\right]^{5/3}}{D_L(z_k)^2}
    \left[\frac{(1+\cos^2\iota_k)^2}{4} + \cos^2 \iota_k\right]
    \right\}.
\end{equation}
In this form, $\Omega_\gw$ can be computed a function of the source parameters for each individual merger.
Using the posterior medians from the GWTC-3 catalog~\cite{abbott2023gwtc3} for the chirp masses, redshifts, and inclinations of the 42 BBHs involved in our work (with $T_{\rm{obs}}=2$ years given $\sim4$ months of O1, $\sim9$ months of O2, and $\sim 11$ months of O3), we found that their contribution is $\sim 2.3 \times 10^{-11}$ at $25$ Hz.
Compared with the most recent estimate for the total $\Omega_\gw \sim 6.9\times 10^{-10}$ at $25$ Hz~\cite{KAGRA:2021gwtc3population}, the resolved mergers contribute $\sim 3.3\%$ to the expected total energy density. 
This means that the resolved mergers' contribution is $\sim0.2\%$ of the GWTC-3 upper limit result~\cite{abbottUpperLimitsIsotropic2021} that we considered in our analysis ($\sim 1.2 \times 10^{-8}$) and hence is largely negligible for the purposes of our inference.

Furthermore, the upper limit is a factor of $\sim 17 \times$ higher than the expected total $\Omega_\gw$~\cite{KAGRA:2021gwtc3population} ($\sim 1.2 \times 10^{-8}$ versus $\sim 6.9\times 10^{-10}$).
This implies that the upper limit data will result in conservative (higher) measurements of $H_0$.
This is because the unresolved contribution to $\Omega_\gw$ will always be lower than the total (resolved plus unresolved), and hence the upper limits (which are on the total $\Omega_\gw$, not just the unresolved component) will favor values of $\Omega_\gw$ higher than what these unresolved mergers could produce alone.
Given this, there is more posterior support for lower values of $H_0$ (i.e., a worse measurement, in the same way that considering only BBHs leads to conservative results when compared to including BNSs and NSBHs).

\subsection{Population models}

We describe the mass and redshift population models used in our inference, which we selected in order to compare with previous results~\citep{abbott2023cceh}. For a description of the \texttt{PowerLaw+Peak} mass model, we refer to existing sources \citep{talbotMeasuringBlackHoleMass2018,theligoscientificcollaborationBinaryBlackHole2019,talbot2024gwpopulation} and its implementation via equations B.10-12 in~\cite{mastrogiovanniICAROGWPythonPackage2023}. For the redshift model, we use a functional form of the Madau-Dickinson model~\citep{madau2014cosmic} as parametrized in~\cite{callisterShoutsMurmursCombining2020}:
\begin{equation}
    \mathcal{R}(z) = [1\ +(1\ +z_{p})^{-\gamma-\kappa}]\frac{(1\ +z)^{\gamma}}{1\ +\left(\frac{1+z}{1+z_{p}}\right)^{\gamma+\kappa}} R_0 \label{eq:md-model}.
\end{equation}
As noted in~\cite{callisterShoutsMurmursCombining2020}, this parametrization subsumes some astrophysical effects on the BBH population, such as formation-to-merger time delays, without needing to explicitly model them.

Table~\ref{tab:priors} contains descriptions of these population hyperparameters, as well as the uniform priors used for all population and cosmological parameters in the inference.

\begin{table}[]
    \centering
    \begin{tabular}{ccc} \hline
     {\bf Parameter} & {\bf Description} & {\bf Prior} \\ \hline\hline
     %%%%%%%%%%%%%%%%%%%%%%%%%%%%%%%%%%%%%%%
    & \multicolumn{1}{c}{\textit{Mass model parameters}}     & \\\hline
    %%%%%%%%%%%%%%%%%%%%%%%%%%%%%%%%%%%%%%%
    $\alpha$ & Index of power law primary mass & $\mathcal{U}(1.5, 12)$\\
    $\beta$ & Index of power law secondary mass & $\mathcal{U}(-4, 12)$\\
    $m_{min}$ & Minimum mass & $\mathcal{U}(2 M_\odot, 10 M_\odot)$\\
    $m_{max}$ & Maximum mass & $\mathcal{U}(50 M_\odot, 200 M_\odot)$\\
    $\lambda_g$ & Fraction of events in Gaussian & $\mathcal{U}(0, 1)$\\
    $\mu_g$ & Peak of Gaussian & $\mathcal{U}(20 M_\odot, 50 M_\odot)$\\
    $\sigma_g$ & Width of Gaussian & $\mathcal{U}(0.4 M_\odot, 10 M_\odot)$\\
    $\delta_m$ & Range of tapering function & $\mathcal{U}(0 M_\odot, 10 M_\odot)$\\\hline
    %%%%%%%%%%%%%%%%%%%%%%%%%%%%%%%%%%%%%%%
    & \multicolumn{1}{c}{\textit{Population hyperparameters}}     & \\\hline
    %%%%%%%%%%%%%%%%%%%%%%%%%%%%%%%%%%%%%%%
    $R_0$ & The BBH merger rate at $z=0$ & $\mathcal{U}(0, 100)$\\
    $\gamma$ & Index for the power law before $z_p$ & $\mathcal{U}(0, 12)$\\
    $\kappa$ & Index for the power law after $z_p$ & $\mathcal{U}(0, 6)$\\
    $z_p$ & Peak redshift& $\mathcal{U}(0, 4)$\\\hline
    %%%%%%%%%%%%%%%%%%%%%%%%%%%%%%%%%%%%%%%
    & \multicolumn{1}{c}{\textit{Cosmological parameters}}     & \\\hline
    %%%%%%%%%%%%%%%%%%%%%%%%%%%%%%%%%%%%%%%
    $\Omega_{m,0}$ & Present matter energy density & $\mathcal{U}(0, 1)$\\
    $w_0$ & Dark energy equation of state parameter& $\mathcal{U}(-3, 0)$\\
     $H_0$ & \textbf{Hubble constant} & $\mathcal{U}(10, 200)$\\\hline\hline
    \end{tabular}
    
    \caption{A summary of the parameters considered and the uniform priors $\mathcal{U}$ used for parameter inference. The population models involve the eight-parameter {\tt PowerLaw+Peak} mass model~\citep{talbotMeasuringBlackHoleMass2018,theligoscientificcollaborationBinaryBlackHole2019,talbot2024gwpopulation} and the four-parameter Madau-Dickinson redshift distribution~\citep{madau2014cosmic}, yielding a total of 15 parameters when combined with three cosmological parameters.
    }
    \label{tab:priors}
\end{table}

\subsection{Analysis validation}
For all results of $\mathcal{L}_\FG$, $\mathcal{L}_\BG$, and $\mathcal{L}_\joint$, we ensured convergence by checking consistency between results obtained using lower (2048) and higher (3072) counts of live points, as is conventional in nested sampling analysis. 

The kernel density estimators (KDEs) used in our analysis were each a standard gaussian-kernel KDE. These were generated with \texttt{SciPy} and its default settings in~\texttt{scipy.stats.gaussian\_kde} \footnote{\href{https://docs.scipy.org/doc/scipy/reference/generated/scipy.stats.gaussian_kde.html}{https://docs.scipy.org/doc/scipy/reference/generated/scipy.stats.gaussian\_kde.html}}, which automatically select the bandwidth scale based on the dataset according to one of two default bandwidth selection methods.
To check for consistency, we used each of these two KDE bandwidth selection methods (the ``Scott's Rule'' and Silverman methods) for the analysis and found no appreciable differences in the results.

Generating the KDE for $\mathcal{L}_\BG$ required boundary effect remediation, given that the $\Omega_\gw$ upper limit posterior from~\cite{abbottUpperLimitsIsotropic2021} peaks at $\sim 0$, which would cause a naive KDE approach to create an anomalously-vanishing probability as $\Omega_\gw \rightarrow 0$.
We remedied this by mirroring the underlying posterior distribution and generating the KDE from that while accounting for renormalization, as has been performed previously (e.g., described in Appendix B of~\cite{perkins2021improved}).
This boundary effect was not present for the 15 hyperparameter posteriors from~\cite{abbott2023cceh}, as mirroring either did not affect the results, or distorted the KDE and thus made it \textit{less} representative of the posterior distribution.

Finally, if our KDE method is to be consistent with the full hierarchical likelihood evaluation utilized by standard population analysis methods, the BBH spectral siren results of \cite{abbott2023cceh} should be recovered given that we adopted their population hyperparameter posteriors to constuct the $\mathcal{L}_\FG$ KDE. We found that the results of $\mathcal{L}_\FG$ are directly comparable to the $H_0$ results obtained via the BBH spectral siren method of \cite{abbott2023cceh}, whose maximum a-posteriori probability and $68.3\%$ highest density interval is reported as $46^{+49}_{-26} \, \rm km~s^{-1}~Mpc^{-1}$. We recover a result consistent with theirs: $57^{+43}_{-35} \, \rm km~s^{-1}~Mpc^{-1}$. 
We similarly obtain consistent results for the 14 other hyperparameters in~\cite{abbott2023cceh}, indicating that the posterior-generated KDE is a reasonable proxy for a full hierarchical population analysis approach.

\subsection{Other parameter results}\label{subapp:15param}
\begin{figure*}[htp]
  \includegraphics[width=\textwidth]{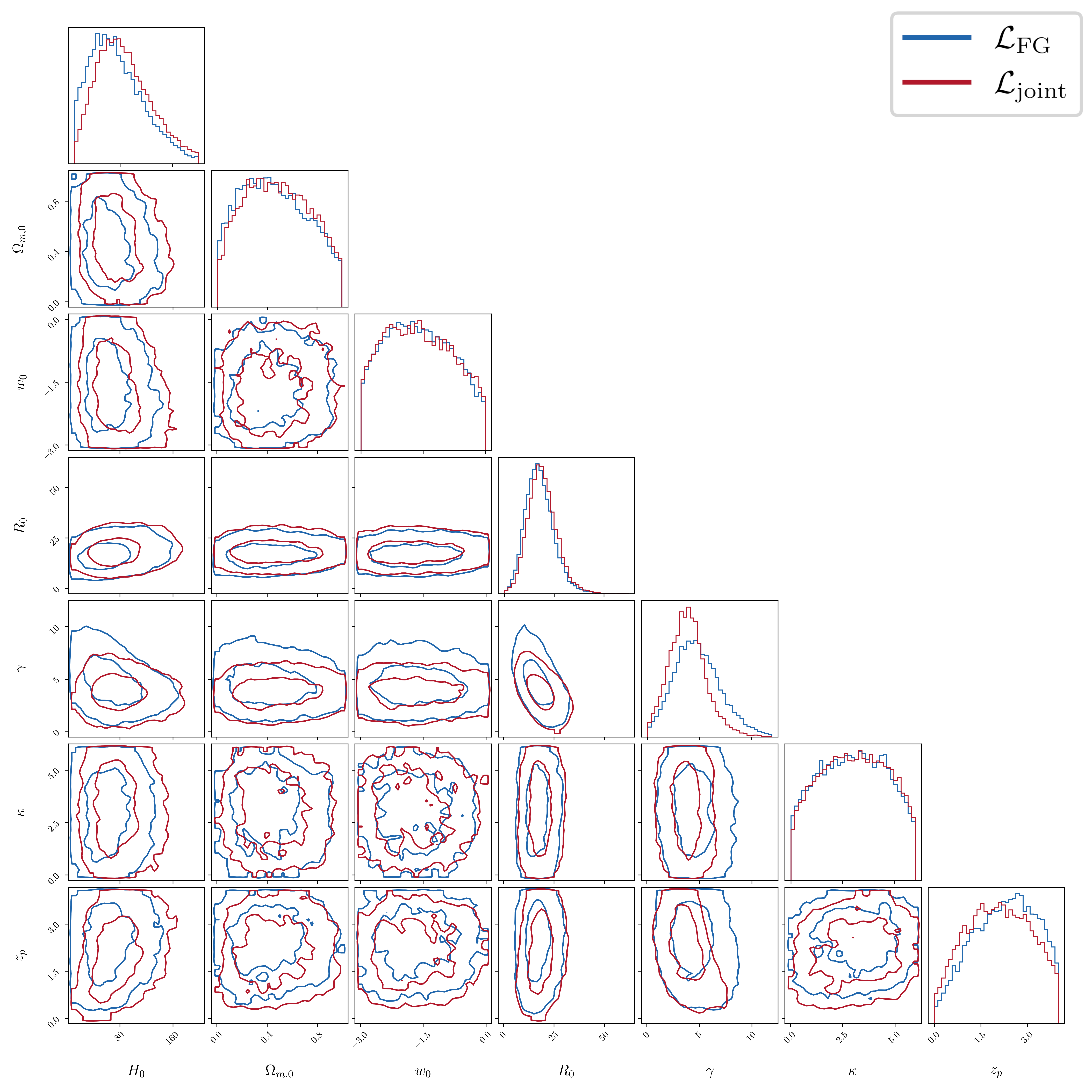}
  \caption{Redshift model hyperparameters, together with cosmological parameters, as inferred via the BBH spectral siren ($\mathcal{L}_\mathrm{FG}$, blue) and stochastic siren ($\mathcal{L}_\mathrm{joint}$, red) methods. Observe that most hyperparameters do not demonstrate an improvement with the current non-detection of the BBH GWB, with the exception of $\gamma$ as elaborated in the text.}
  \label{fig:redshift-infer}
\end{figure*}

We show our inference results for the redshift model and cosmological parameters in Fig.~\ref{fig:redshift-infer}. Note that in general, measurements of most hyperparameters are not greatly altered by the inclusion of the GWB non-detection.

However, observe that the redshift model hyperparameter $\gamma$ (the exponent governing the lower-redshift merger rate, as in Eq.~\eqref{eq:md-model}) has some marginal improvement, namely a tendency toward lower value. This is sensible given that the upper limit on $\Omega_\gw$ will favor lower values of $\gamma$ since larger $\gamma$ implies more mergers, and thus a larger $\Omega_\gw$. There is hence some correlation between $H_0$ and these parameters; the findings of \cite{ferraiuolo2025inferring} are consistent with this result.
We also note that while the $\mathcal{L}_\joint$ posteriors for $\Omega_{m,0}$ and $z_p$ are shifted slightly relative to $\mathcal{L}_\FG$, the posteriors themselves are too broad to offer informative constraints.

\begin{figure*}[htp]
  \includegraphics[width=\textwidth]{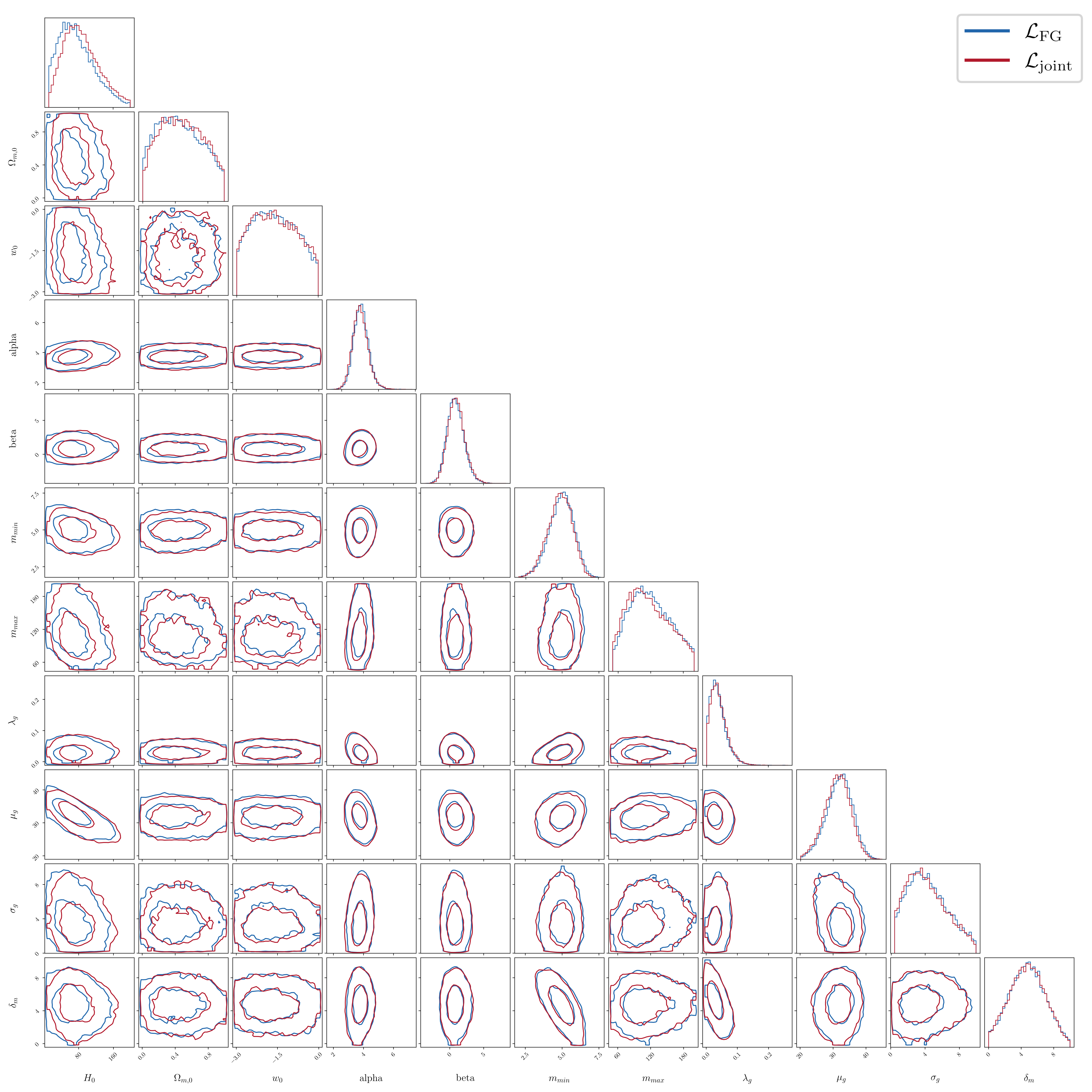}
  \caption{Mass model hyperparameters, together with cosmological parameters, as inferred via the BBH spectral siren ($\mathcal{L}_\mathrm{FG}$, blue) and stochastic siren ($\mathcal{L}_\mathrm{joint}$, red) methods. Observe that effectively no mass hyperparameter constraints demonstrate an improvement with the current non-detection of the BBH GWB.}
  \label{fig:mass-infer}
\end{figure*}

Our parameter inference also considered eight mass model hyperparameters, but the posteriors are effectively unchanged by the inclusion of the GWB (see Fig.~\ref{fig:mass-infer}), indicating no improvement in constraining these parameters.

\section{SNR of the GWB and Projections for Future Detectors} 
\label{app:SNR}
The optimal SNR of the GW energy density $\Omega_{\gw}$ in a given detector network is given by~\citep{allenDetectingStochasticBackground1999}
\begin{equation}
\langle {\rm{SNR}} \rangle_{\text{opt}} = \sqrt{(\gamma \Omega_{\gw} | \gamma \Omega_{\gw})}\, ,
\label{eqn:SNR_opt}
\end{equation}
where $\gamma$ is the overlap reduction function and the inner product is defined in the main text.
We can then write the SNR as 
\begin{equation}
    \langle {\rm{SNR}} \rangle_{\text{opt}} = \sqrt{2T \left( \frac{3H_0^2}{10\pi^2} \right)^2 \int_0^\infty df \frac{\gamma^2 \Omega_\gw^2}{f^6 P_1(f) P_2(f)} }\, ,
\end{equation}   
noting how it scales with the Hubble constant ($\Omega_{\gw} \sim H_0^{-3} \Rightarrow {\rm{SNR}} \sim H_0^{-1}$) and with observing time (${\rm{SNR}} \sim \sqrt{T}$).
From this scaling, it is clear that a longer observing time increases the chance of GWB detection, and that $H_0$ impacts the time to detection that a non-detection implies larger values of $H_0$). 

\begin{figure}
    \centering
    \includegraphics[width=0.33\linewidth]{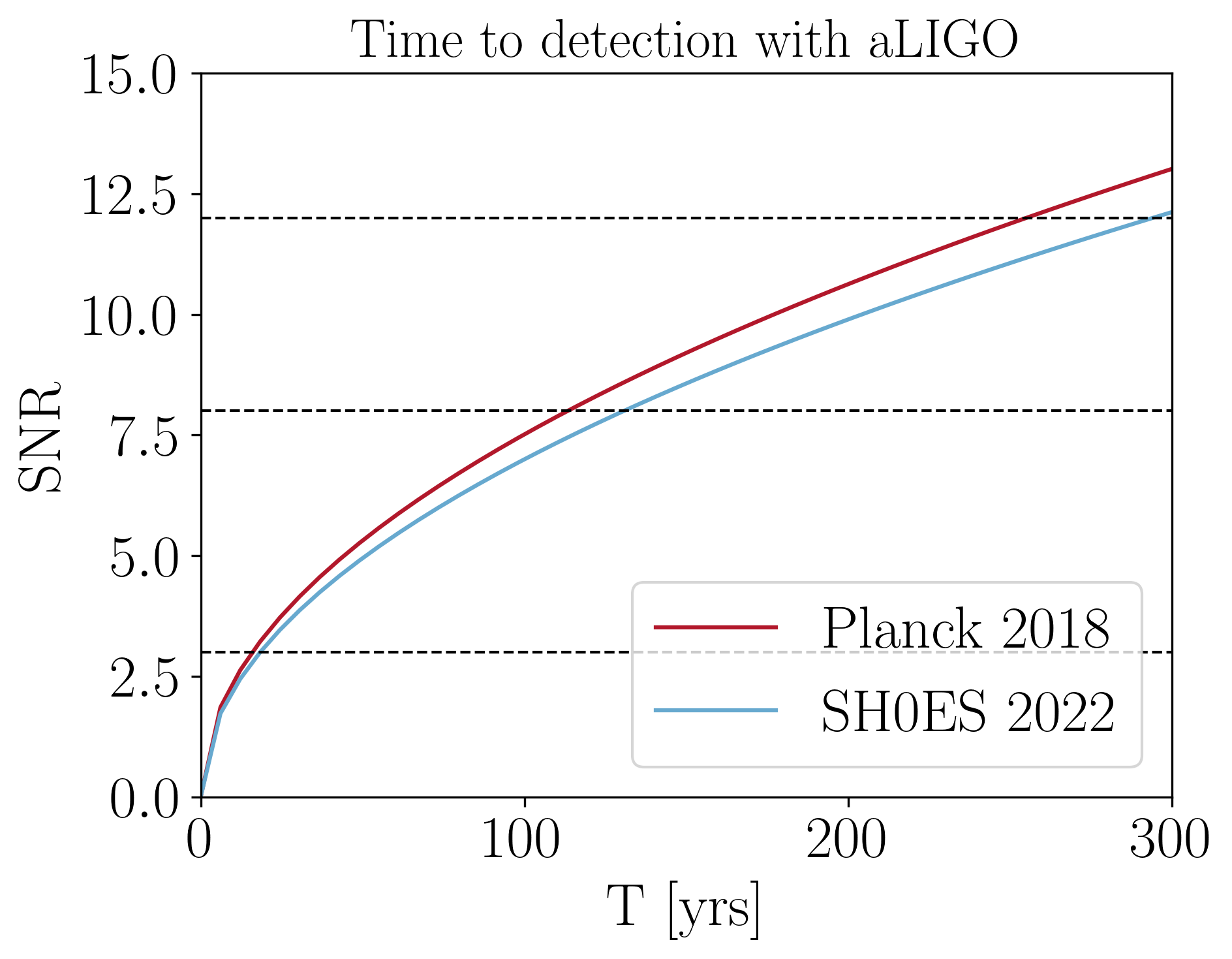}
    \includegraphics[width=0.325\linewidth]{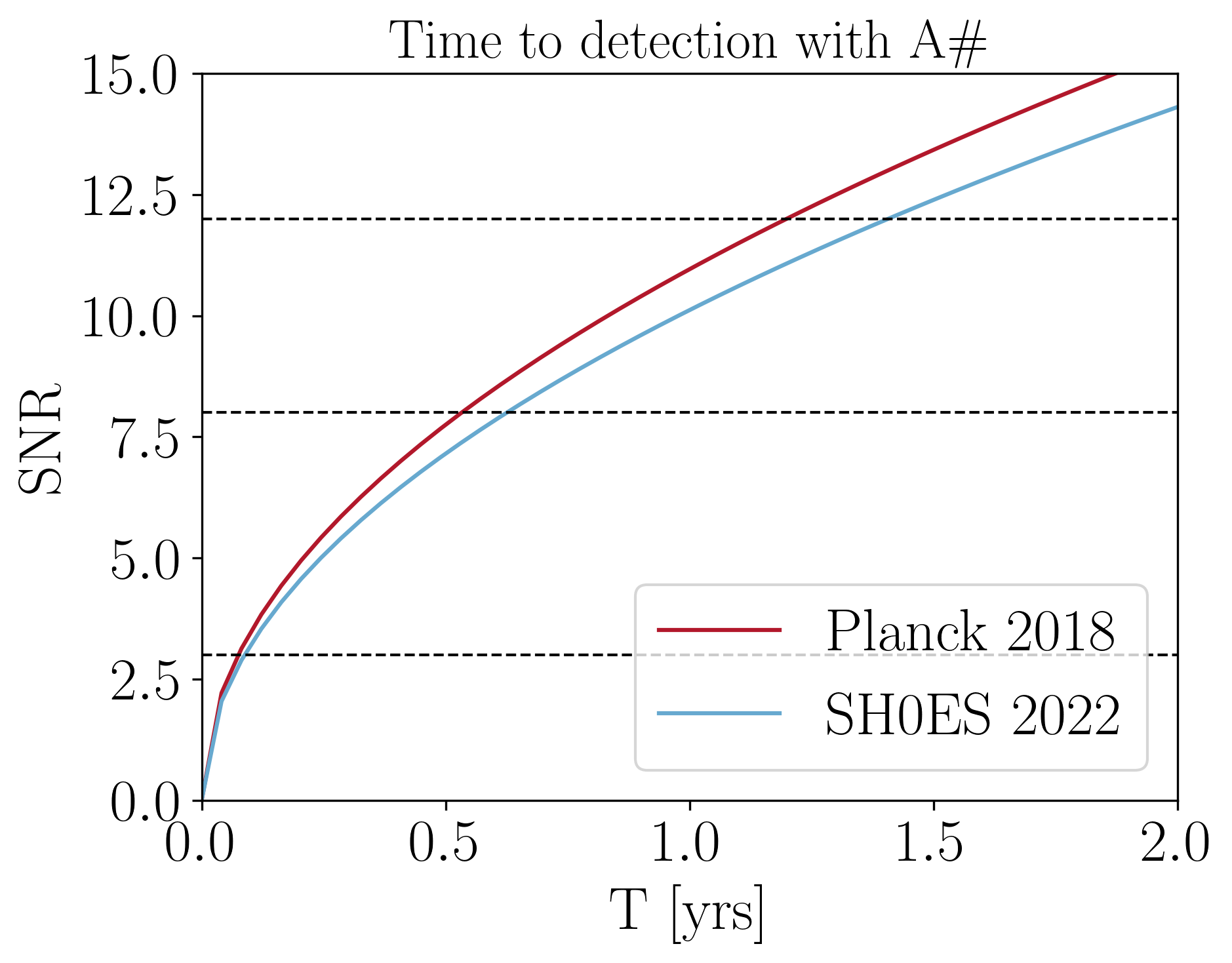}
    \includegraphics[width=0.33\linewidth]{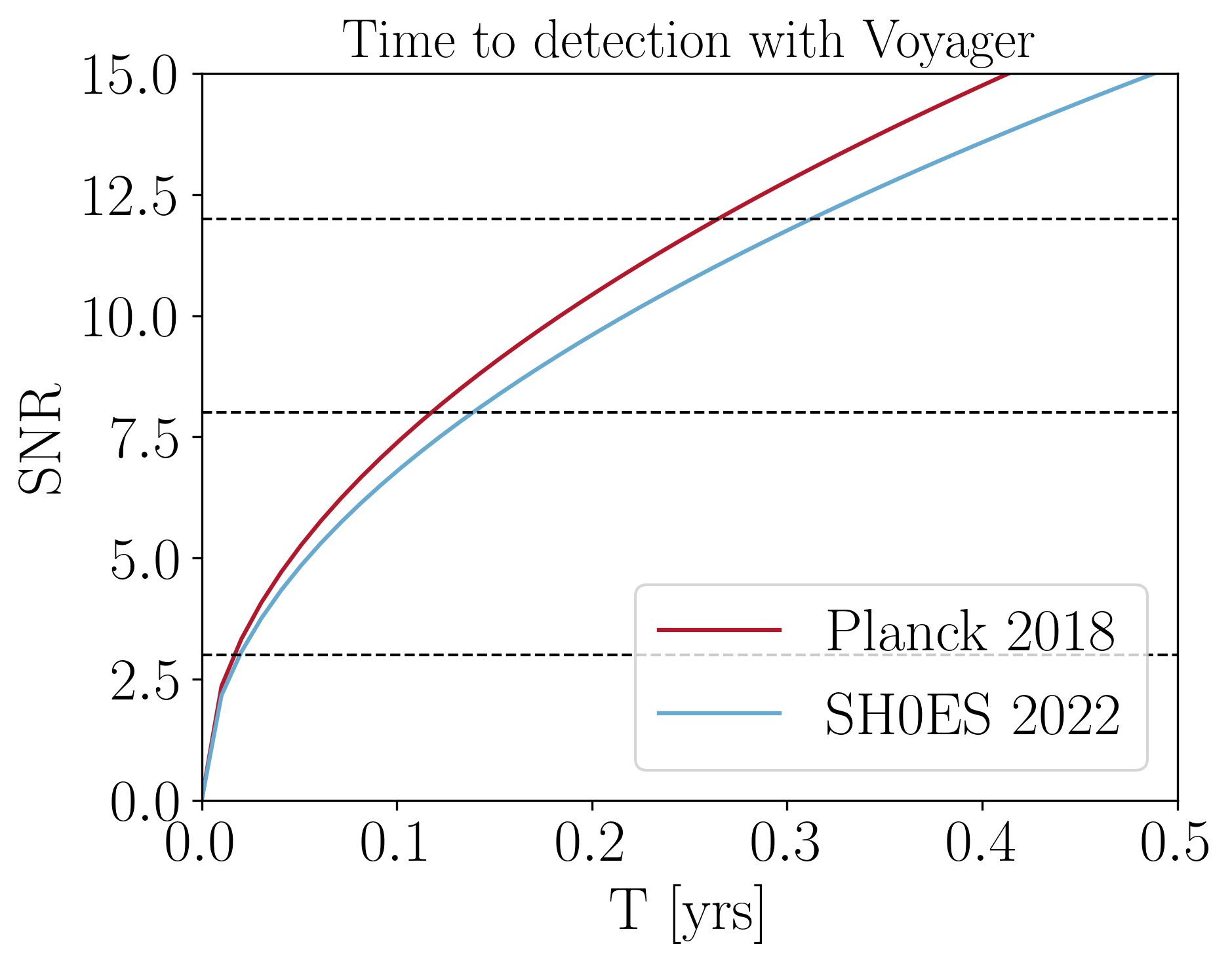}
    \caption{The time to detect a GWB with a detector network consisting of LIGO-Hanford, LIGO-Livingston, and Virgo with the sensitivity of aLIGO, A\#, or Voyager. In all plots, the dashed lines represent SNR values of three, eight, and twelve respectively, and results are shown for both the Planck~\citep{aghanim2020planck} and SH0ES~\citep{Riess:2021jrx} values of $H_0$ to give an example for how time to detection changes with $H_0$. Note that the time to detection is very different across these different networks, so these plots have a different scaling on the x-axis.}
    \label{fig:SNR1d}
\end{figure}

To demonstrate this scaling behavior, we make projections of the time required for GWB detection using future detector networks (Fig.~\ref{fig:SNR1d}).
Note that as in the rest of this work, we are considering only the BBH contribution to $\Omega_{\gw}$. This means that our energy spectrum may be an underestimate, implying that our time-to-detection projections are conservative.
Here, we pin the mass and redshift hyperparameters to their mean posterior values obtained from~\cite{abbott2023cceh}, which are $\alpha = 3.4$, $\beta=1.08$, $m_{\mathrm{min}}=5.08$, $m_{\mathrm{max}}=86.85$, $\lambda_g=0.04$, $\mu_g=33.73$, $\sigma_g =3.56$, and $\delta_m =4.83$ for the mass model, and $R_0=17.975$, $\gamma =2.7$, $\kappa=2.9$, and $z_p=1.9$ for the redshift distribution. For cosmological parameters, we used $\Omega_{m,0} =0.3065$, $w_0=-1$, $H_0=\{67.4,73\}$ from~\cite{aghanim2020planck, Riess:2021jrx}. Note that the value of $\Omega_{m,0}$ is taken from the Planck 2015~\cite{ade2016planck} instead of Planck 2018 in order to match LVK work~\citep{abbott2023cceh}.

We consider three different cases with the sensitivities predicted for advanced LIGO (aLIGO), A\#~\citep{fritschel2022asharp}, and Voyager~\citep{VoyagerAdhikariCryogenicSiliconInterferometer2020}. 
In all cases, we conservatively assume a three-detector network composed of LIGO-Hanford, LIGO-Livingston, and Virgo, all with the same sensitivity. 
Adding the SNR between each pair of detectors in quadrature, we obtain the network SNR:
\begin{equation}
\langle {\rm{SNR}} \rangle_{\text{net}} = \sqrt{{\rm{SNR}}_{\text{HL}}^2 + {\rm{SNR}}_{\text{HV}}^2 + {\rm{SNR}}_{\text{LV}}^2}\,.
\end{equation}
With these estimates, we find the time to detection in aLIGO too long to be practical (on the order of centuries of observation needed for an SNR of eight).
However,  A\# and Voyager are more promising, with both networks capable of detecting the GWB with the assumed properties within one year. 
The factor of $\sim10$ improvement in sensitivity between aLIGO and A\# is responsible for this dramatic improvement in time to detection.  

\begin{figure}
%    \centering
    \includegraphics[width=0.45\linewidth]{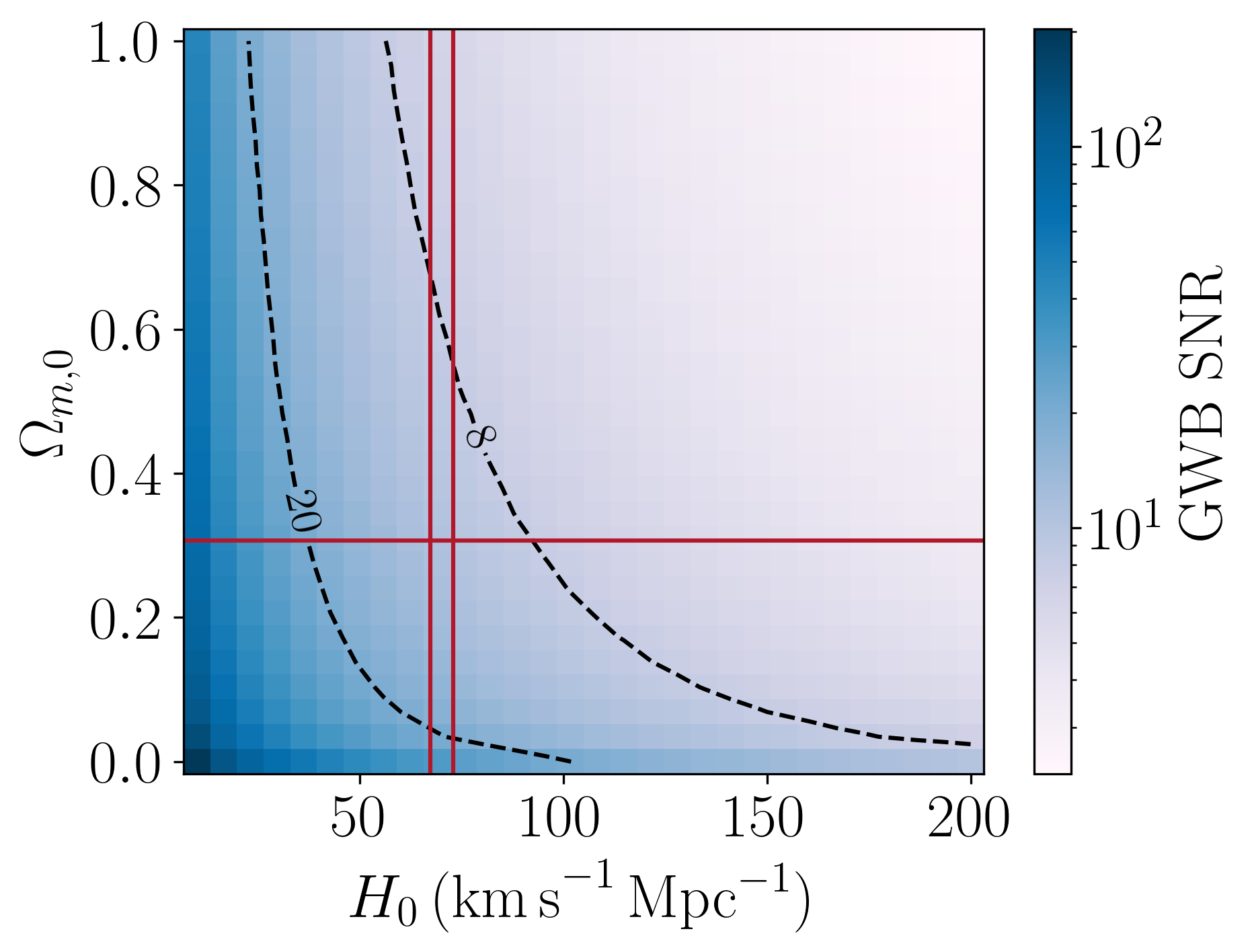}
    \includegraphics[width=0.45\linewidth]{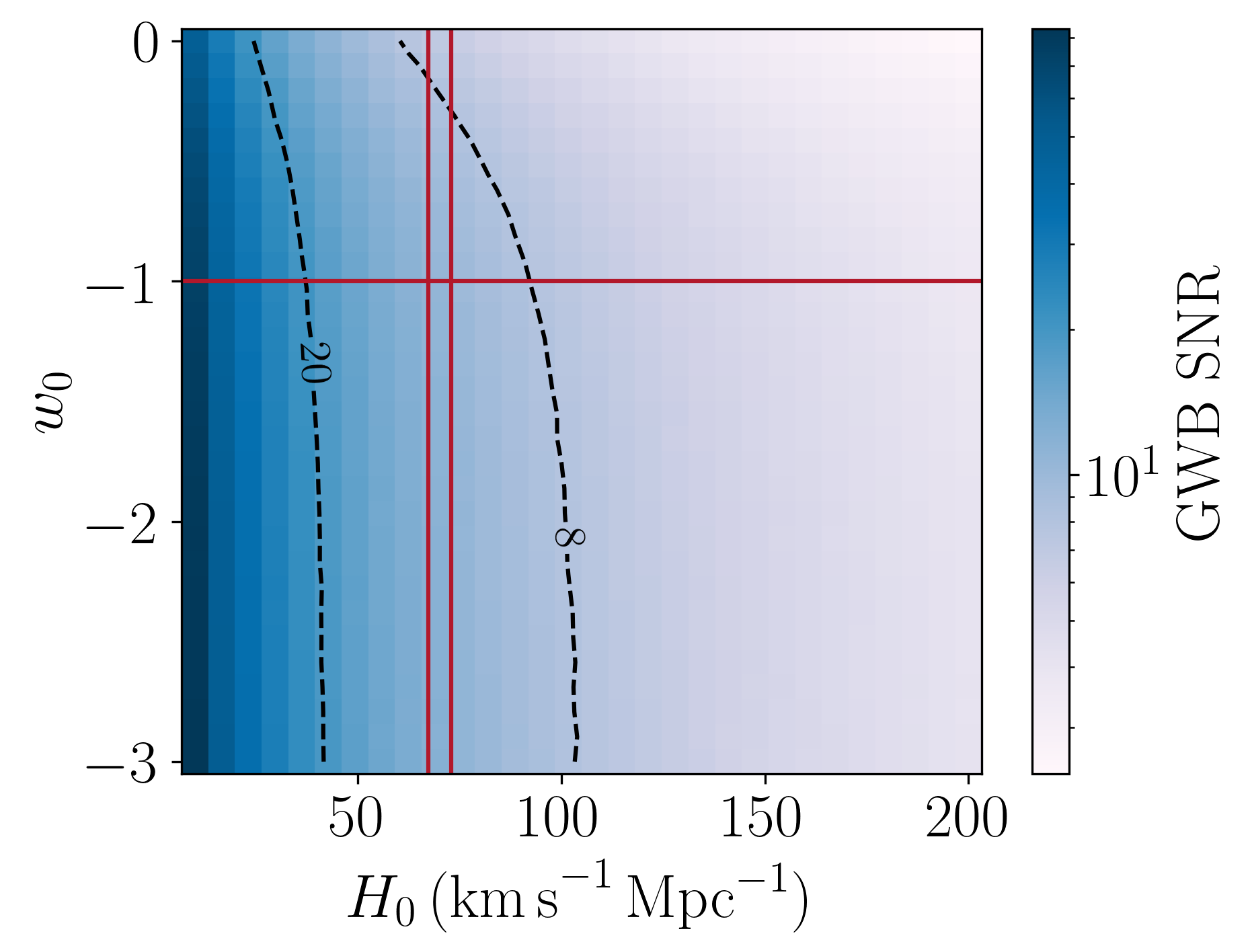}
    \includegraphics[width=0.45\linewidth]{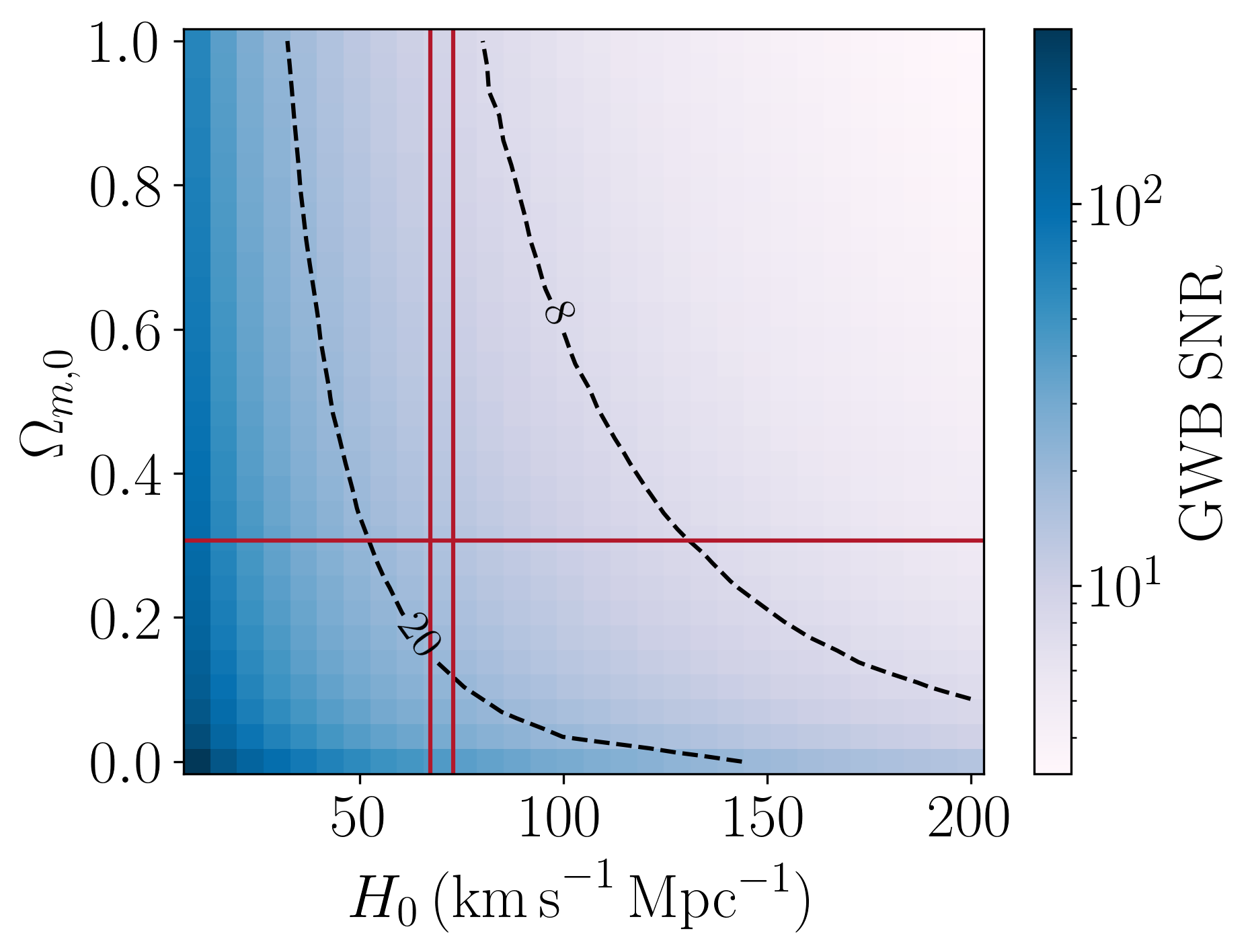}
    \includegraphics[width=0.45\linewidth]{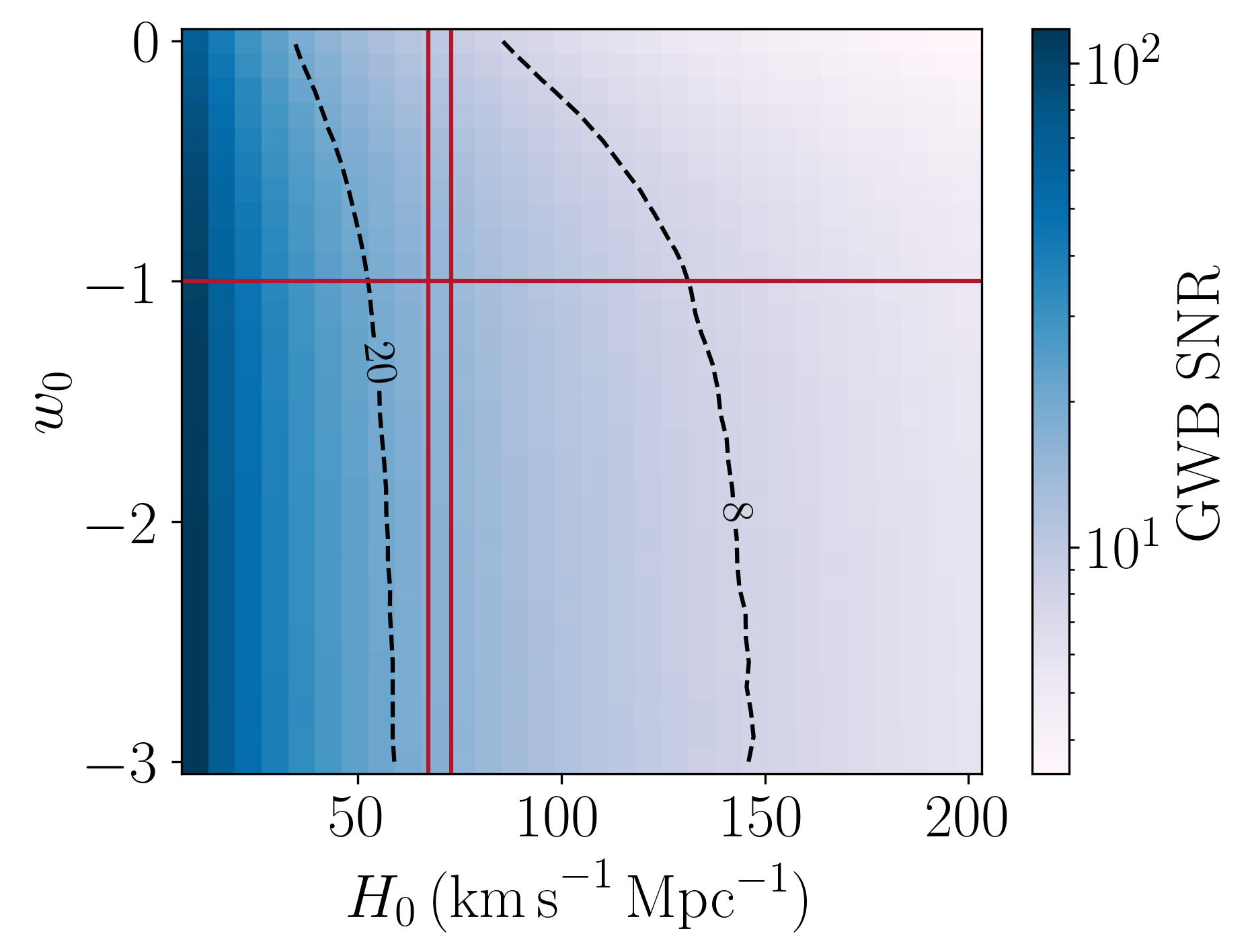}
    \\ % hard-coding the sizes and spaces since the automatic formatting is bad
    \hspace{-1.35cm}
    \includegraphics[width=0.378\linewidth]{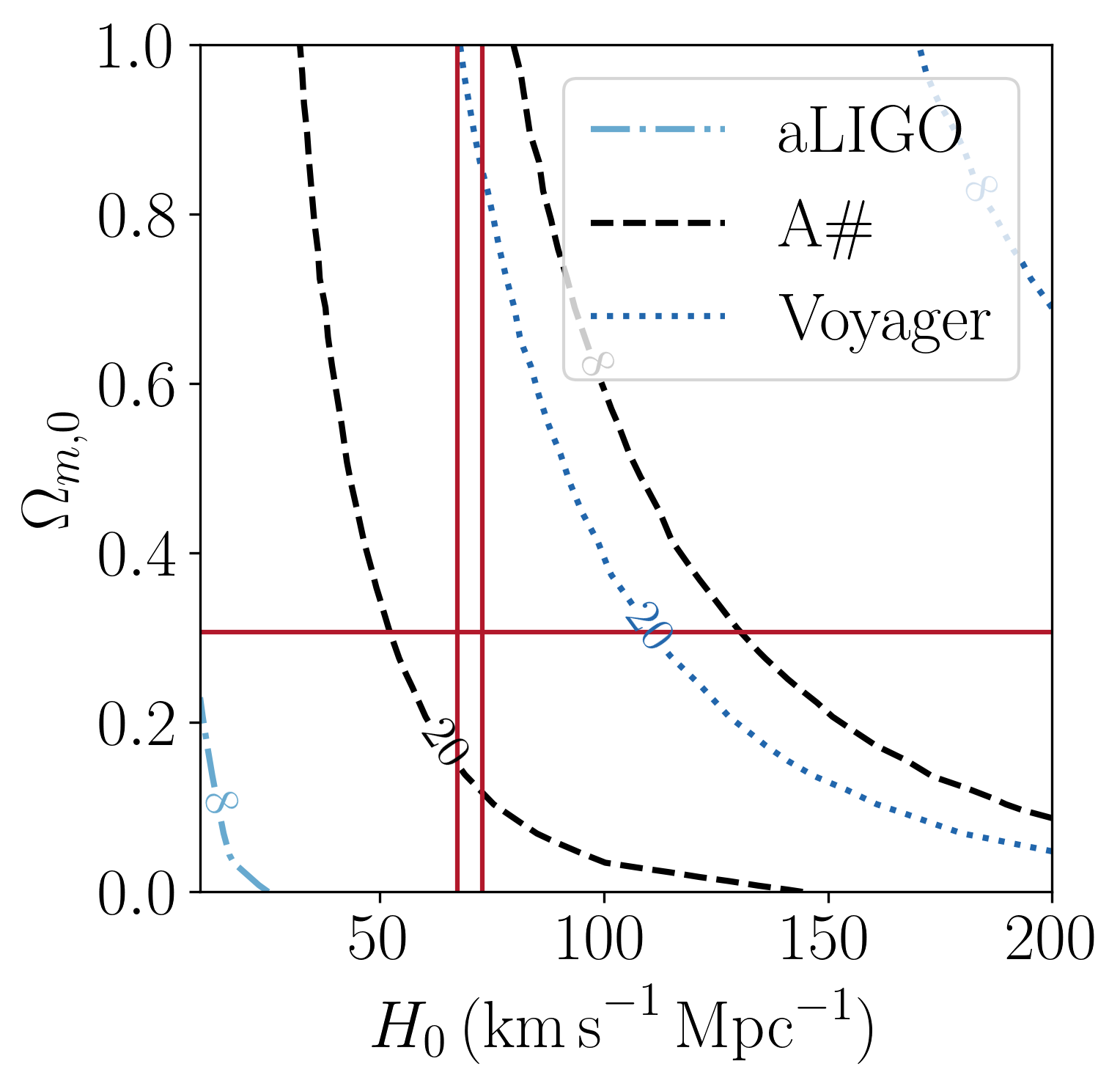}
    \hspace{0.9cm}
    \includegraphics[width=0.395\linewidth]{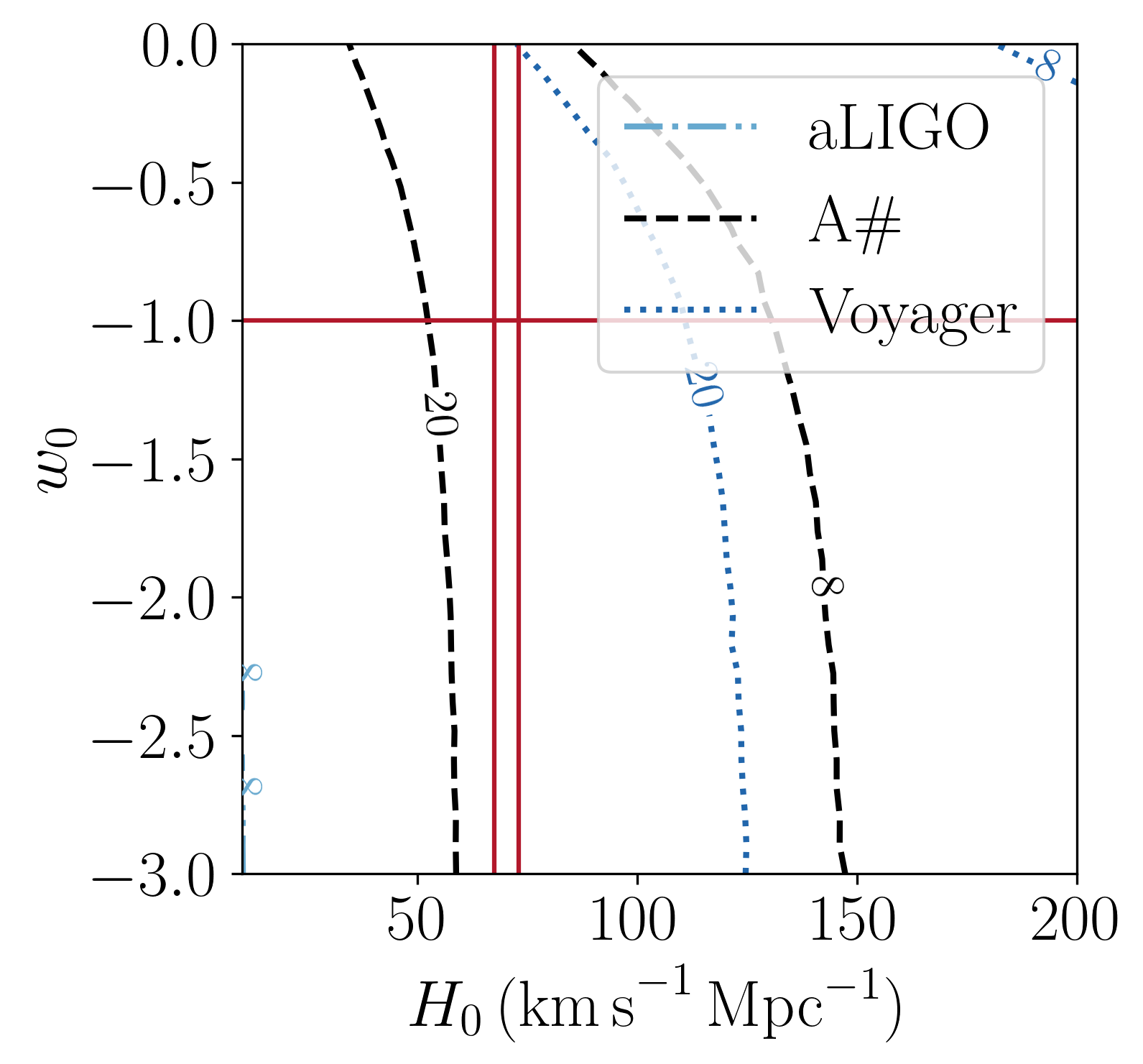}
    \caption{SNR of the GWB for various future detector networks as a function of two cosmological parameters ($H_0$ with either $\Omega_{m,0}$ or $w_0$), with the other 13 (hyper)parameters fixed to values denoted in the text.
    Top (middle) plots: a network of three A\# detectors after one year (two years) of observation, with SNR shown as a heatmap. Contour lines illustrate constant values of $\mathrm{SNR}=8$ and $\mathrm{SNR}=20$, while the red horizontal and vertical lines indicate existing cosmological parameter measurements. Observe that regions of parameter space that are detectable (areas with higher SNR) expand with increasing observing time, such that the GWB would be detectable within 1-2 years.
    Bottom: the same as the previous plots, but for networks with aLIGO, A\#, and Voyager sensitivities after two years of observation. Observe that there are effectively no values of these parameters that would make the GWB measurable in aLIGO after only two years.
    Analogous plots that pair $H_0$ with population hyperparameters are shown in the remainder of this text.}
    \label{fig:SNR_2d}
\end{figure}

The estimates in Fig.~\ref{fig:SNR1d} were performed assuming specific values for each of the parameters, yet it is informative to consider how the SNR can change with different combinations of the parameters. We hence show the relation between two parameters at a time (with all other parameters held fixed) in Fig.~\ref{fig:SNR_2d}, with SNR as a function of the cosmological parameters $\{H_0, \Omega_{m,0}, w_0\}$ represented via either a heatmap or contour lines (for SNR values of 8 and 20). 
These plots illustrate which regions of parameter space could be ruled out by a non-detection of the GWB, as any combination of parameters that corresponds to an SNR larger than eight (beyond the contour line) would be measurable.

\begin{figure}
%    \centering
    \includegraphics[width=0.45\linewidth]{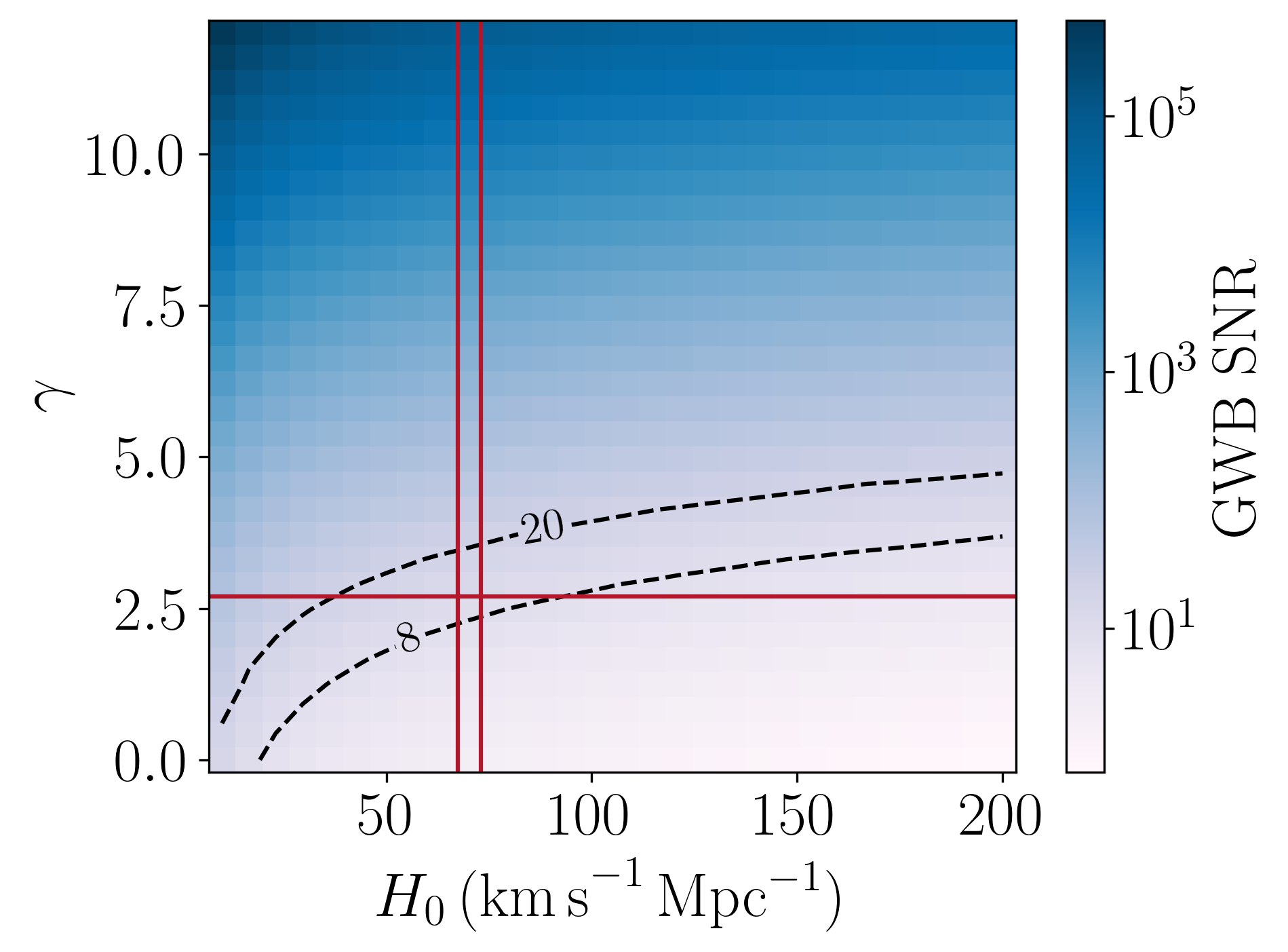}
    \includegraphics[width=0.42\linewidth]{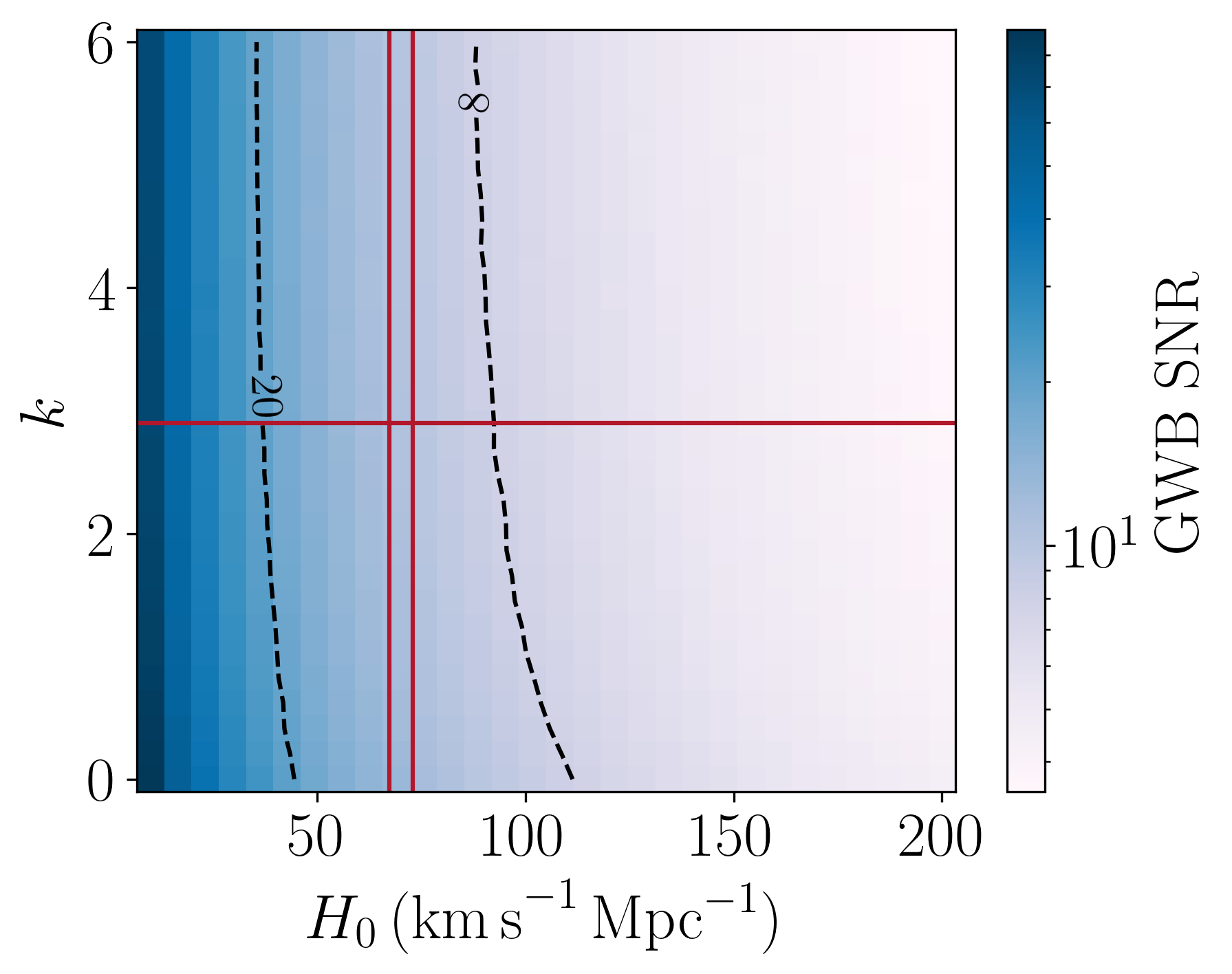}
    \includegraphics[width=0.45\linewidth]{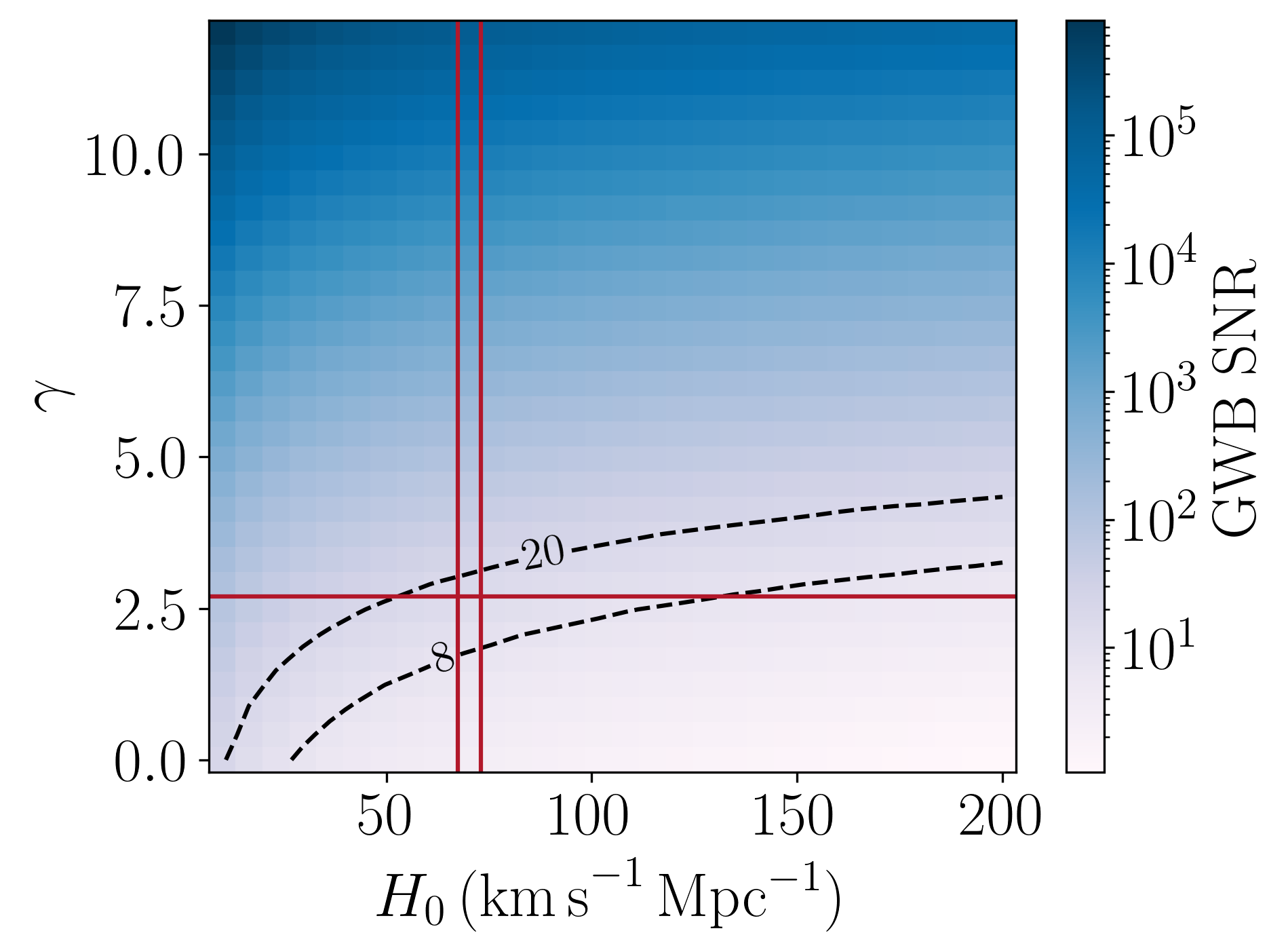}
    \includegraphics[width=0.42\linewidth]{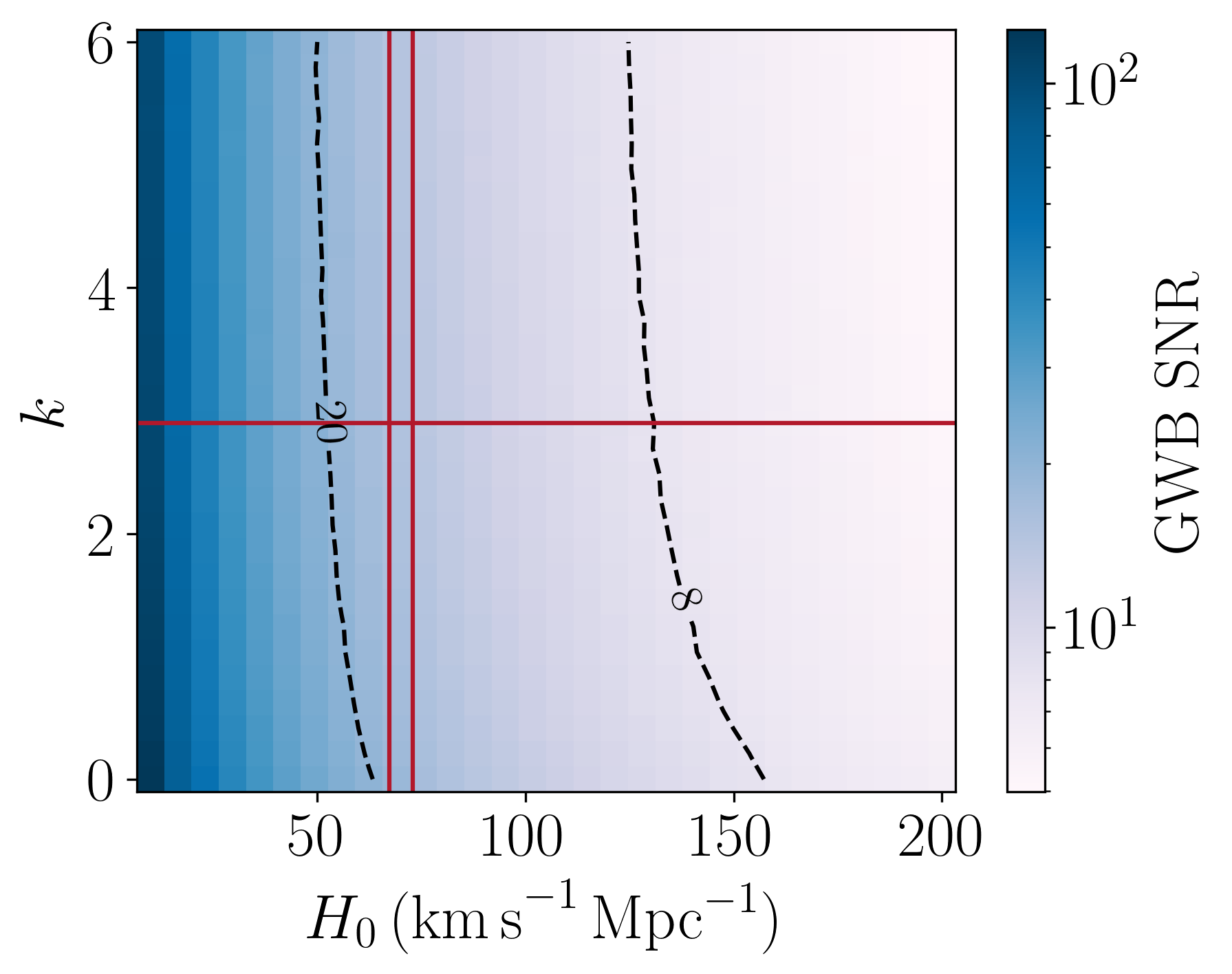}
    \\
    \hspace{-1.35cm}
    \includegraphics[width=0.378\linewidth]{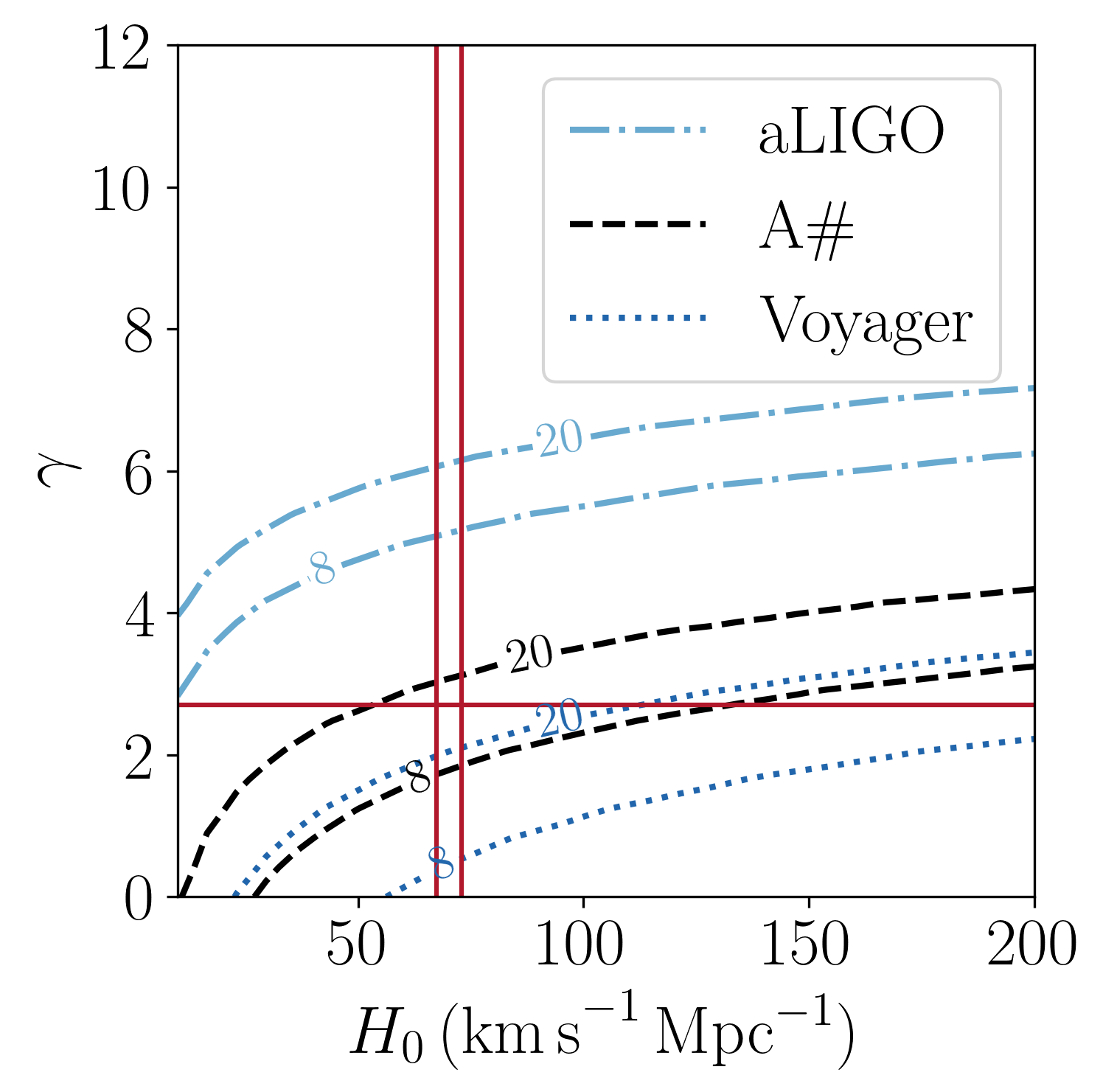}
    \hspace{0.9cm}
    \includegraphics[width=0.378\linewidth]{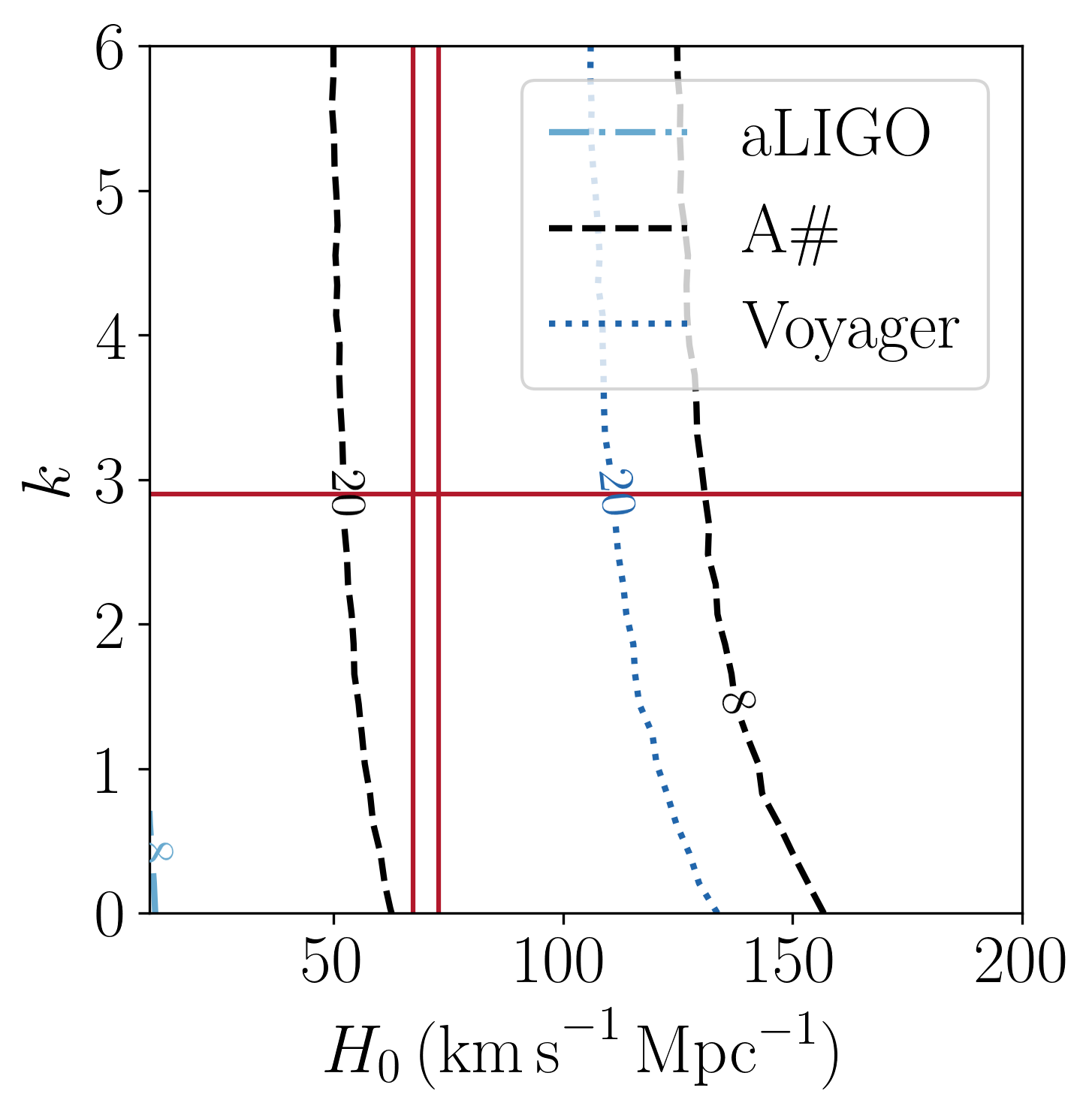}
    \caption{SNR of the GWB for various future detector networks as a function of two parameters ($H_0$ with either $\gamma$ or $k$), with the other 13 (hyper)parameters fixed to values denoted in the text. Refer to Fig.~\ref{fig:SNR_2d} for more explanation of the subplots.}
    \label{fig:SNR_2d_pop1}
\end{figure}

\begin{figure}
%    \centering
    \includegraphics[width=0.45\linewidth]{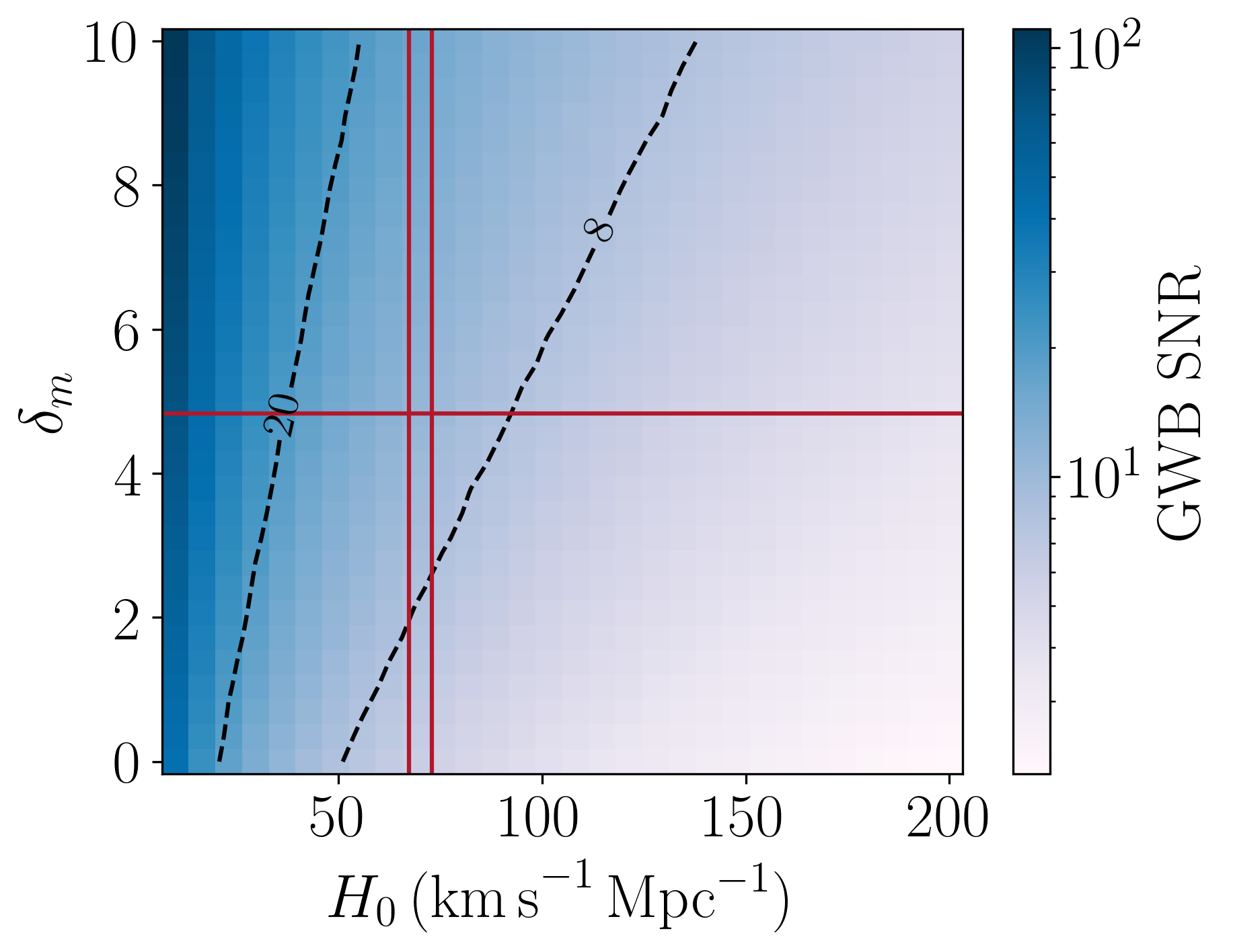}
    \includegraphics[width=0.45\linewidth]{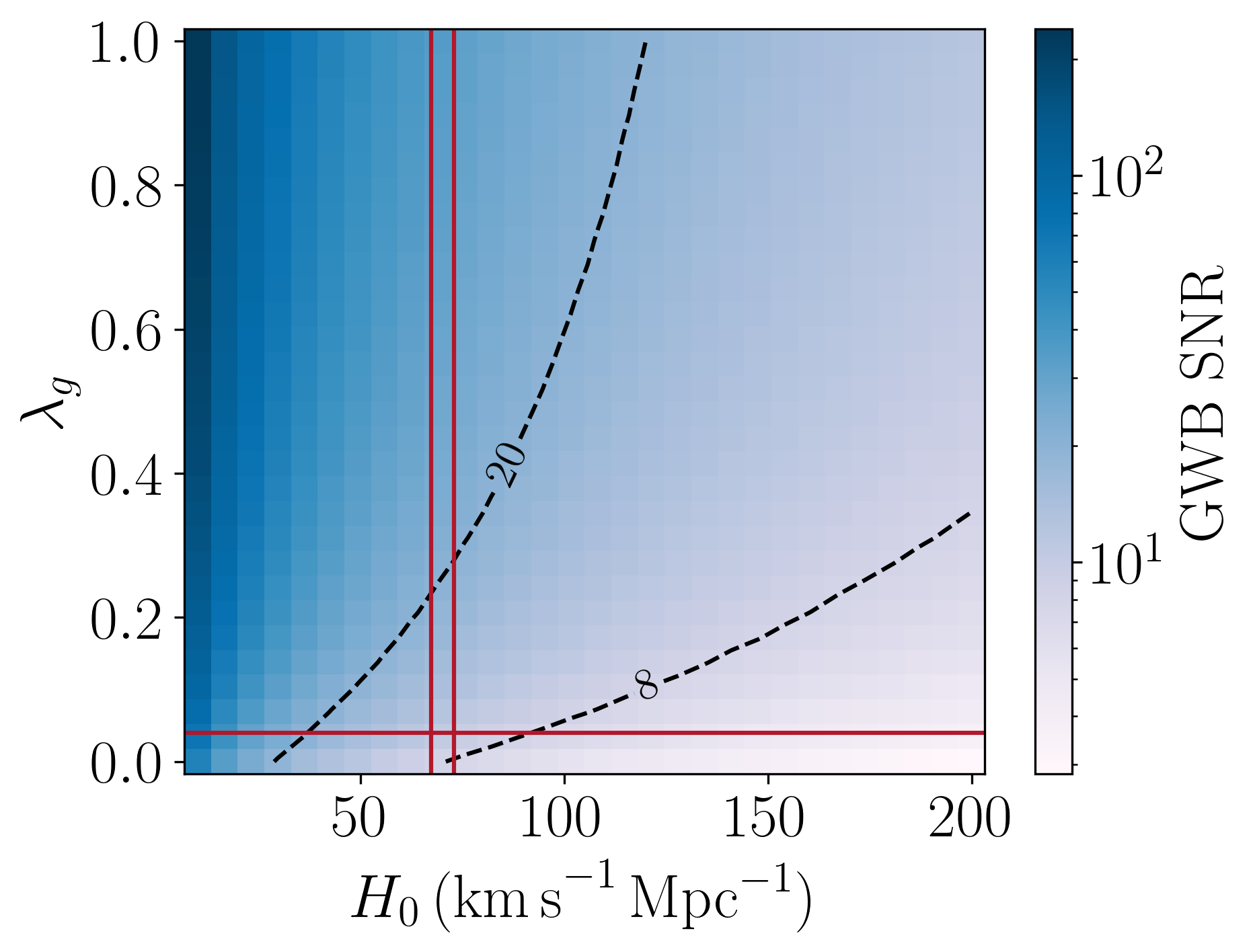}
    \includegraphics[width=0.45\linewidth]{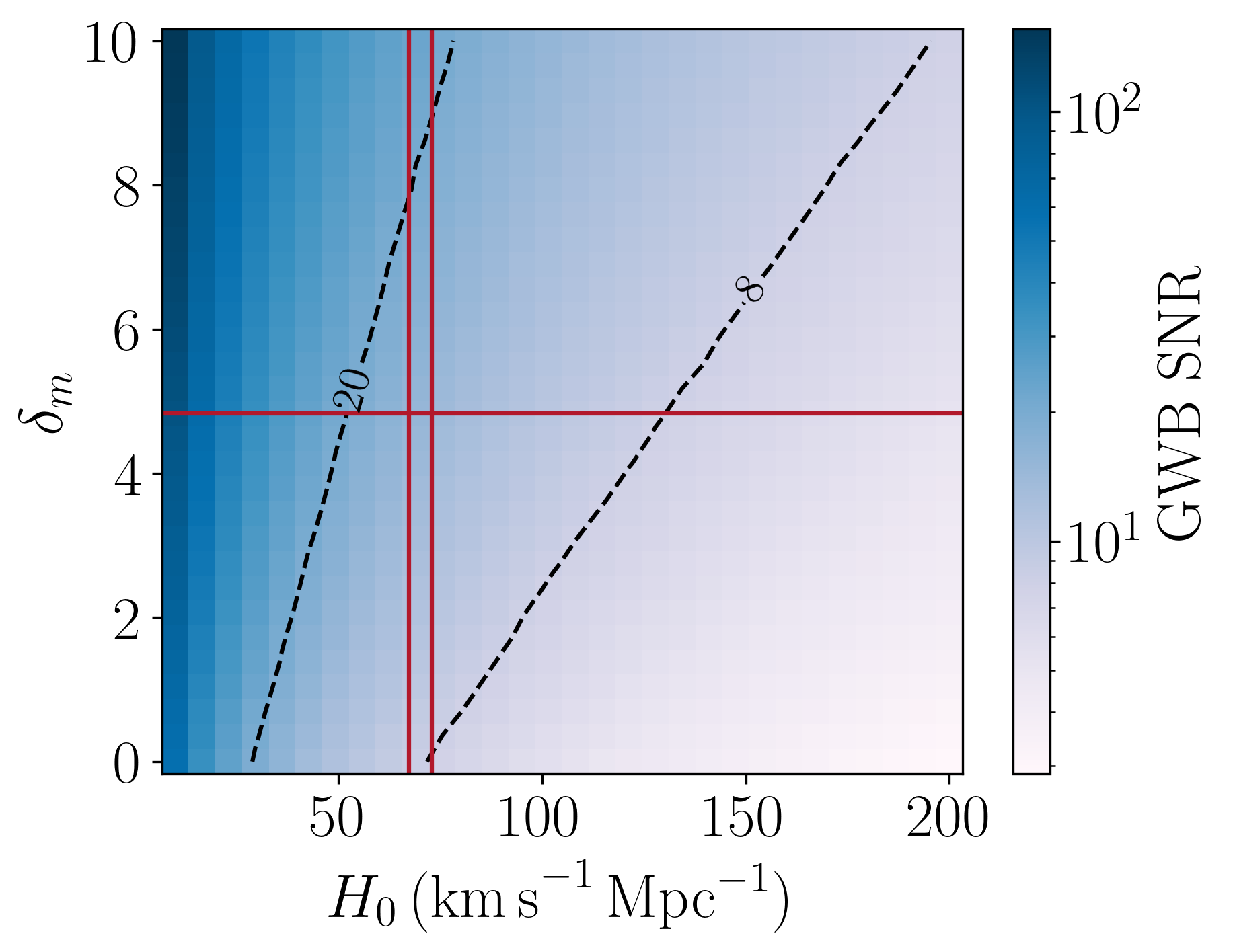}
    \includegraphics[width=0.45\linewidth]{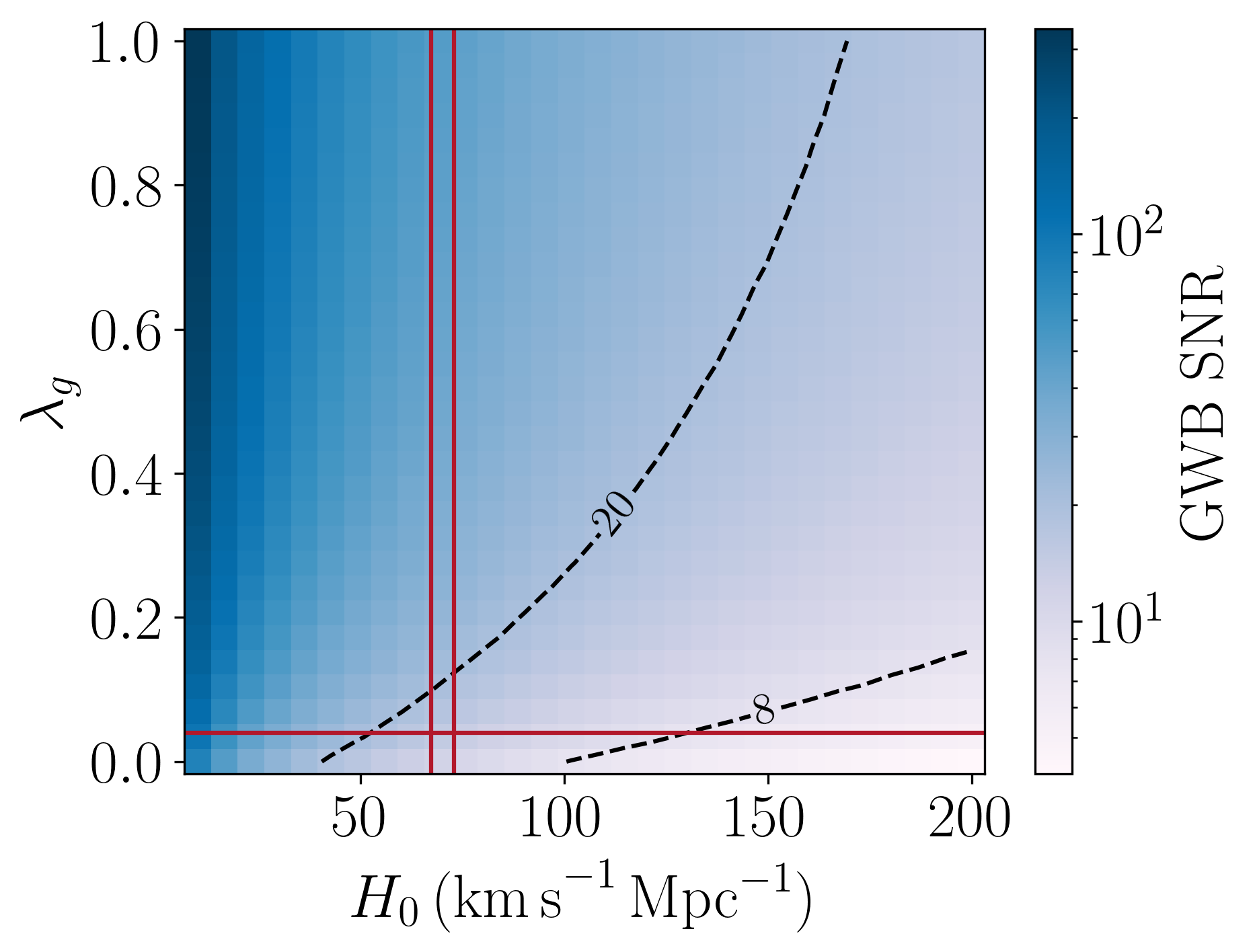}
    \\
    \hspace{-1.35cm}
    \includegraphics[width=0.378\linewidth]{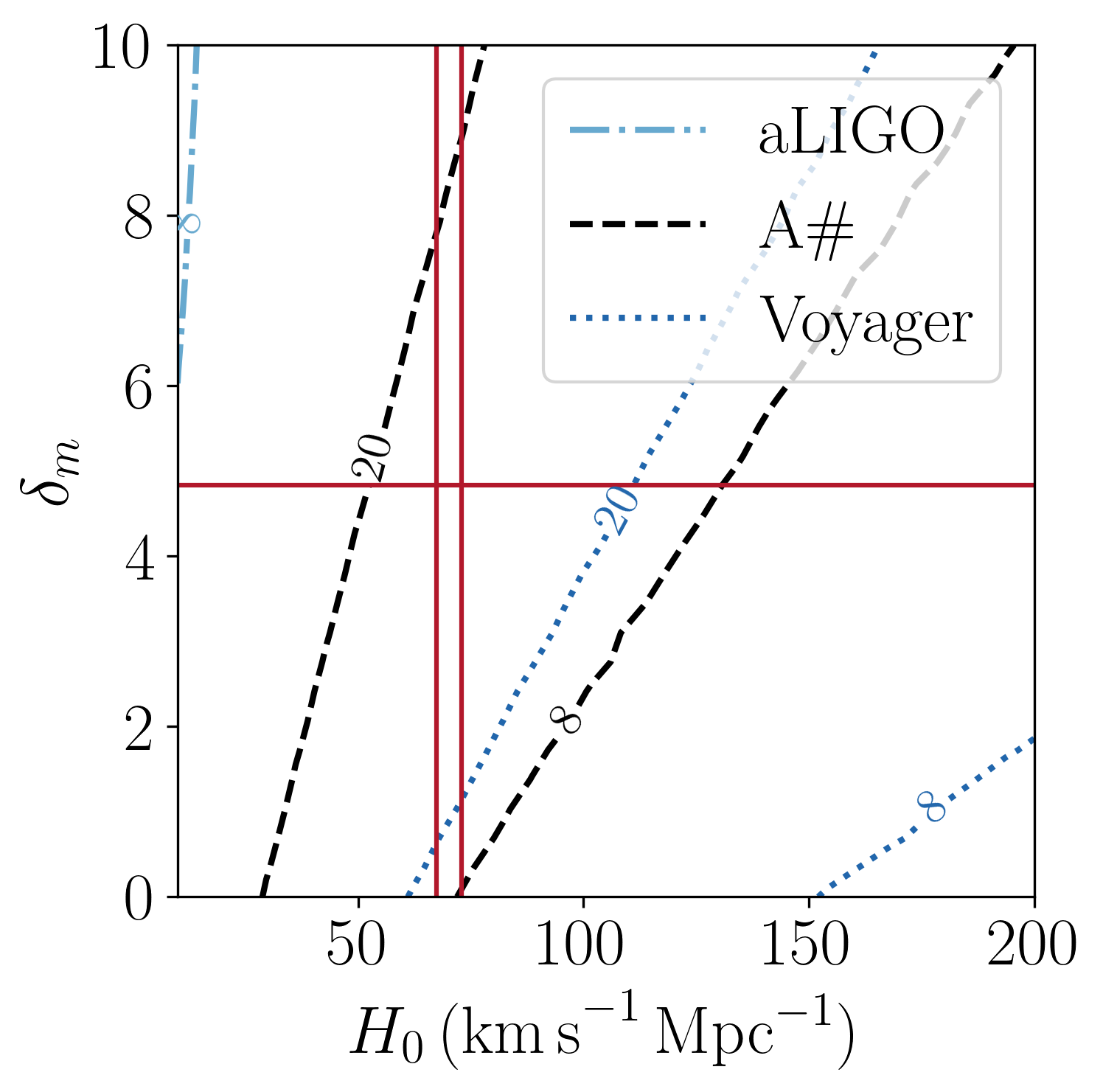}
    \hspace{0.9cm}
    \includegraphics[width=0.395\linewidth]{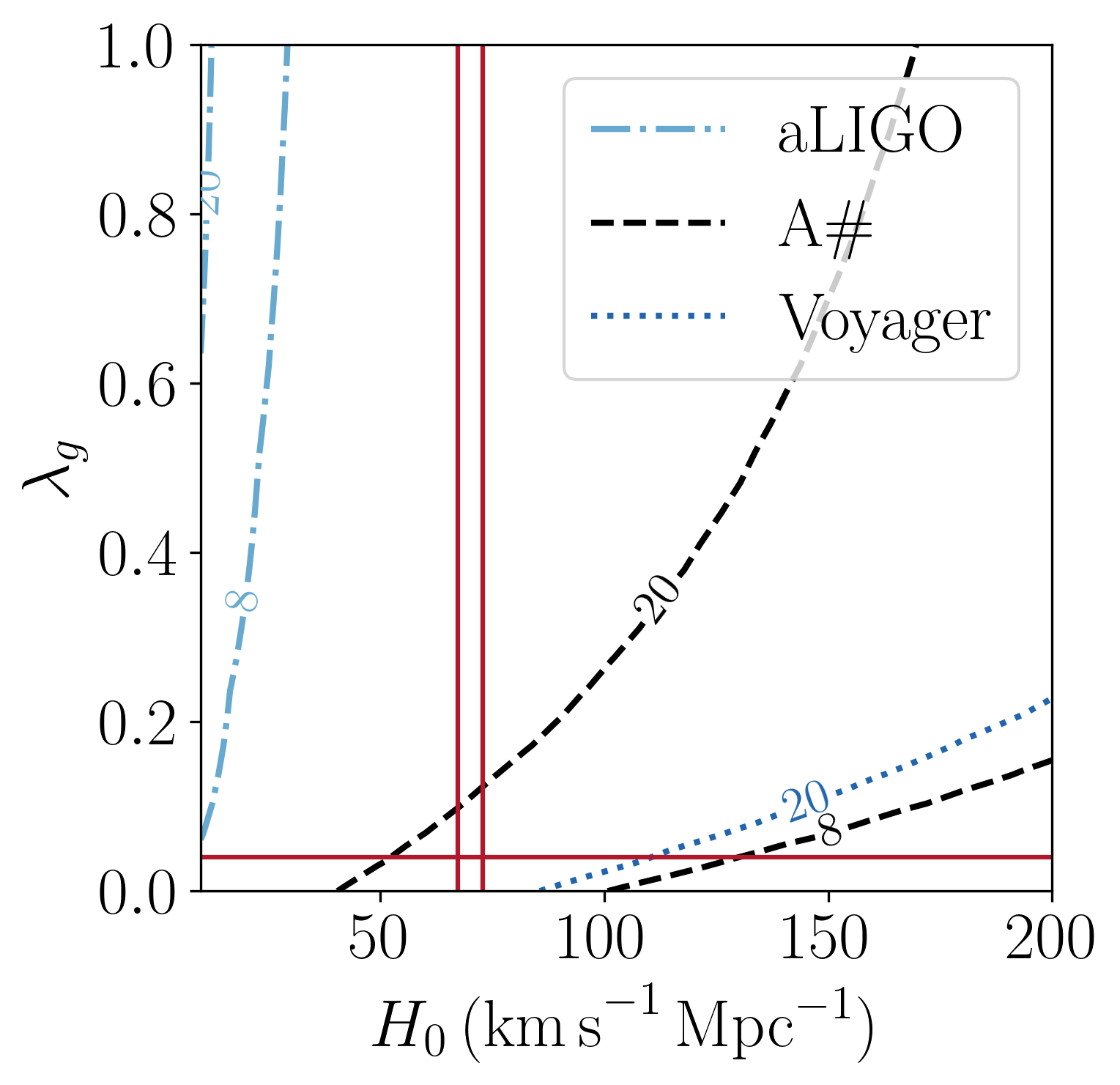}
    \caption{SNR of the GWB for various future detector networks as a function of two parameters ($H_0$ with either $\delta_m$ or $\lambda_g$), with the other 13 (hyper)parameters fixed to values denoted in the text. Refer to Fig.~\ref{fig:SNR_2d} for more explanation of the subplots.}
    \label{fig:SNR_2d_pop2}
\end{figure}

\begin{figure}
%    \centering
    \includegraphics[width=0.45\linewidth]{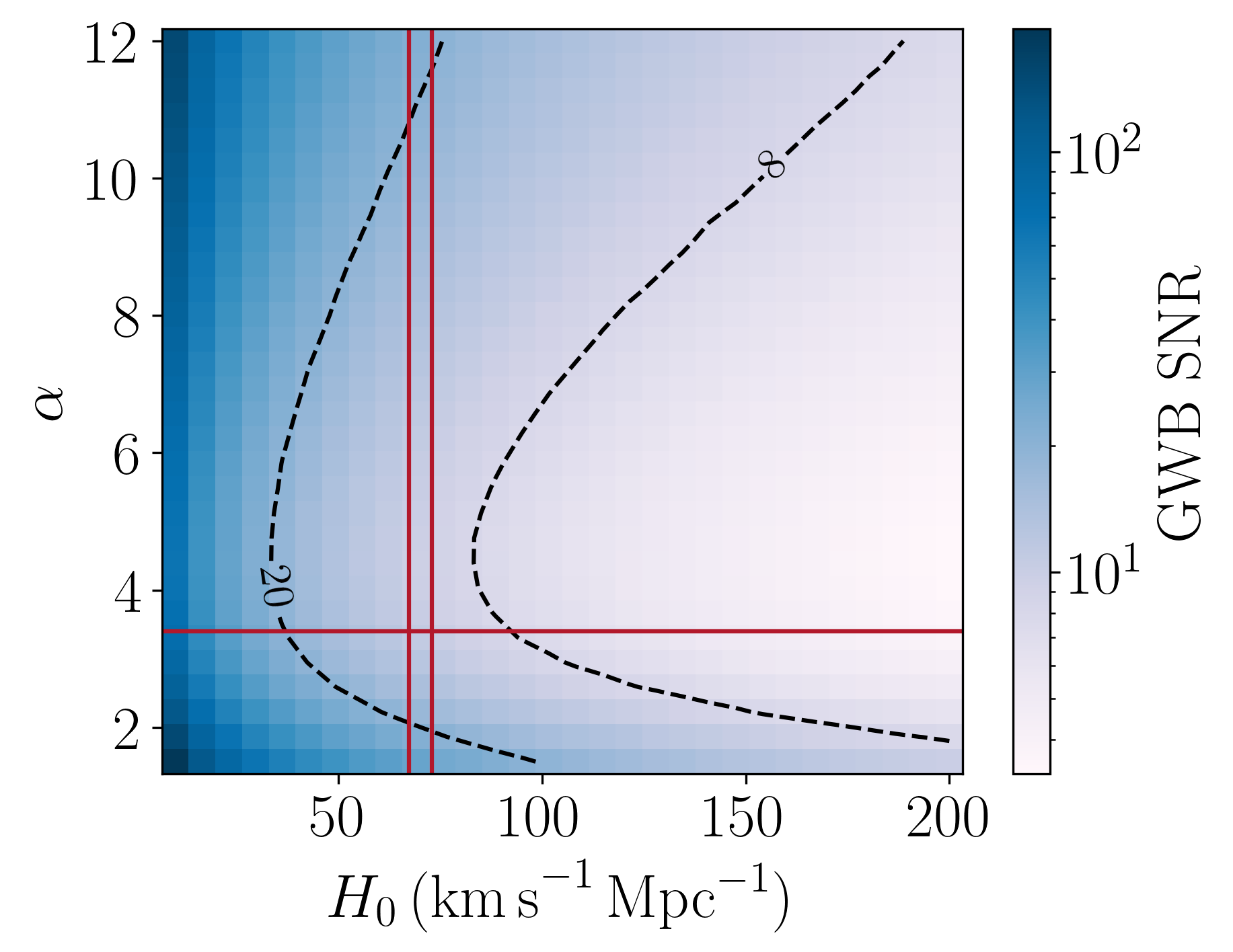}
    \includegraphics[width=0.45\linewidth]{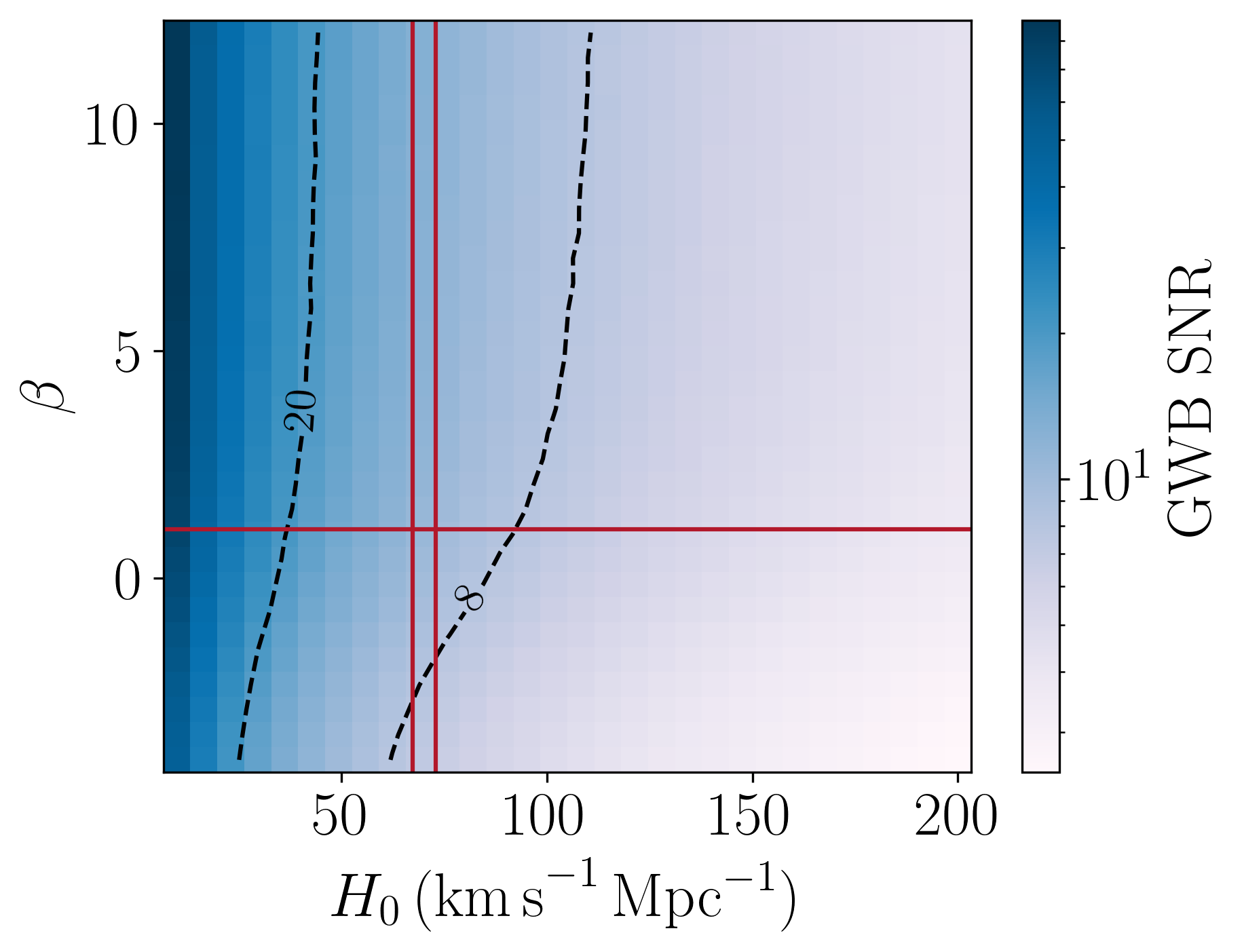}
    \includegraphics[width=0.45\linewidth]{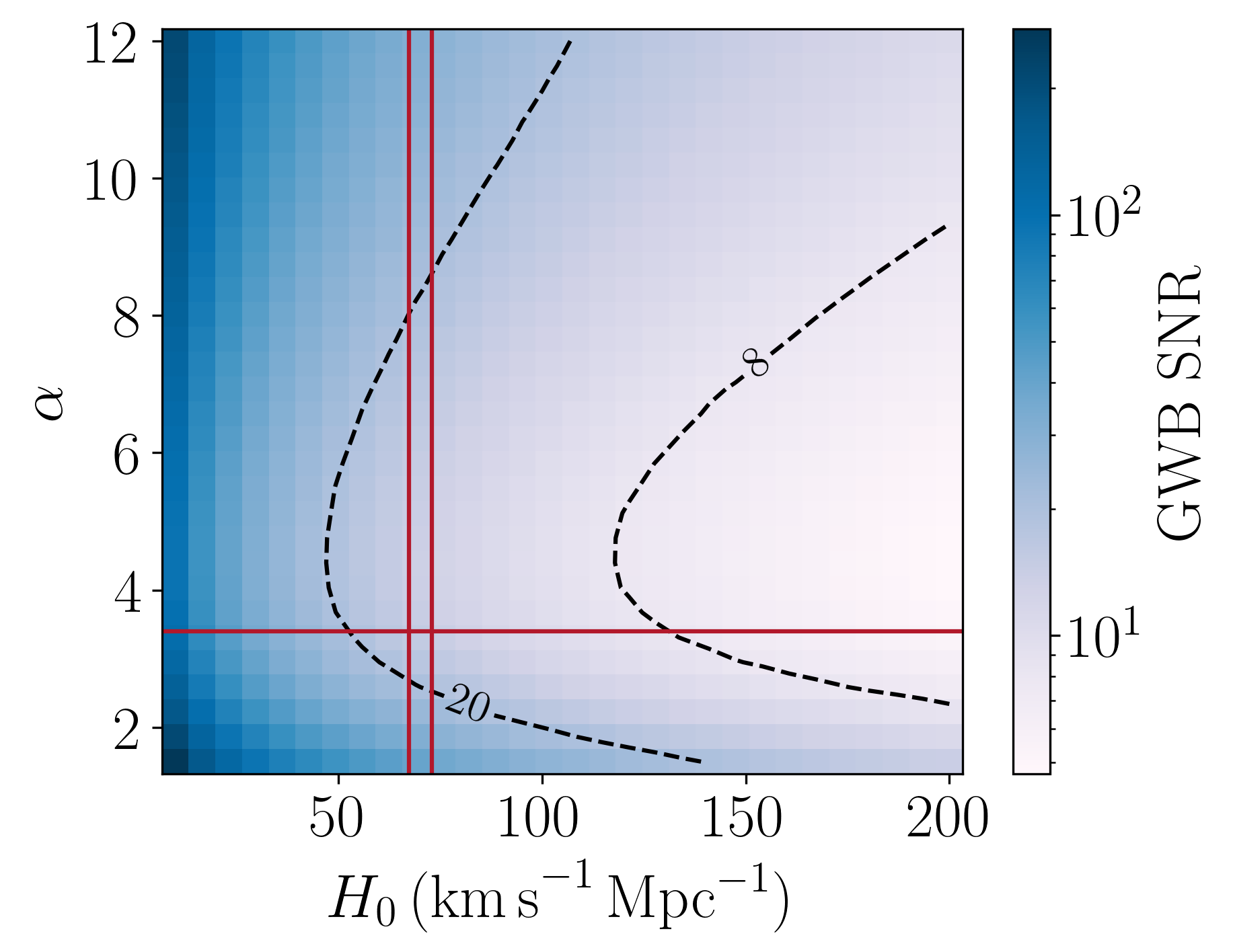}
    \includegraphics[width=0.45\linewidth]{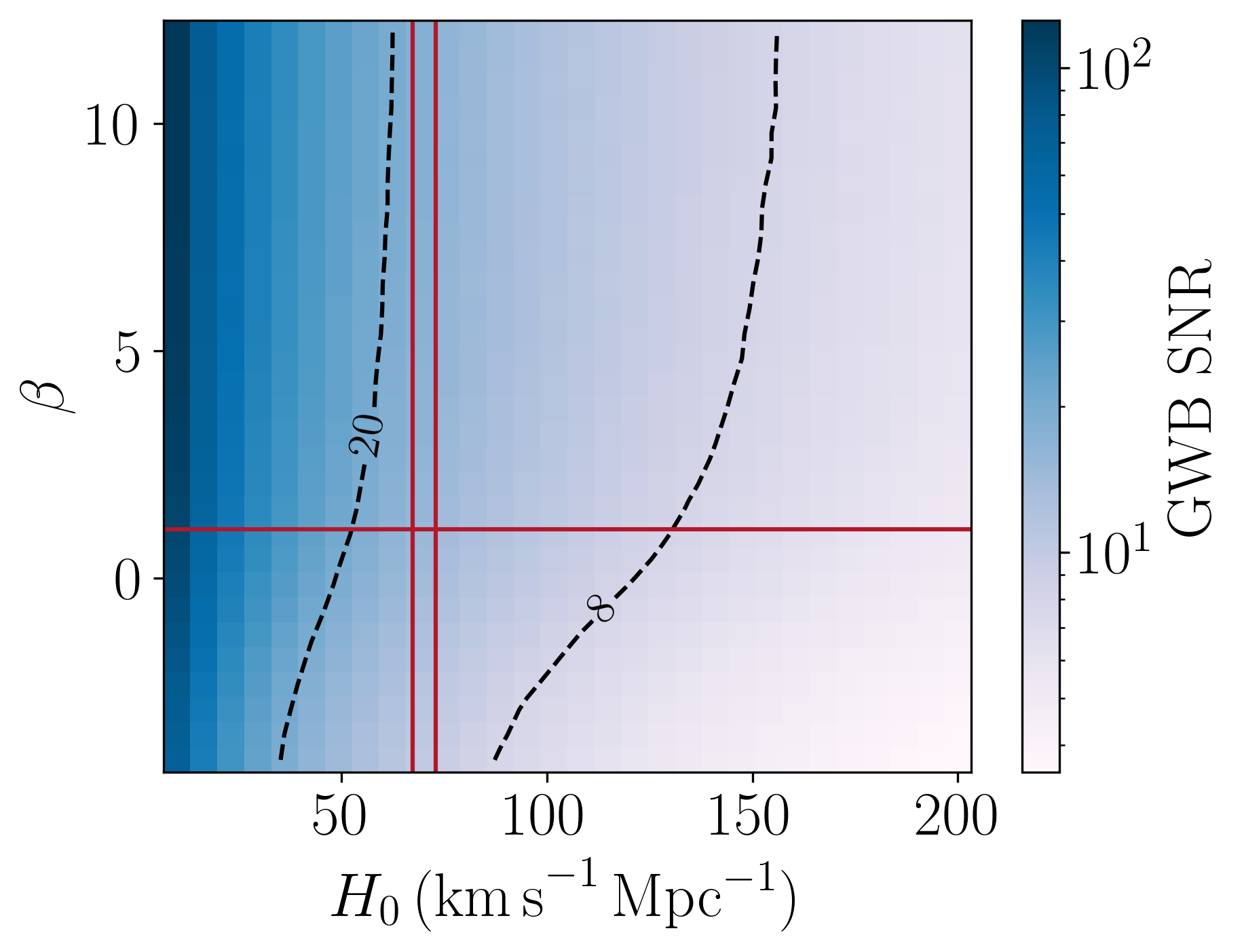}
    \\
    \hspace{-1.35cm}
    \includegraphics[width=0.378\linewidth]{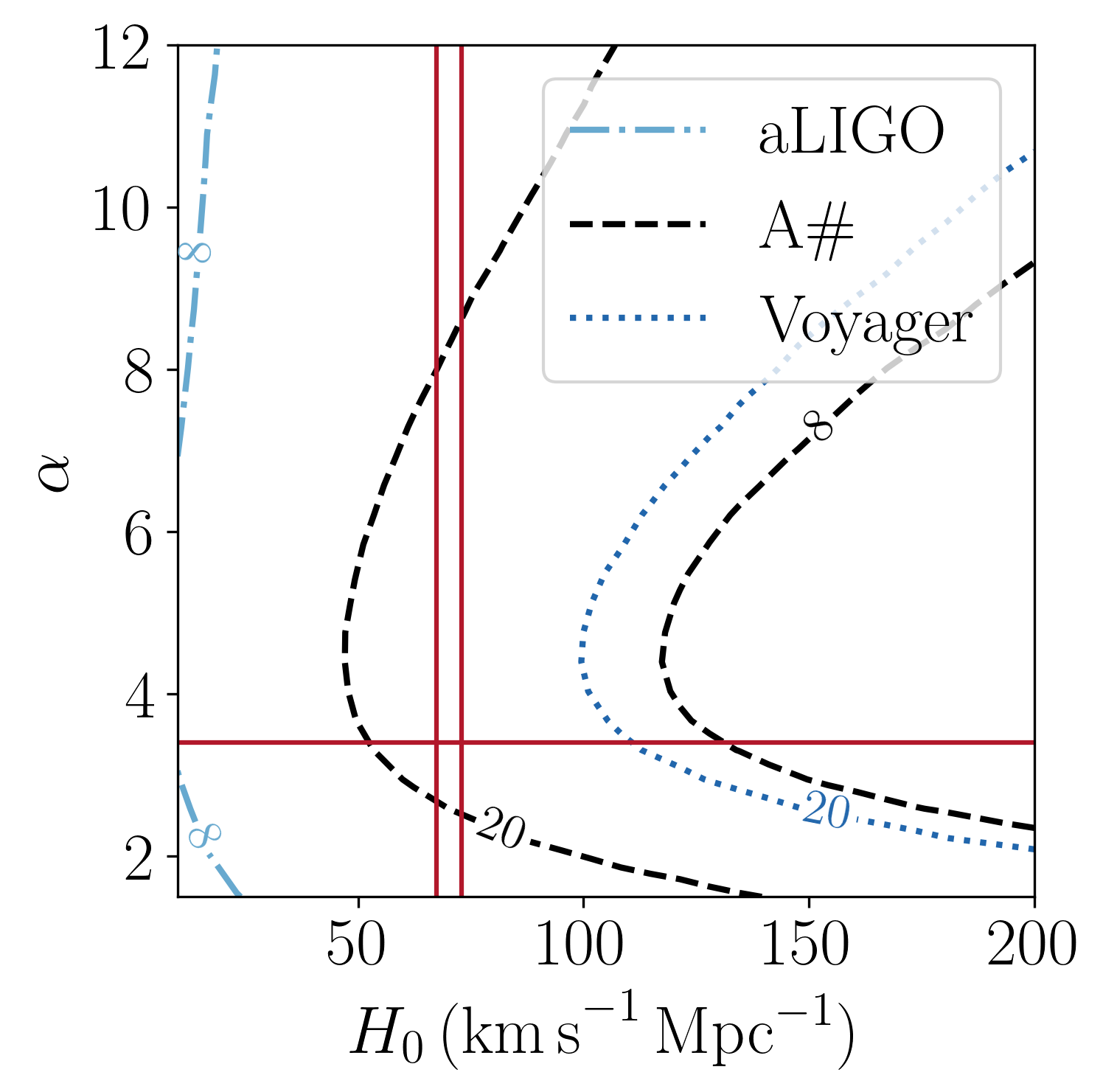}
    \hspace{0.9cm}
    \includegraphics[width=0.378\linewidth]{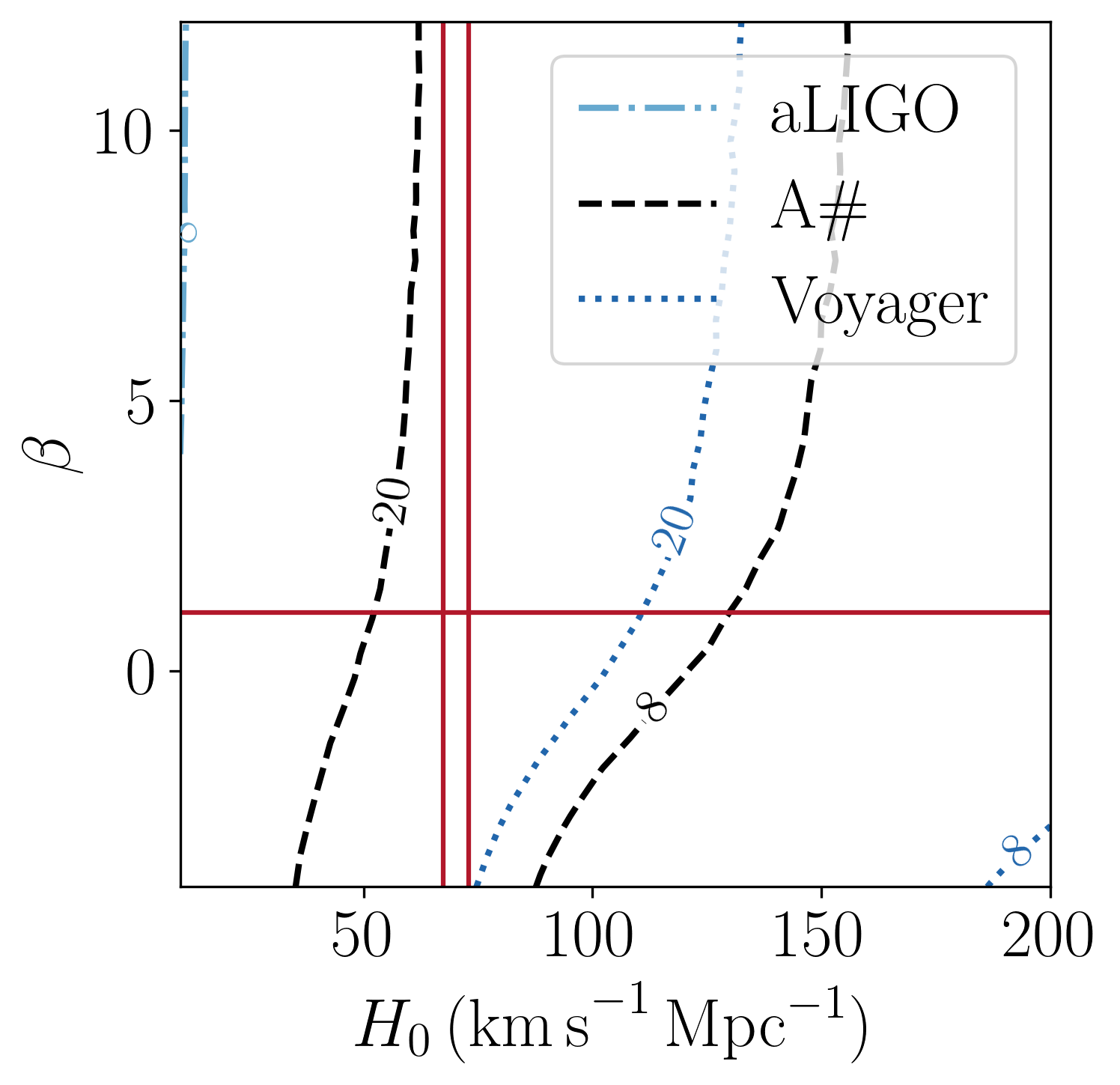}
    \caption{SNR of the GWB for various future detector networks as a function of two parameters ($H_0$ with either $\alpha$ or $\beta$), with the other 13 (hyper)parameters fixed to values denoted in the text. Refer to Fig.~\ref{fig:SNR_2d} for more explanation of the subplots.}
    \label{fig:SNR_2d_pop3}
\end{figure}

\begin{figure}
%    \centering
    \includegraphics[width=0.45\linewidth]{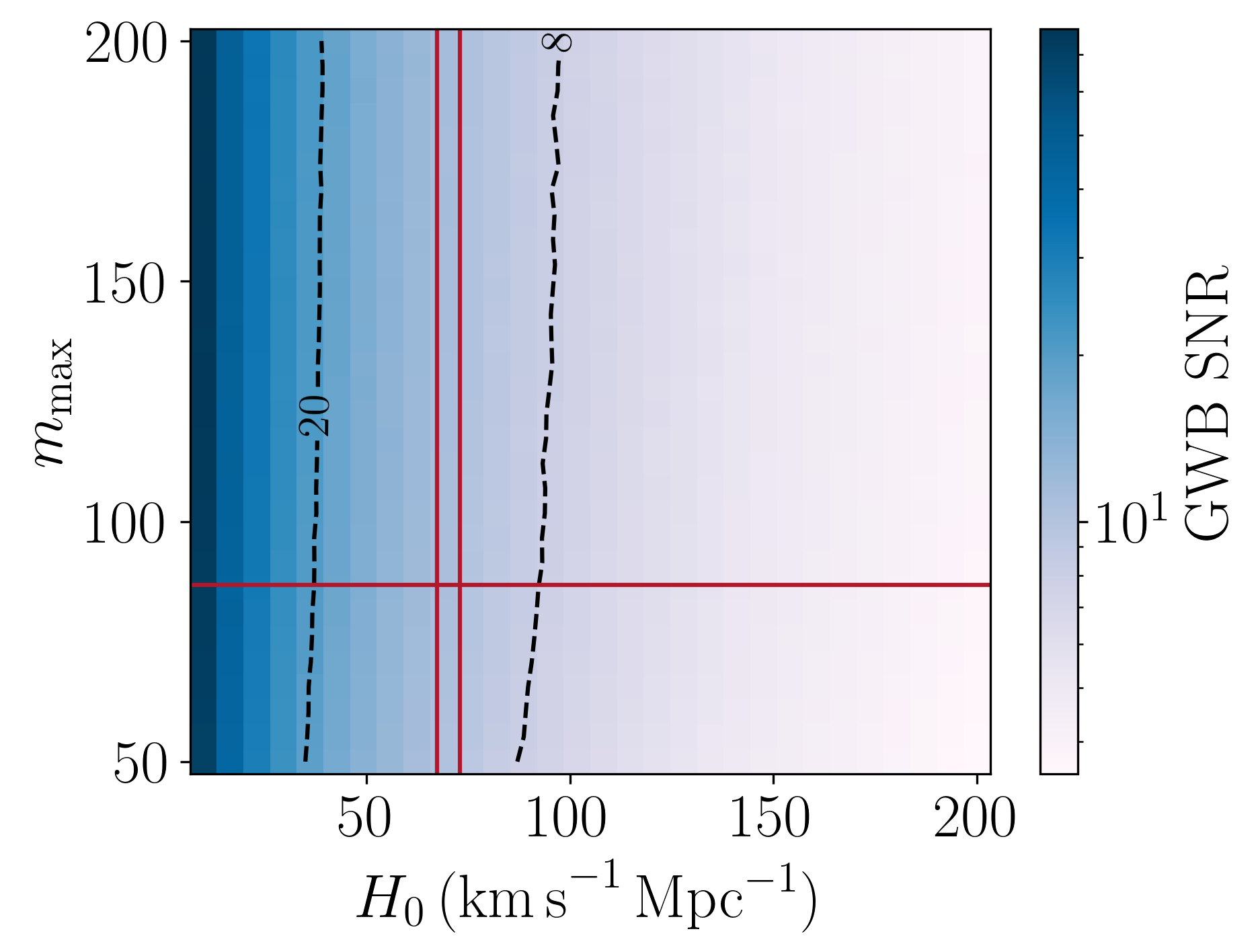}
    \includegraphics[width=0.45\linewidth]{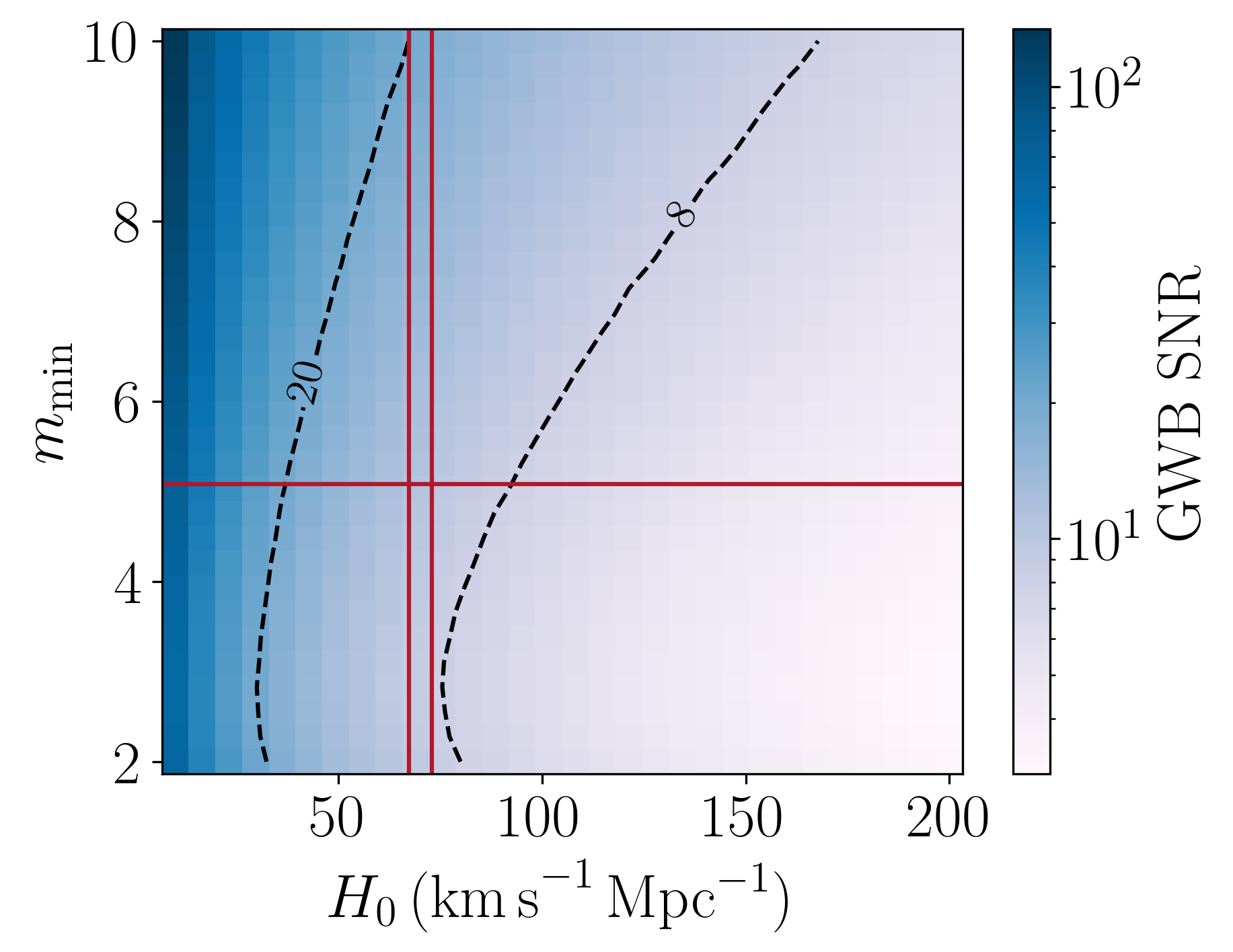}
    \includegraphics[width=0.45\linewidth]{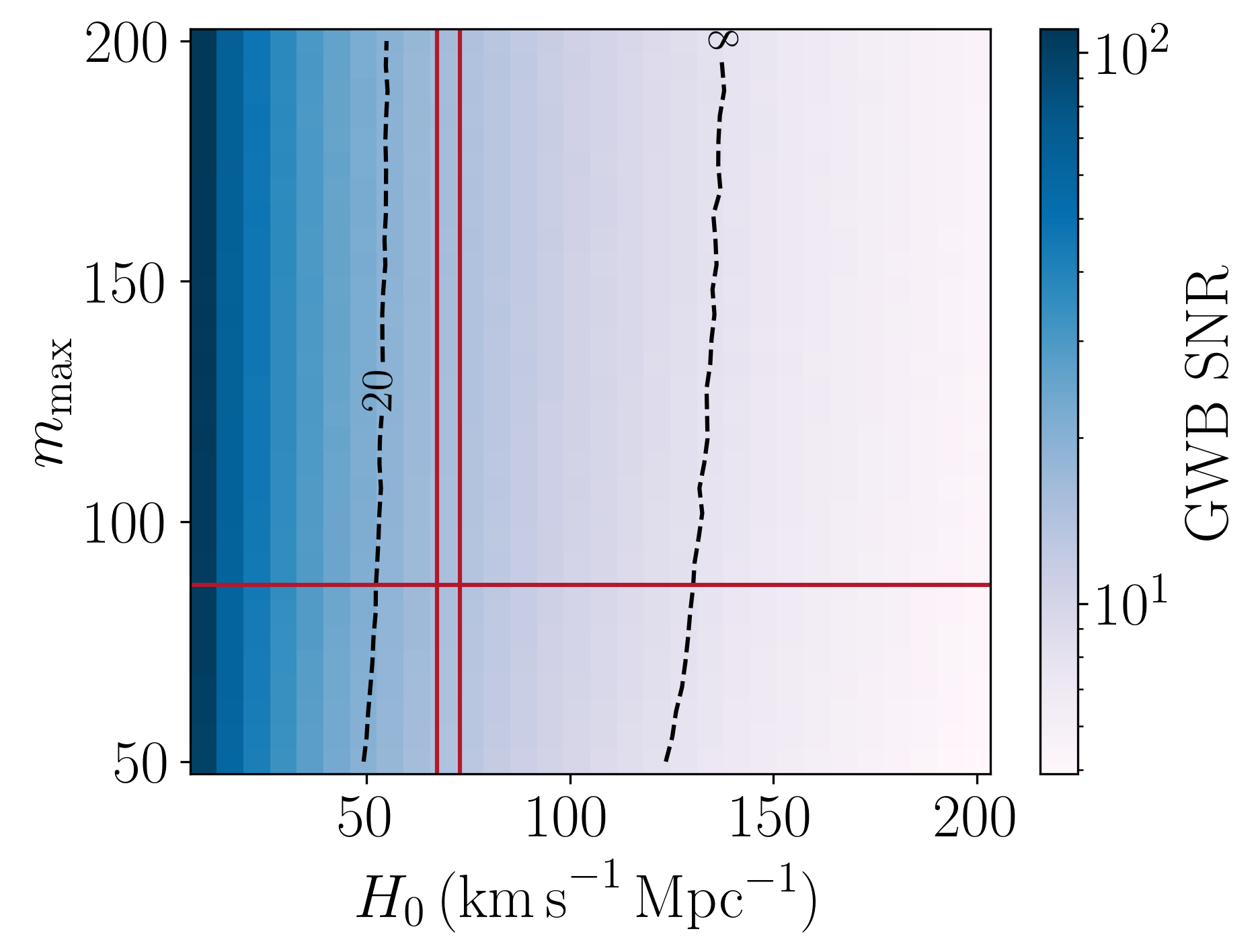}
    \includegraphics[width=0.45\linewidth]{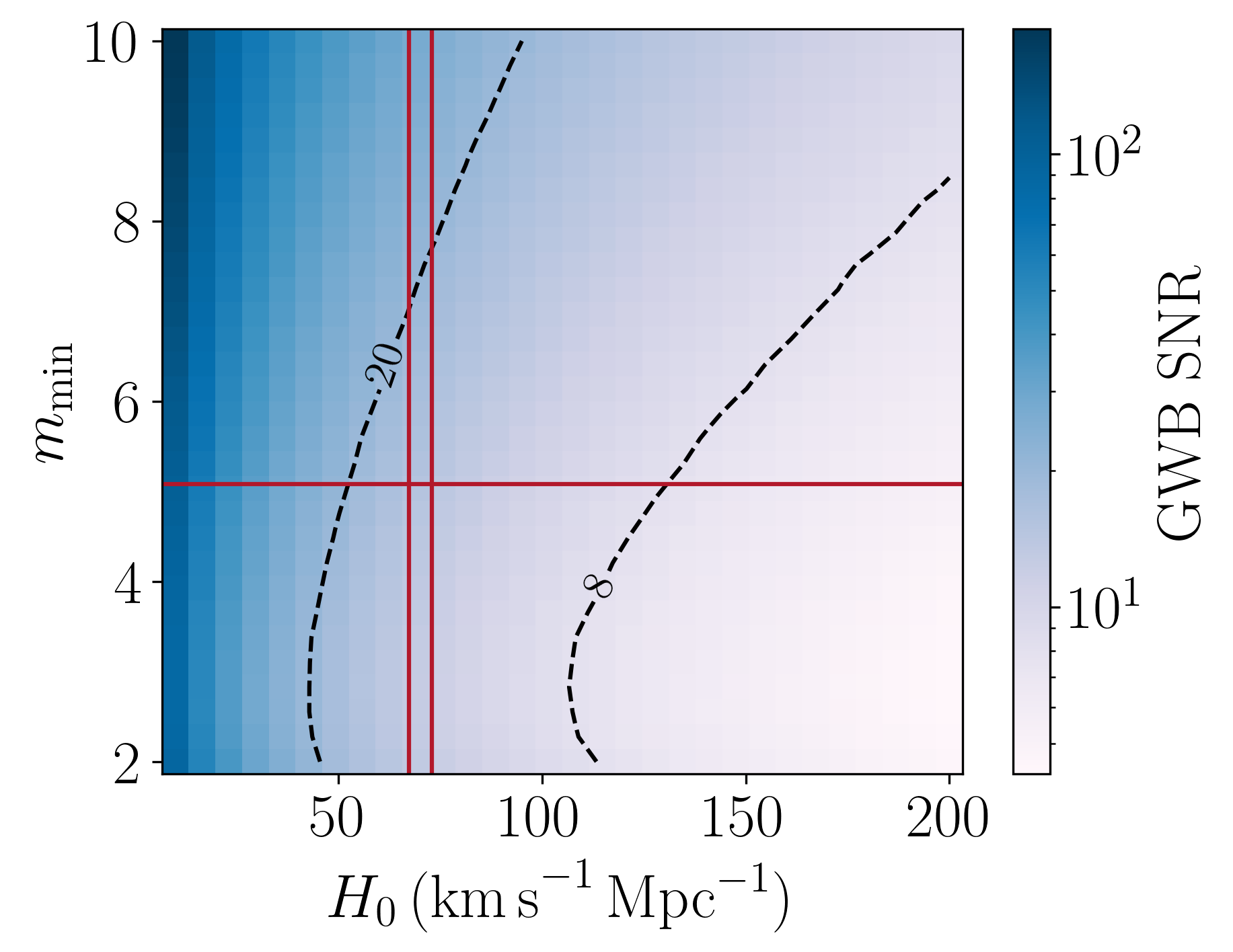}
    \\
    \hspace{-1.35cm}
    \includegraphics[width=0.378\linewidth]{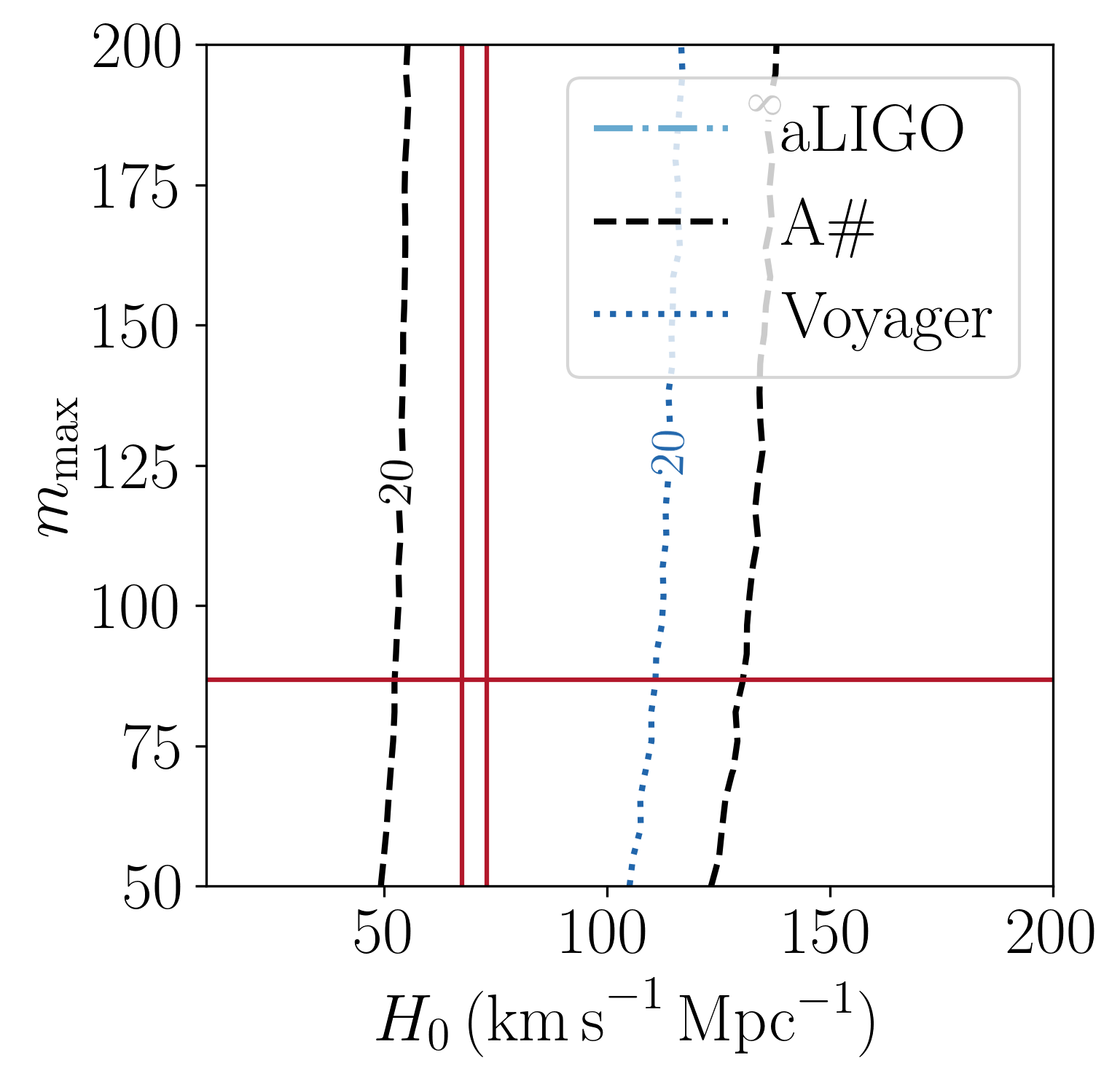}
    \hspace{0.9cm}
    \includegraphics[width=0.378\linewidth]{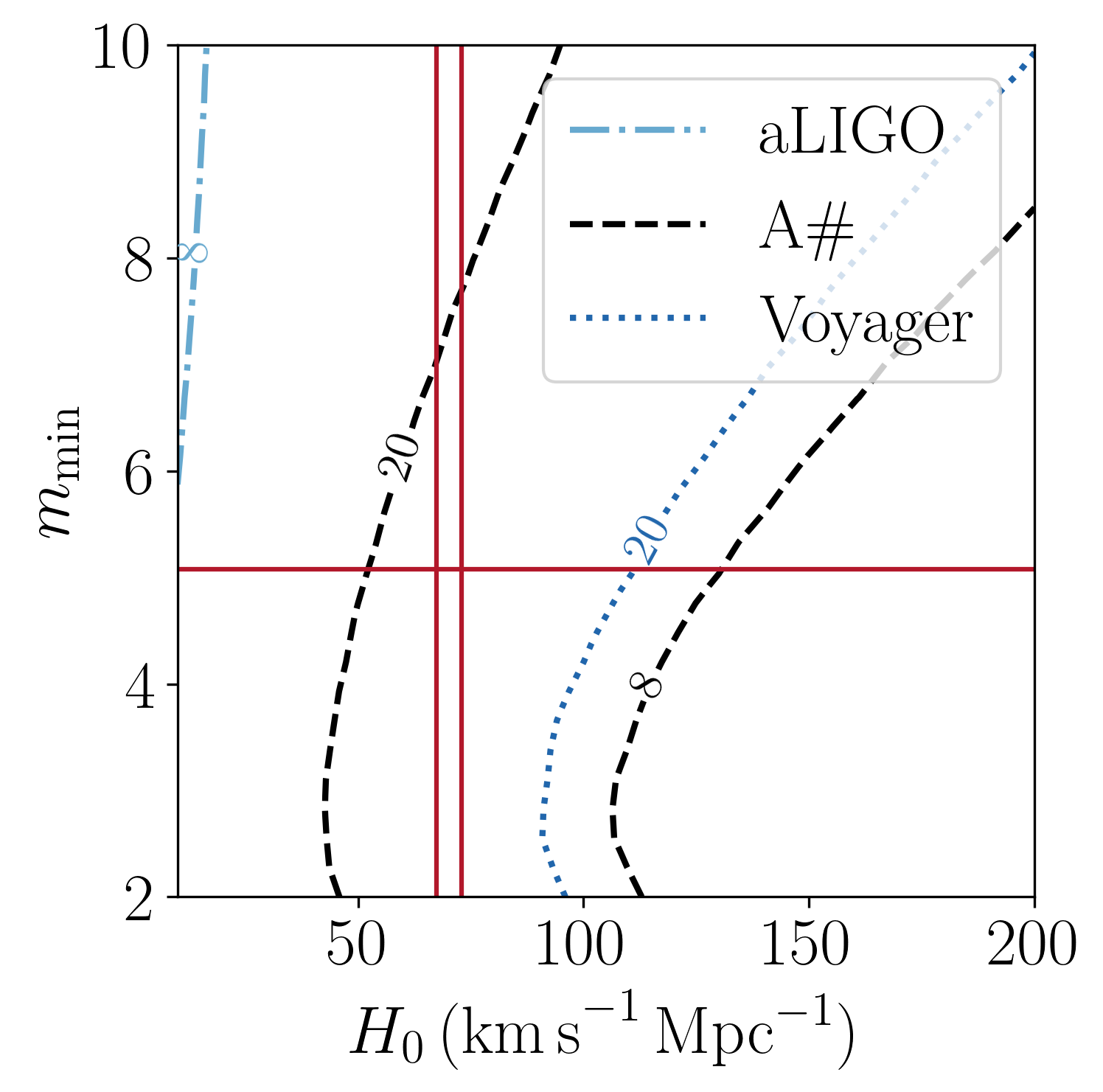}
    \caption{SNR of the GWB for various future detector networks as a function of two parameters ($H_0$ with either $m_{\rm{max}}$ or $m_{\rm{min}}$), with the other 13 (hyper)parameters fixed to values denoted in the text. Refer to Fig.~\ref{fig:SNR_2d} for more explanation of the subplots.}
    \label{fig:SNR_2d_pop4}
\end{figure}

\begin{figure}
%    \centering
    \includegraphics[width=0.45\linewidth]{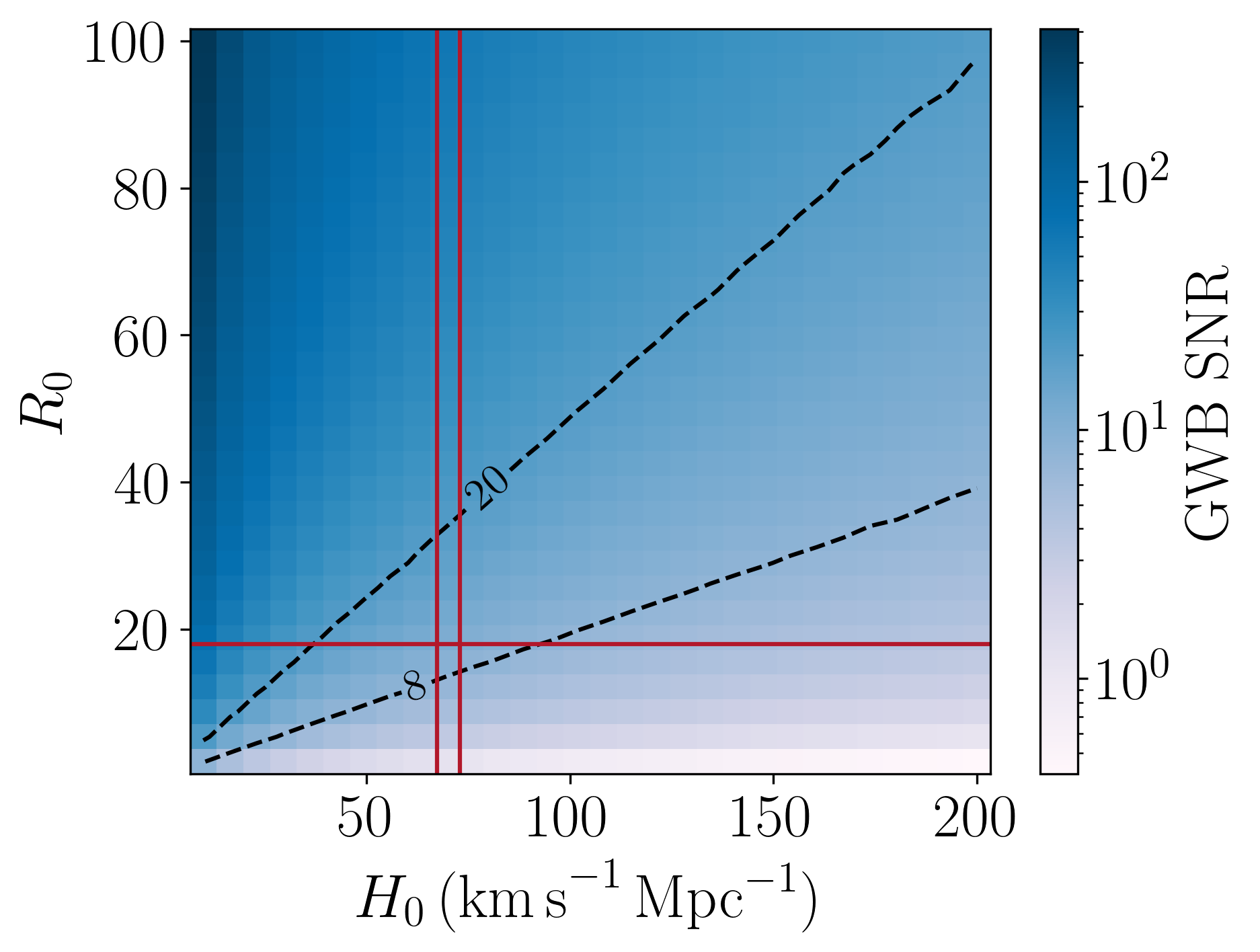}
    \includegraphics[width=0.42\linewidth]{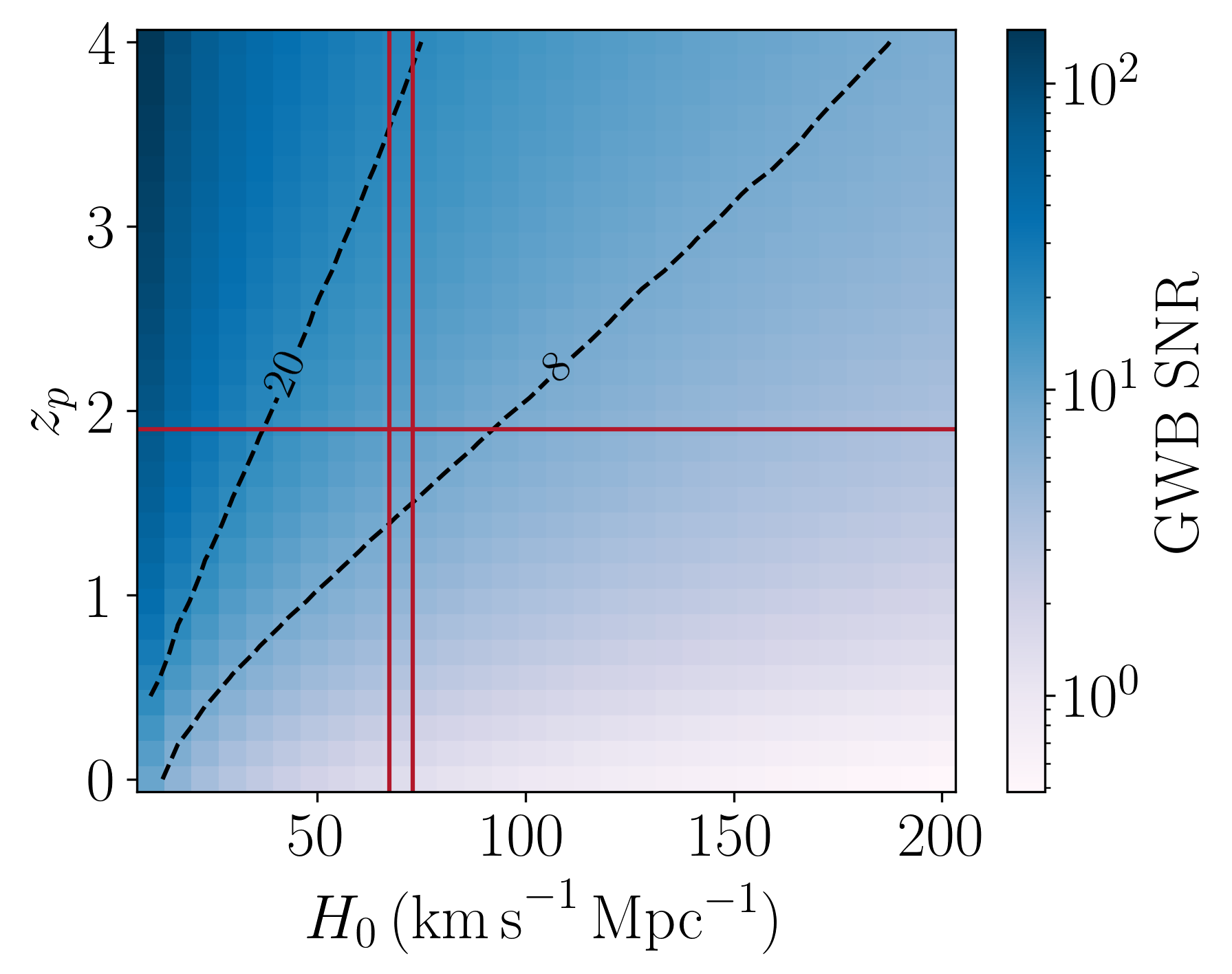}
    \includegraphics[width=0.45\linewidth]{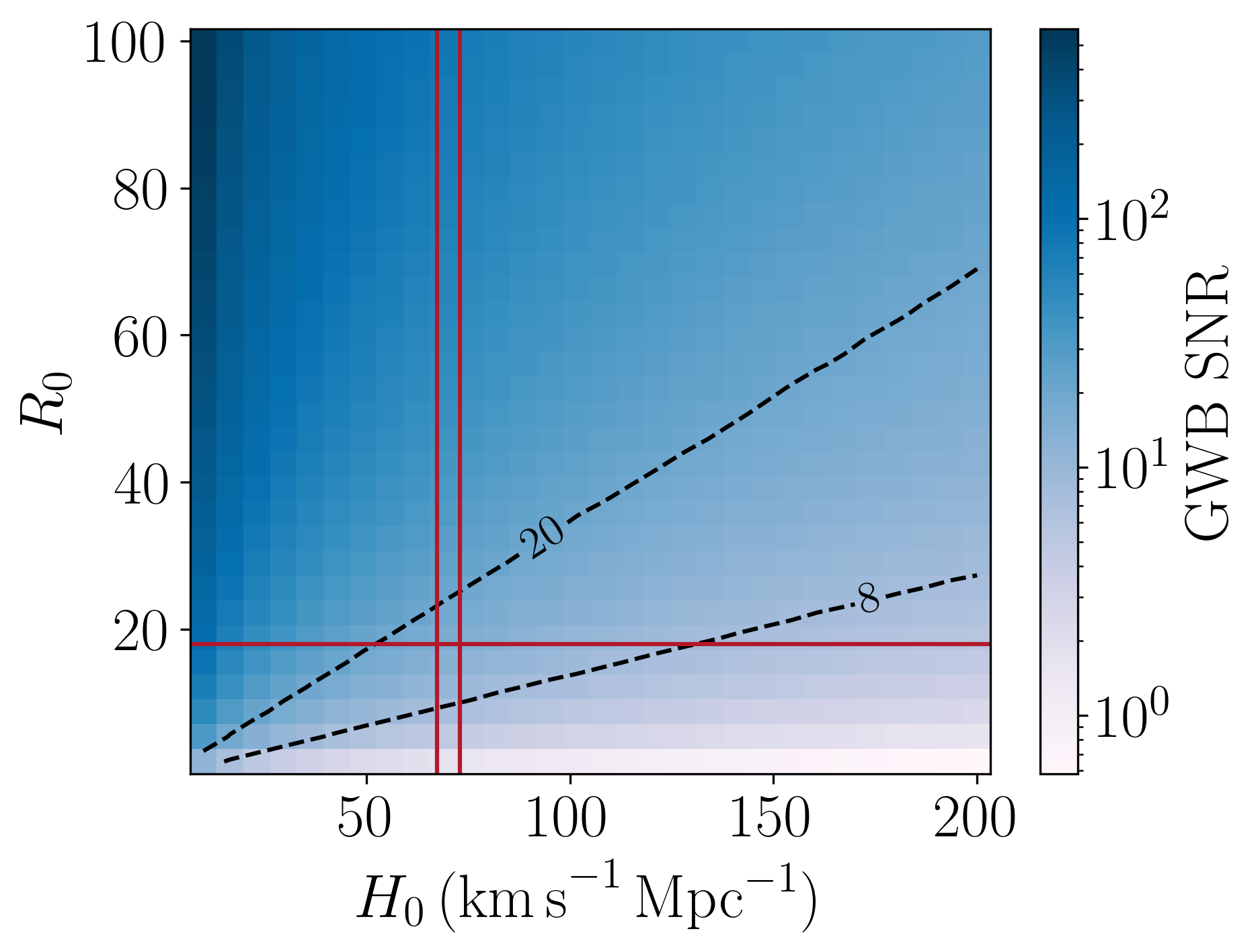}
    \includegraphics[width=0.42\linewidth]{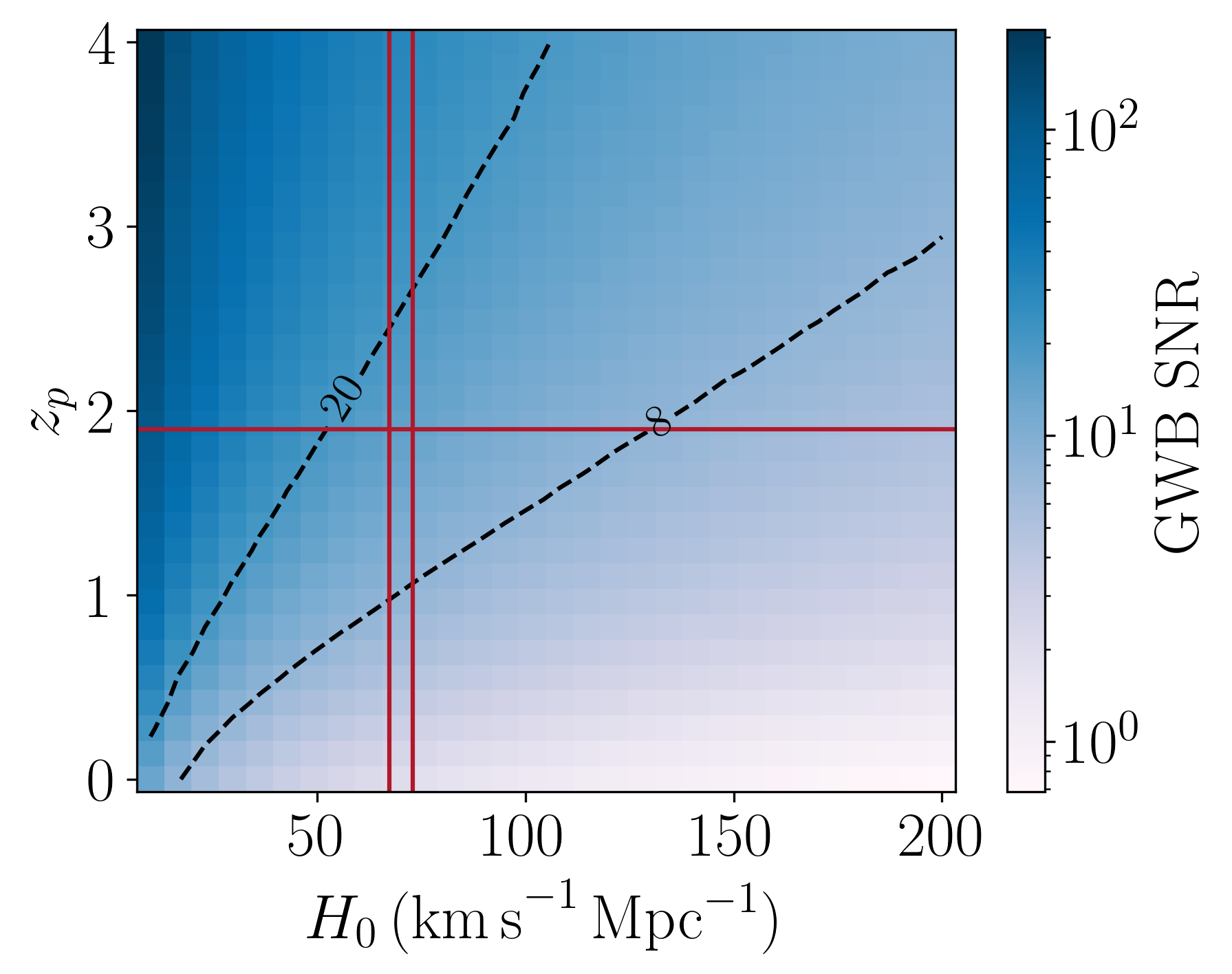}
    \\
    \hspace{-1.35cm}
    \includegraphics[width=0.378\linewidth]{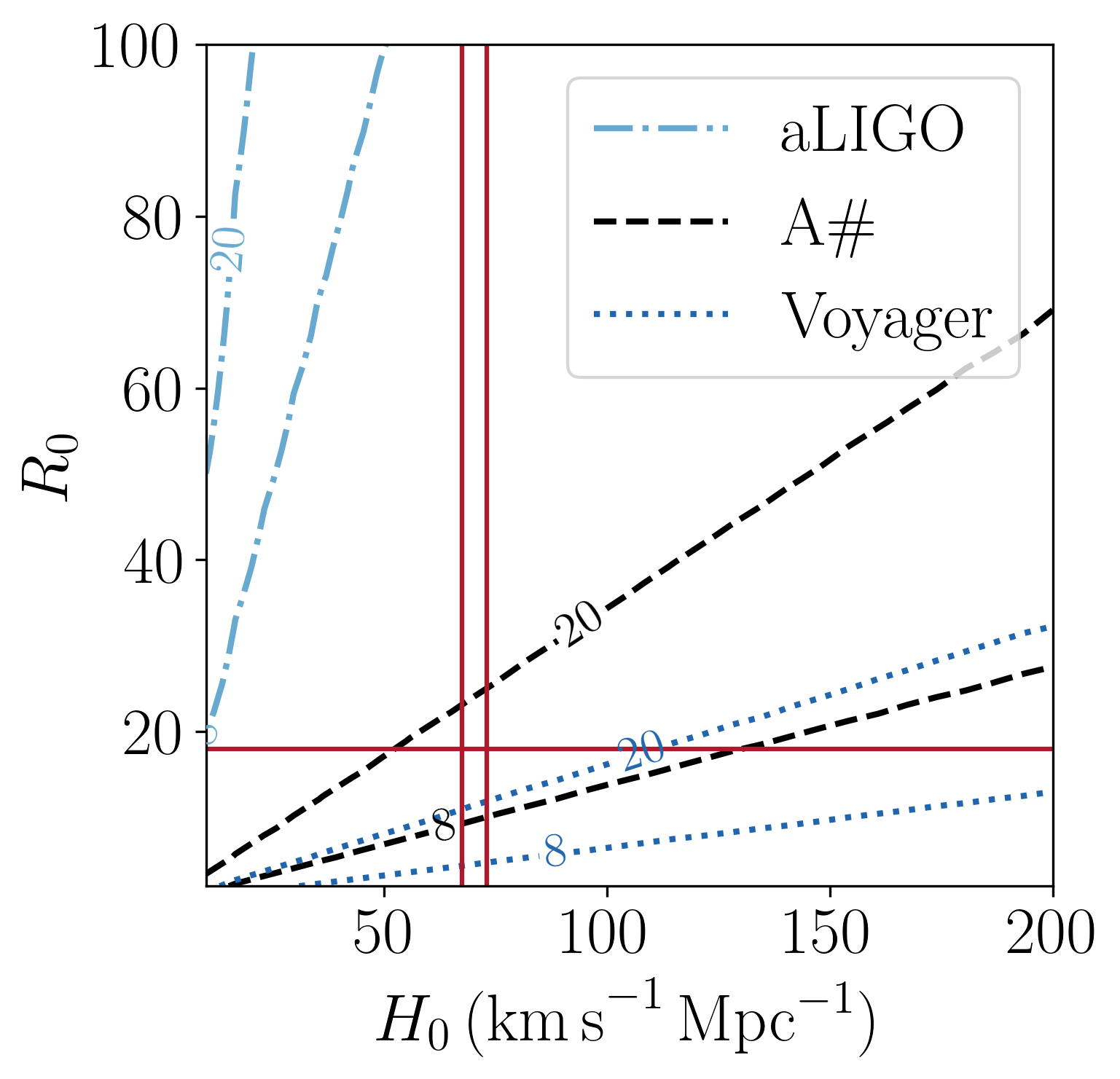}
    \hspace{0.9cm}
    \includegraphics[width=0.378\linewidth]{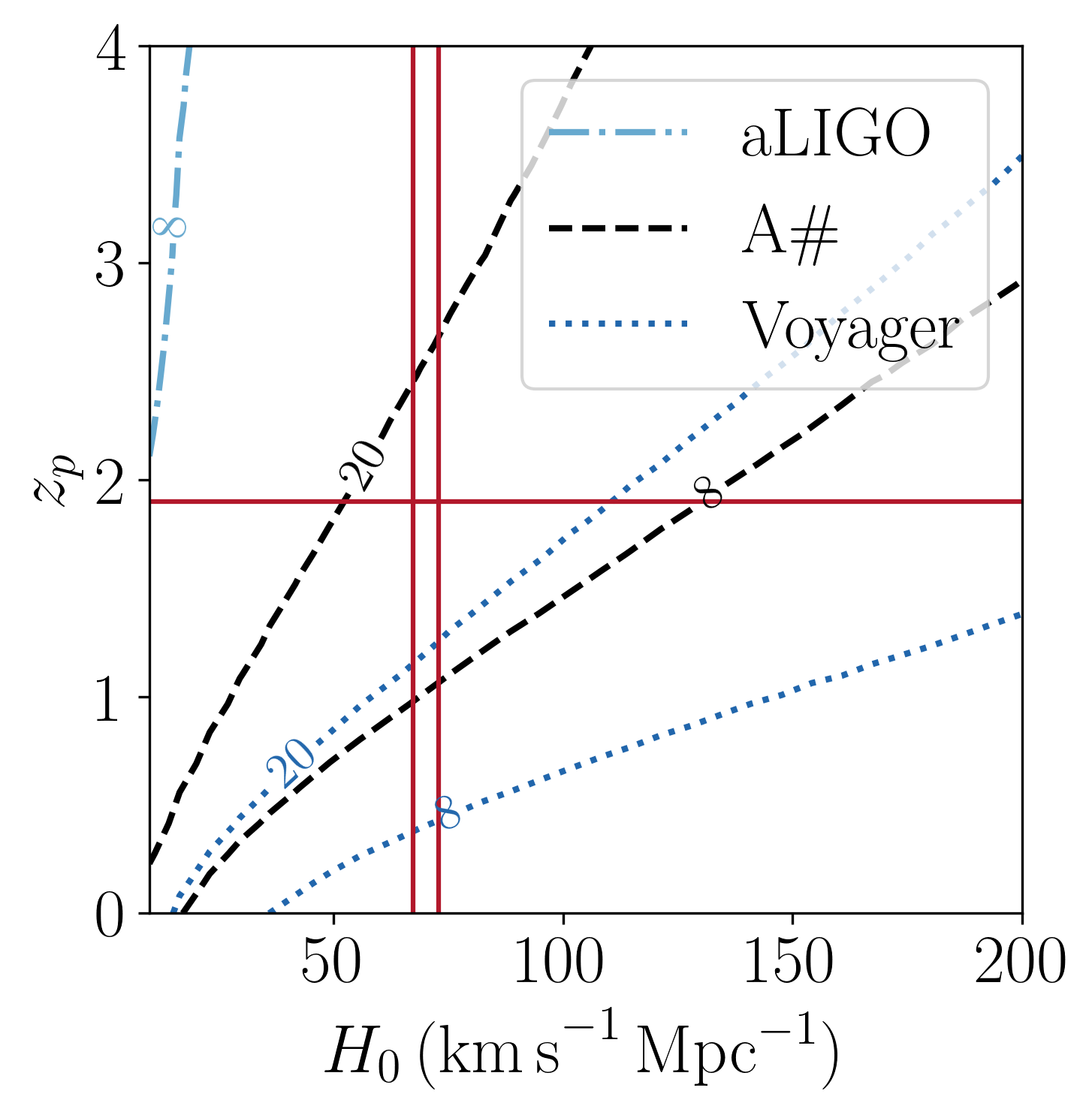}
    \caption{SNR of the GWB for various future detector networks as a function of two parameters ($H_0$ with either $R_0$ or $z_p$), with the other 13 (hyper)parameters fixed to values denoted in the text. Refer to Fig.~\ref{fig:SNR_2d} for more explanation of the subplots.}
    \label{fig:SNR_2d_sfr1}
\end{figure}

\begin{figure}
%    \centering
    \includegraphics[width=0.45\linewidth]{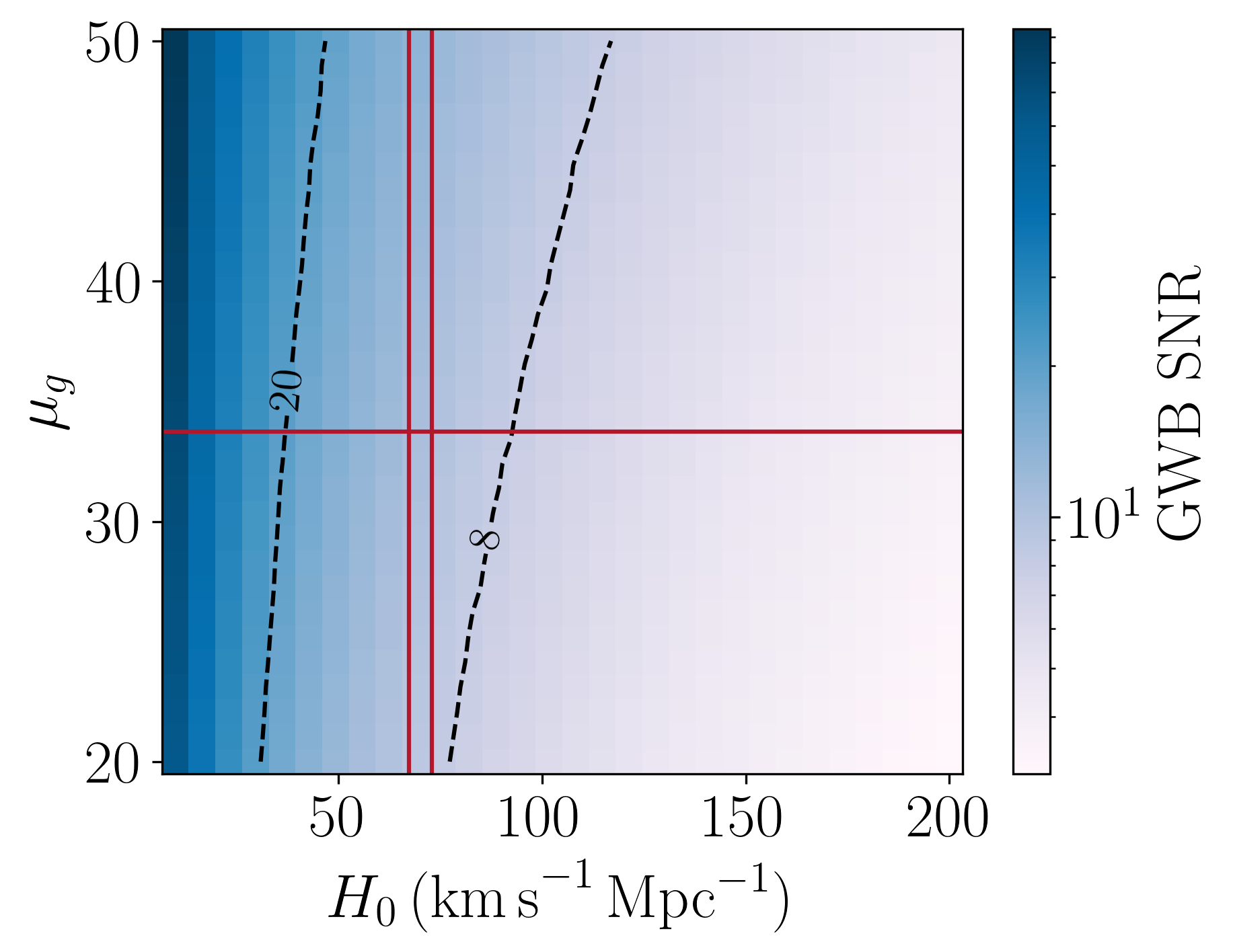}
    \includegraphics[width=0.45\linewidth]{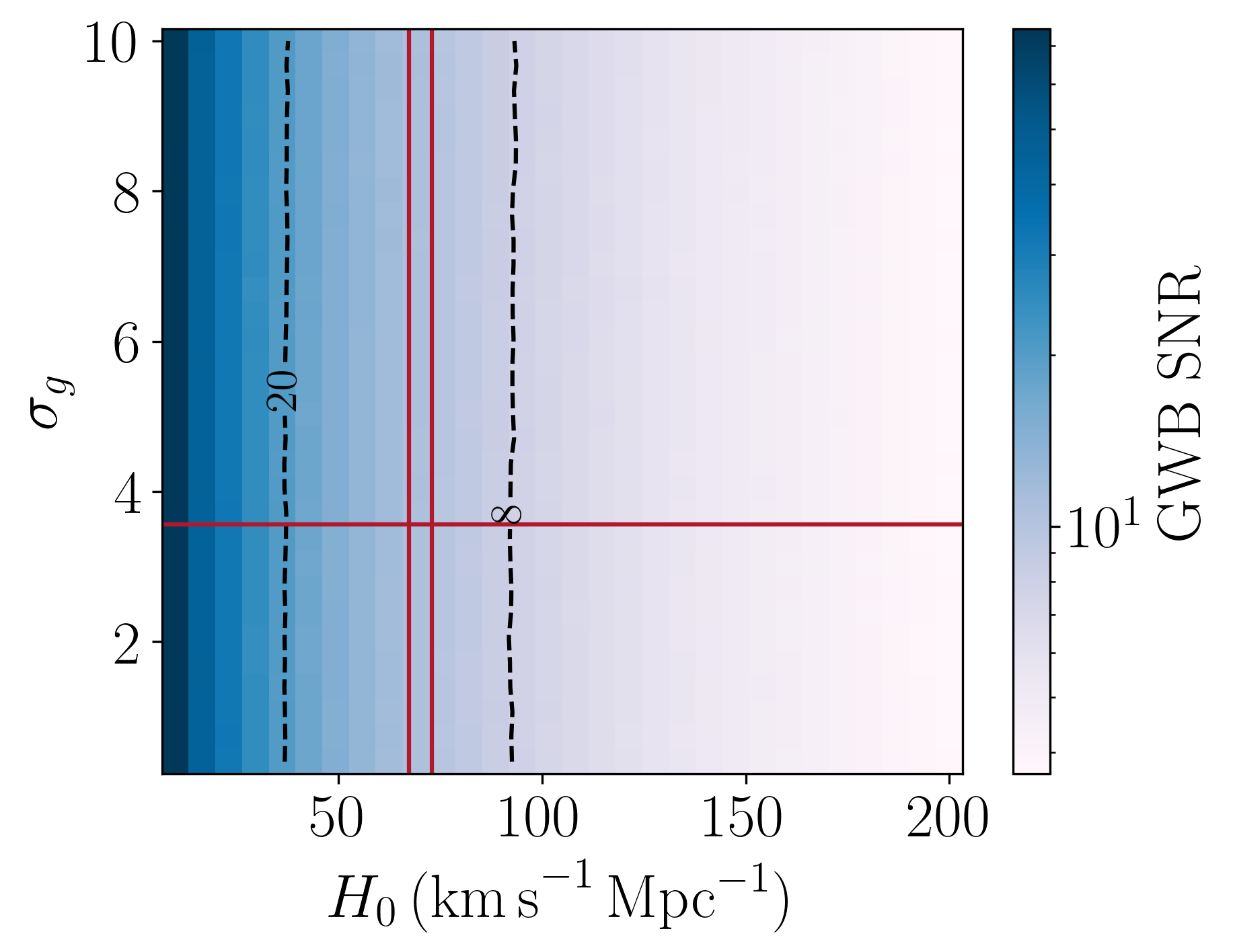}
    \includegraphics[width=0.45\linewidth]{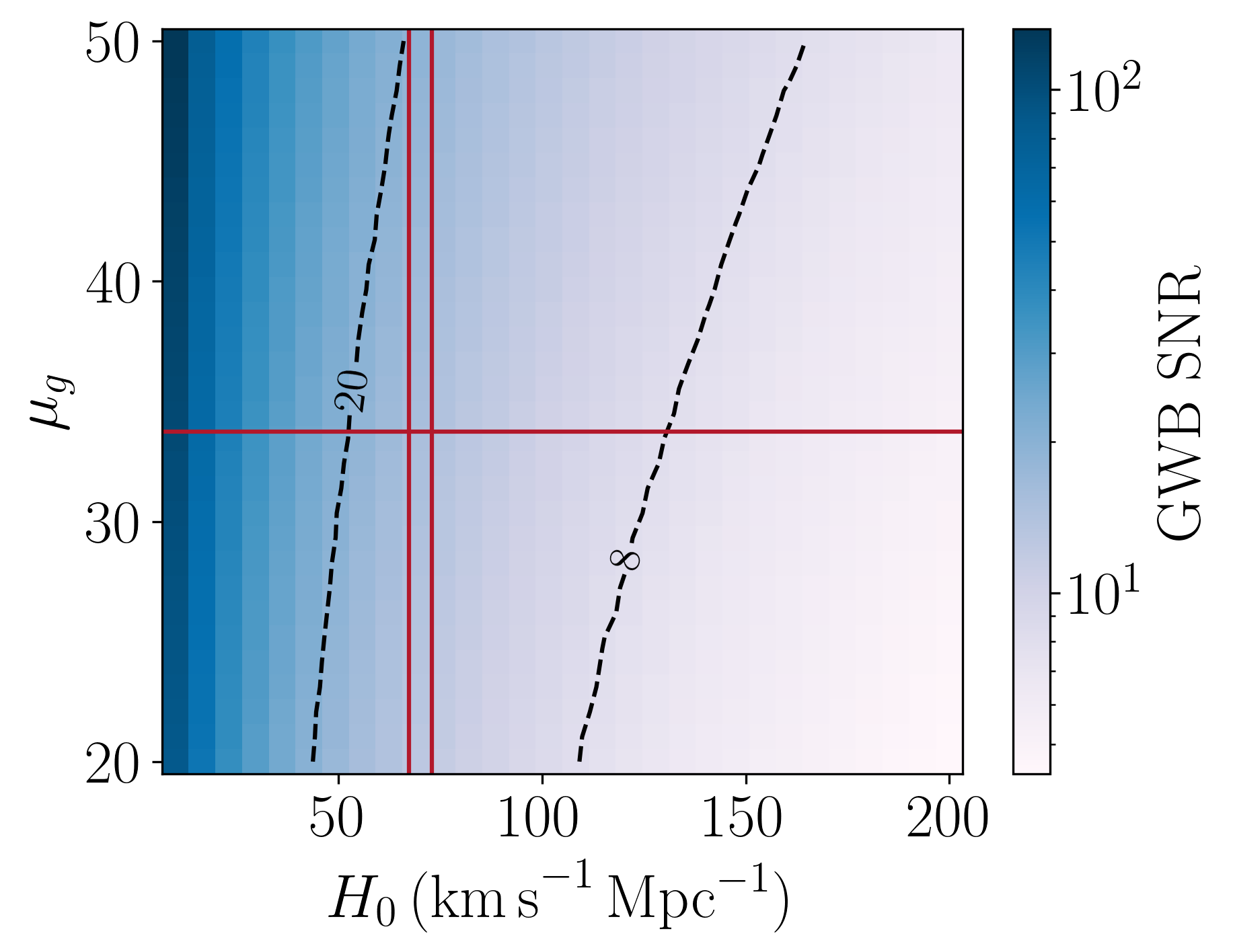}
    \includegraphics[width=0.45\linewidth]{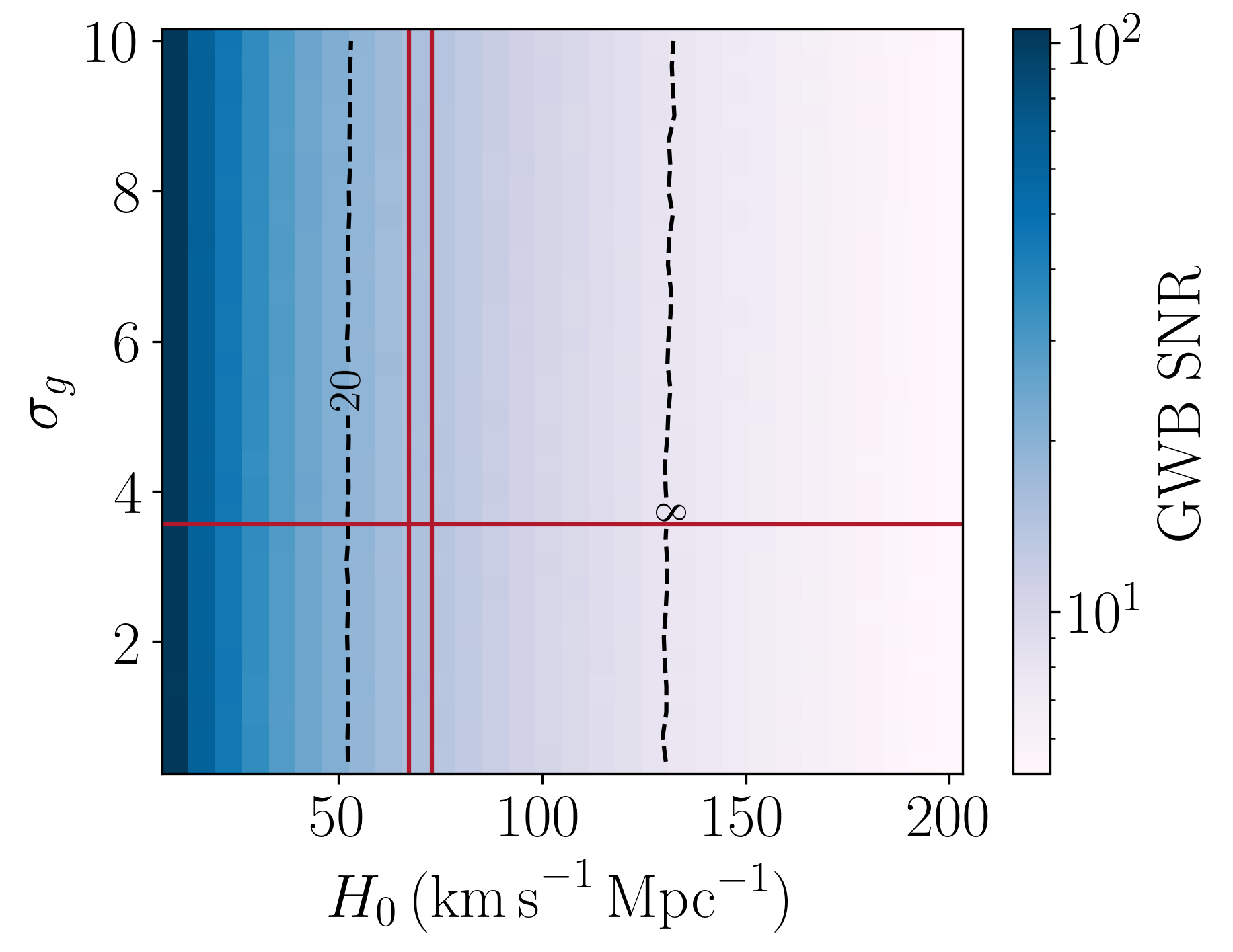}
    \\
    \hspace{-1.35cm}
    \includegraphics[width=0.378\linewidth]{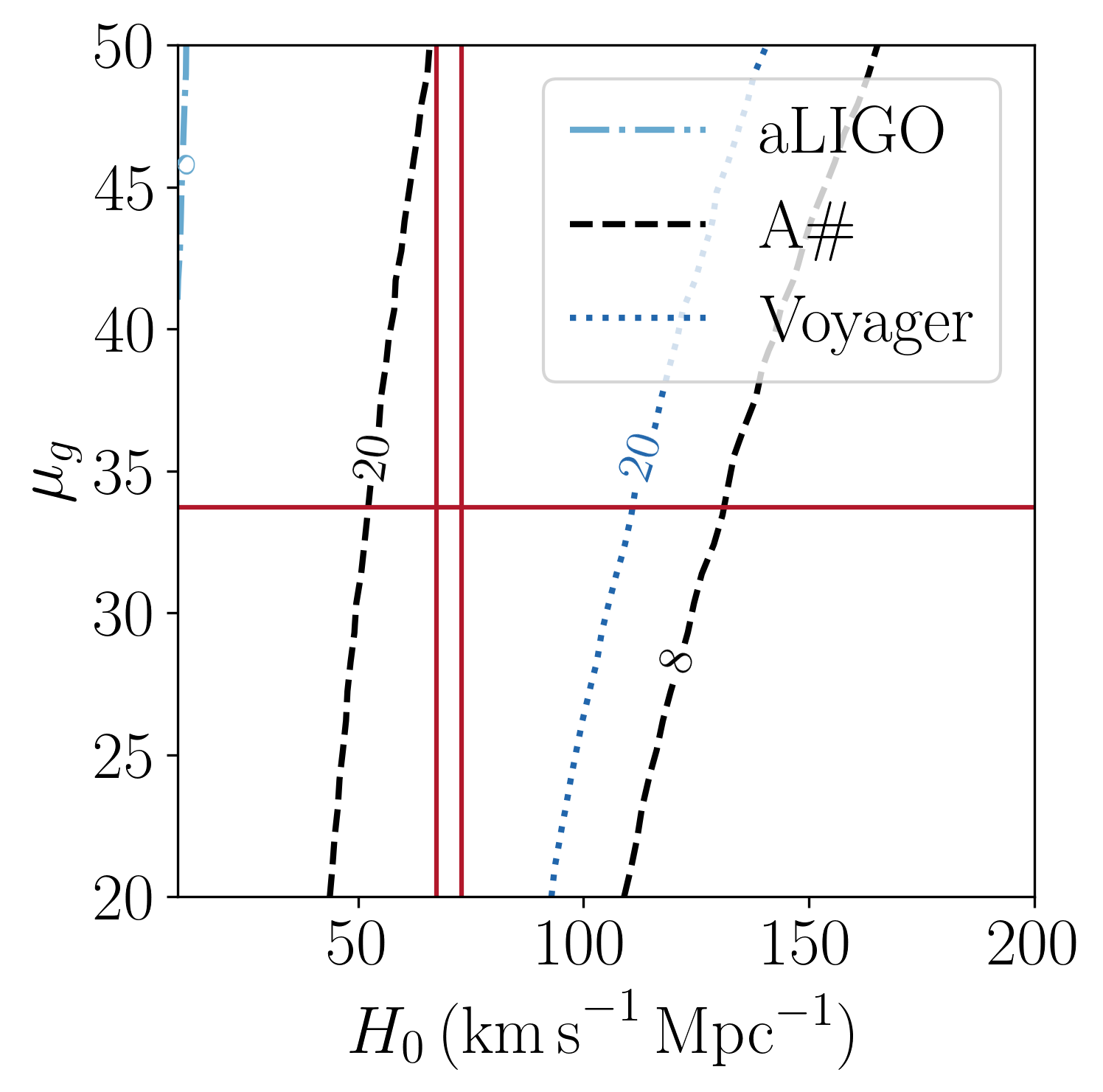}
    \hspace{0.9cm}
    \includegraphics[width=0.378\linewidth]{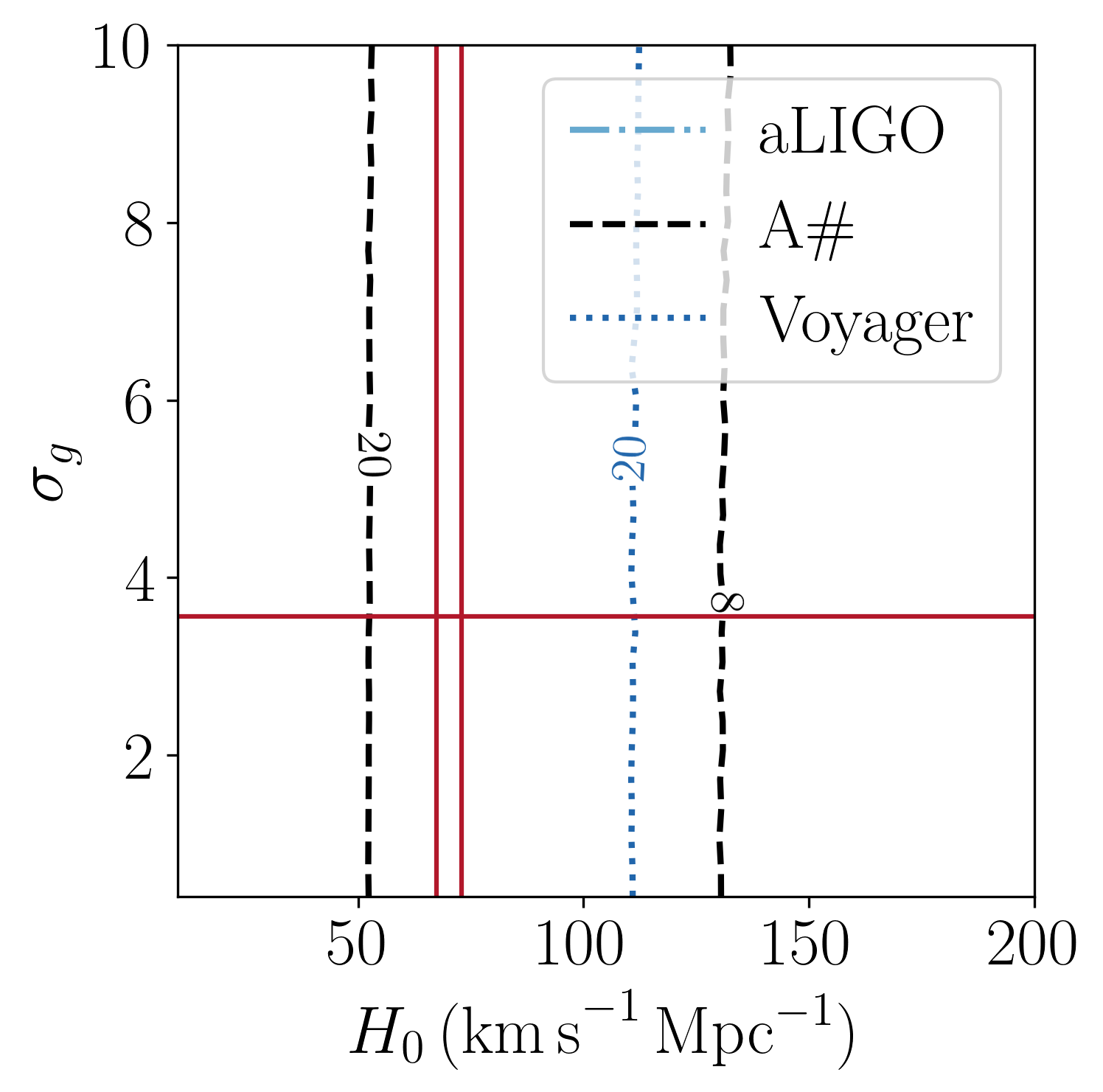}
    \caption{SNR of the GWB for various future detector networks as a function of two parameters ($H_0$ with either $\mu_g$ or $\sigma_g$), with the other 13 (hyper)parameters fixed to values denoted in the text. Refer to Fig.~\ref{fig:SNR_2d} for more explanation of the subplots.}
    \label{fig:SNR_2d_sfr2}
\end{figure}

The top (middle) two plots show A\# detectors after one year (two years) of observation to demonstrate the increase in SNR over time. An animation illustrating this change in SNR over time is available in the supplemental archive file.
The horizontal and vertical lines respectively correspond to the best-fit parameter values and the $H_0$ tension values~\citep{aghanim2020planck,Riess:2021jrx}.
The bottom two plots of Fig.~\ref{fig:SNR_2d} analogously show the SNR of the GWB, but for various networks of detectors observing for a two-year period. Figure~\ref{fig:SNR_2d} is consistent with Fig.~\ref{fig:SNR1d}, demonstrating that after two years of observation, aLIGO would not confidently detect a GWB with these parameters, while A\# and Voyager would measure it with SNR $\gtrsim20$.
For completeness, additional versions of Fig.~\ref{fig:SNR_2d} that show SNR as a function of other parameters can be found below (Figs.~\ref{fig:SNR_2d_pop1} -- \ref{fig:SNR_2d_sfr2}).

\bibliography{bibliography}

\end{document}

%% file: authors.tex
\author{Bryce~Cousins} 
\email[Corresponding author: ]{brycec2@illinois.edu}
\affiliation{Department of Physics, University of Illinois Urbana-Champaign, Urbana, IL 61801, USA.}
\affiliation{Illinois Center for Advanced Studies of the Universe, University of Illinois Urbana-Champaign, Urbana, IL 61801, USA.}
% 0000-0002-7026-1340

\author{Kristen~Schumacher} 
\affiliation{Department of Physics, University of Illinois Urbana-Champaign, Urbana, IL 61801, USA.}
\affiliation{Illinois Center for Advanced Studies of the Universe, University of Illinois Urbana-Champaign, Urbana, IL 61801, USA.}
%\email{kes13@illinois.edu}
% 0009-0002-8573-5075

\author{Adrian~Ka-Wai~Chung} 
\affiliation{Department of Physics, University of Illinois Urbana-Champaign, Urbana, IL 61801, USA.}
\affiliation{Illinois Center for Advanced Studies of the Universe, University of Illinois Urbana-Champaign, Urbana, IL 61801, USA.}
%\email{akwchung@illinois.edu}
% 0000-0003-2020-3254

\author{Colm~Talbot} 
\affiliation{Kavli Institute for Cosmological Physics, The University of Chicago, Chicago, IL 60637, USA.}
%\email{colmt@uchicago.edu}

\author{Thomas~Callister} 
\affiliation{Kavli Institute for Cosmological Physics, The University of Chicago, Chicago, IL 60637, USA.}
%\email{thomas.a.callister@gmail.com}

\author{Daniel~E.~Holz} 
\affiliation{Kavli Institute for Cosmological Physics, The University of Chicago, Chicago, IL 60637, USA.}
\affiliation{Department of Physics, Department of Astronomy \& Astrophysics, Enrico Fermi Institute, University of Chicago, Chicago, IL 60637, USA.}
%\affiliation{Department of Astronomy \& Astrophysics, The University of Chicago, Chicago, IL 60637, USA.}
%\affiliation{Enrico Fermi Institute, The University of Chicago, Chicago, IL 60637, USA.}
%\email{holz@uchicago.edu}
% 0000-0002-0175-5064

\author{Nicol\'as Yunes} 
\affiliation{Department of Physics, University of Illinois Urbana-Champaign, Urbana, IL 61801, USA.}
\affiliation{Illinois Center for Advanced Studies of the Universe, University of Illinois Urbana-Champaign, Urbana, IL 61801, USA.}
%\email{nyunes@illinois.edu}

%% file: bibliography.bib
@misc{Cousins2025prl_suppmat,
  title = {{Supplemental Material for ``The Stochastic Siren: Astrophysical Gravitational-Wave Background Measurements of the Hubble Constant''}},
  author = "{Cousins, B. and Schumacher, K. and Chung, AKW and Talbot, C. and Callister, T. and Holz, D. and Yunes, N.}",
  year = {2026},
  publisher = {Physical Review Letters},
  howpublished = {\url{https://doi.org/10.1103/4lzh-bm7y}},
}

@article{renzini2024projections,
  title={Projections of the uncertainty on the compact binary population background using popstock},
  author={Renzini, Arianna I and Golomb, Jacob},
  journal={Astronomy \& Astrophysics},
  volume={691},
  pages={A238},
  year={2024},
  publisher={EDP Sciences}
}

@article{meacher2015mock,
  title={Mock data and science challenge for detecting an astrophysical stochastic gravitational-wave background with Advanced LIGO and Advanced Virgo},
  author={Meacher, Duncan and Coughlin, Michael and Morris, Sean and Regimbau, Tania and Christensen, Nelson and Kandhasamy, Shivaraj and Mandic, Vuk and Romano, Joseph D and Thrane, Eric},
  journal={Physical Review D},
  volume={92},
  number={6},
  pages={063002},
  year={2015},
  publisher={APS}
}

@article{perkins2021improved,
  title={Improved gravitational-wave constraints on higher-order curvature theories of gravity},
  author={Perkins, Scott E and Nair, Remya and Silva, Hector O and Yunes, Nicolas},
  journal={Physical Review D},
  volume={104},
  number={2},
  pages={024060},
  year={2021},
  publisher={APS}
}

@article{antoniadis2023pta1,
  title={The second data release from the European Pulsar Timing Array-III. Search for gravitational wave signals},
  author={Antoniadis, John and Arumugam, P and Arumugam, S and Babak, S and Bagchi, M and Nielsen, A-S Bak and Bassa, CG and Bathula, A and Berthereau, A and Bonetti, M and others},
  journal={Astronomy \& Astrophysics},
  volume={678},
  pages={A50},
  year={2023},
  publisher={EDP Sciences}
}

@article{reardon2023pta2,
  title={Search for an isotropic gravitational-wave background with the Parkes Pulsar Timing Array},
  author={Reardon, Daniel J and Zic, Andrew and Shannon, Ryan M and Hobbs, George B and Bailes, Matthew and Di Marco, Valentina and Kapur, Agastya and Rogers, Axl F and Thrane, Eric and Askew, Jacob and others},
  journal={The Astrophysical Journal Letters},
  volume={951},
  number={1},
  pages={L6},
  year={2023},
  publisher={IOP Publishing}
}

@article{xu2023pta3,
  title={Searching for the nano-hertz stochastic gravitational wave background with the Chinese pulsar timing array data release I},
  author={Xu, Heng and Chen, Siyuan and Guo, Yanjun and Jiang, Jinchen and Wang, Bojun and Xu, Jiangwei and Xue, Zihan and Caballero, R Nicolas and Yuan, Jianping and Xu, Yonghua and others},
  journal={Research in Astronomy and Astrophysics},
  volume={23},
  number={7},
  pages={075024},
  year={2023},
  publisher={IOP Publishing}
}

@article{essick2024ensuring,
  title={Ensuring consistency between noise and detection in hierarchical Bayesian inference},
  author={Essick, Reed and Fishbach, Maya},
  journal={The Astrophysical Journal},
  volume={962},
  number={2},
  pages={169},
  year={2024},
  publisher={IOP Publishing}
}

@article{Farr2019selection,
doi = {10.3847/2515-5172/ab1d5f},
url = {https://dx.doi.org/10.3847/2515-5172/ab1d5f},
year = {2019},
month = {may},
publisher = {The American Astronomical Society},
volume = {3},
number = {5},
pages = {66},
author = {Farr, Will M.},
title = {Accuracy Requirements for Empirically Measured Selection Functions},
journal = {Research Notes of the AAS}
}

@article{linderExploringExpansionHistory2003,
  title = {Exploring the {{Expansion History}} of the {{Universe}}},
  author = {Linder, Eric V.},
  year = {2003},
  month = mar,
  journal = {Physical Review Letters},
  volume = {90},
  number = {9},
  eprint = {astro-ph/0208512},
  pages = {091301},
  issn = {0031-9007, 1079-7114},
  doi = {10.1103/PhysRevLett.90.091301},
  urldate = {2024-07-03},
  archiveprefix = {arXiv},
  langid = {english}
}

@article{zhuGravitationalWaveBackground2013,
  title = {On the Gravitational Wave Background from Compact Binary Coalescences in the Band of Ground-Based Interferometers},
  author = {Zhu, Xing-Jiang and Howell, Eric J. and Blair, David G. and Zhu, Zong-Hong},
  year = {2013},
  month = may,
  journal = {Monthly Notices of the Royal Astronomical Society},
  volume = {431},
  number = {1},
  eprint = {1209.0595},
  primaryclass = {astro-ph, physics:gr-qc},
  pages = {882--899},
  issn = {0035-8711, 1365-2966},
  doi = {10.1093/mnras/stt207},
  urldate = {2024-01-24},
  archiveprefix = {arXiv},
  langid = {english}
}

@ARTICLE{astropy2018,
       author = {{Astropy Collaboration} and others ({Astropy Contributors})},
        title = "{The Astropy Project: Building an Open-science Project and Status of the v2.0 Core Package}",
      journal = {The Astronomical Journal},
         year = 2018,
        month = sep,
       volume = {156},
       number = {3},
          eid = {123},
        pages = {123},
          doi = {10.3847/1538-3881/aabc4f},
archivePrefix = {arXiv},
       eprint = {1801.02634},
 primaryClass = {astro-ph.IM},
       adsurl = {https://ui.adsabs.harvard.edu/abs/2018AJ....156..123A}
}

@ARTICLE{astropy2013,
       author = {{Astropy Collaboration} and others},
        title = "{Astropy: A community Python package for astronomy}",
      journal = {Astronomy and Astrophysics},
         year = "2013",
        month = "Oct",
       volume = {558},
          eid = {A33},
        pages = {A33},
          doi = {10.1051/0004-6361/201322068},
archivePrefix = {arXiv},
       eprint = {1307.6212},
 primaryClass = {astro-ph.IM},
       adsurl = {https://ui.adsabs.harvard.edu/abs/2013A&A...558A..33A}
}

@article{hunter2007matplotlib,
  title={Matplotlib: A 2D graphics environment},
  author={Hunter, John D},
  journal={Computing in science \& engineering},
  volume={9},
  number={03},
  pages={90--95},
  year={2007},
  publisher={IEEE Computer Society}
}

@article{harris2020numpy,
  title={Array programming with NumPy},
  author={{Harris, Charles R} and others},
  journal={Nature},
  volume={585},
  number={7825},
  pages={357--362},
  year={2020},
  publisher={Nature Publishing Group UK London}
}

@article{virtanen2020scipy,
  title={SciPy 1.0: fundamental algorithms for scientific computing in Python},
  author={{Virtanen, Pauli} and others},
  journal={Nature methods},
  volume={17},
  number={3},
  pages={261--272},
  year={2020},
  publisher={Nature Publishing Group}
}

@ARTICLE{astropy2022,
       author = {{Astropy Collaboration} and others},
        title = "{The Astropy Project: Sustaining and Growing a Community-oriented Open-source Project and the Latest Major Release (v5.0) of the Core Package}",
      journal = {\apj},
         year = 2022,
        month = aug,
       volume = {935},
       number = {2},
          eid = {167},
        pages = {167},
          doi = {10.3847/1538-4357/ac7c74},
archivePrefix = {arXiv},
       eprint = {2206.14220},
 primaryClass = {astro-ph.IM},
       adsurl = {https://ui.adsabs.harvard.edu/abs/2022ApJ...935..167A}
}

@book{collette2013python,
  title={Python and HDF5: unlocking scientific data},
  author={Collette, Andrew},
  year={2013},
  publisher={" O'Reilly Media, Inc."}
}

@article{abbott2023cceh,
  title = {Constraints on the {{Cosmic Expansion History}} from {{GWTC}}--3},
  author = {{Abbott, R.} and others ({The LIGO Scientific Collaboration}, {the Virgo Collaboration, and the KAGRA Collaboration})},
  year = {2023},
  month = jun,
  journal = {The Astrophysical Journal},
  volume = {949},
  number = {2},
  pages = {76},
  issn = {0004-637X, 1538-4357},
  doi = {10.3847/1538-4357/ac74bb},
  urldate = {2024-05-20},
  langid = {english}
}

@article{abbottUpperLimitsIsotropic2021,
  title = {Upper Limits on the Isotropic Gravitational-Wave Background from {{Advanced LIGO}} and {{Advanced Virgo}}'s Third Observing Run},
  author = {{Abbott, R.} and others and {The LIGO Scientific Collaboration, Virgo Collaboration, and KAGRA Collaboration}},
  year = {2021},
  month = jul,
  journal = {Physical Review D},
  volume = {104},
  number = {2},
  pages = {022004},
  issn = {2470-0010, 2470-0029},
  doi = {10.1103/PhysRevD.104.022004},
  urldate = {2024-02-07},
  langid = {english}
}

@article{allenDetectingStochasticBackground1999,
  title = {Detecting a Stochastic Background of Gravitational Radiation: {{Signal}} Processing Strategies and Sensitivities},
  shorttitle = {Detecting a Stochastic Background of Gravitational Radiation},
  author = {Allen, Bruce and Romano, Joseph D.},
  year = {1999},
  month = mar,
  journal = {Physical Review D},
  volume = {59},
  number = {10},
  eprint = {gr-qc/9710117},
  pages = {102001},
  issn = {0556-2821, 1089-4918},
  doi = {10.1103/PhysRevD.59.102001},
  urldate = {2023-11-08},
  archiveprefix = {arXiv},
  langid = {english}
}

@article{callisterPolarizationBasedTestsGravity2017,
  title = {Polarization-{{Based Tests}} of {{Gravity}} with the {{Stochastic Gravitational-Wave Background}}},
  author = {Callister, Thomas and Biscoveanu, A. Sylvia and Christensen, Nelson and Isi, Maximiliano and Matas, Andrew and Minazzoli, Olivier and Regimbau, Tania and Sakellariadou, Mairi and Tasson, Jay and Thrane, Eric},
  year = {2017},
  month = dec,
  journal = {Physical Review X},
  volume = {7},
  number = {4},
  pages = {041058},
  issn = {2160-3308},
  doi = {10.1103/PhysRevX.7.041058},
  urldate = {2024-02-12},
  langid = {english}
}

@article{callisterShoutsMurmursCombining2020,
  title = {Shouts and {{Murmurs}}: {{Combining Individual Gravitational-wave Sources}} with the {{Stochastic Background}} to {{Measure}} the {{History}} of {{Binary Black Hole Mergers}}},
  shorttitle = {Shouts and {{Murmurs}}},
  author = {Callister, Tom and Fishbach, Maya and Holz, Daniel E. and Farr, Will M.},
  year = {2020},
  month = jun,
  journal = {The Astrophysical Journal},
  volume = {896},
  number = {2},
  pages = {L32},
  issn = {2041-8213},
  doi = {10.3847/2041-8213/ab9743},
  urldate = {2023-10-25},
  langid = {english}
}

@article{christensenParameterEstimationGravitational2022,
  title = {Parameter Estimation with Gravitational Waves},
  author = {Christensen, Nelson and Meyer, Renate},
  year = {2022},
  month = apr,
  journal = {Reviews of Modern Physics},
  volume = {94},
  number = {2},
  eprint = {2204.04449},
  primaryclass = {gr-qc},
  pages = {025001},
  issn = {0034-6861, 1539-0756},
  doi = {10.1103/RevModPhys.94.025001},
  urldate = {2023-11-24},
  archiveprefix = {arXiv},
  langid = {english}
}

@article{ferrariStochasticBackgroundGravitational1999,
  title = {Stochastic Background of Gravitational Waves Generated by a Cosmological Population of Young, Rapidly Rotating Neutron Stars},
  author = {Ferrari, Valeria and Matarrese, Sabino and Schneider, Raffaella},
  year = {1999},
  month = feb,
  journal = {Monthly Notices of the Royal Astronomical Society},
  volume = {303},
  number = {2},
  pages = {258--264},
  issn = {0035-8711, 1365-2966},
  doi = {10.1046/j.1365-8711.1999.02207.x},
  urldate = {2023-09-14},
  langid = {english}
}

@misc{mastrogiovanniICAROGWPythonPackage2023,
  title = {{{ICAROGW}}: {{A}} Python Package for Inference of Astrophysical Population Properties of Noisy, Heterogeneous and Incomplete Observations},
  shorttitle = {{{ICAROGW}}},
  author = {{Mastrogiovanni, Simone} and others},
  year = {2023},
  month = may,
  number = {arXiv:2305.17973},
  eprint = {2305.17973},
  primaryclass = {astro-ph, physics:gr-qc},
  publisher = {arXiv},
  urldate = {2023-11-09},
  archiveprefix = {arXiv},
  langid = {english}
}

@article{perigoisStartrackPredictionsStochastic2021,
  title = {Startrack Predictions of the Stochastic Gravitational-Wave Background from Compact Binary Mergers},
  author = {P{\'e}rigois, C. and Belczynski, C. and Bulik, T. and Regimbau, T.},
  year = {2021},
  month = feb,
  journal = {Physical Review D},
  volume = {103},
  number = {4},
  pages = {043002},
  issn = {2470-0010, 2470-0029},
  doi = {10.1103/PhysRevD.103.043002},
  urldate = {2023-09-13},
  langid = {english}
}

@article{regimbauAstrophysicalGravitationalWave2011,
  title = {The Astrophysical Gravitational Wave Stochastic Background},
  author = {Regimbau, Tania},
  year = {2011},
  month = apr,
  journal = {Research in Astronomy and Astrophysics},
  volume = {11},
  number = {4},
  pages = {369--390},
  issn = {1674-4527},
  doi = {10.1088/1674-4527/11/4/001},
  urldate = {2023-09-13},
  langid = {english}
}

@misc{renziniStochasticGravitationalWaveBackgrounds2022,
  title = {Stochastic {{Gravitational-Wave Backgrounds}}: {{Current Detection Efforts}} and {{Future Prospects}}},
  shorttitle = {Stochastic {{Gravitational-Wave Backgrounds}}},
  author = {Renzini, Arianna I. and Goncharov, Boris and Jenkins, Alexander C. and Meyers, Pat M.},
  year = {2022},
  month = feb,
  number = {arXiv:2202.00178},
  eprint = {2202.00178},
  primaryclass = {gr-qc},
  publisher = {arXiv},
  urldate = {2023-11-08},
  archiveprefix = {arXiv},
  langid = {english}
}

@article{romanoDetectionMethodsStochastic2017,
  title = {Detection Methods for Stochastic Gravitational-Wave Backgrounds: A Unified Treatment},
  shorttitle = {Detection Methods for Stochastic Gravitational-Wave Backgrounds},
  author = {Romano, Joseph D. and Cornish, Neil J.},
  year = {2017},
  month = dec,
  journal = {Living Reviews in Relativity},
  volume = {20},
  number = {1},
  eprint = {1608.06889},
  primaryclass = {gr-qc},
  pages = {2},
  issn = {2367-3613, 1433-8351},
  doi = {10.1007/s41114-017-0004-1},
  urldate = {2024-02-12},
  archiveprefix = {arXiv},
  langid = {english}
}

@article{theligoscientificcollaborationBinaryBlackHole2019,
  title = {Binary {{Black Hole Population Properties Inferred}} from the {{First}} and {{Second Observing Runs}} of {{Advanced LIGO}} and {{Advanced Virgo}}},
  author = {Abbott, B. P. and others and {LIGO Scientific Collaboration and Virgo Collaboration}},
  year = {2019},
  month = sep,
  journal = {The Astrophysical Journal},
  volume = {882},
  number = {2},
  eprint = {1811.12940},
  primaryclass = {astro-ph},
  pages = {L24},
  issn = {2041-8213},
  doi = {10.3847/2041-8213/ab3800},
  urldate = {2023-10-13},
  archiveprefix = {arXiv},
  langid = {english}
}

@ARTICLE{talbotMeasuringBlackHoleMass2018,
       author = {{Talbot}, Colm and {Thrane}, Eric},
        title = "{Measuring the Binary Black Hole Mass Spectrum with an Astrophysically Motivated Parameterization}",
      journal = {\apj},
         year = 2018,
        month = apr,
       volume = {856},
       number = {2},
          eid = {173},
        pages = {173},
          doi = {10.3847/1538-4357/aab34c},
archivePrefix = {arXiv},
       eprint = {1801.02699},
 primaryClass = {astro-ph.HE},
       adsurl = {https://ui.adsabs.harvard.edu/abs/2018ApJ...856..173T}
}

@article{skilling2004nested,
  title={Nested sampling},
  author={Skilling, John},
  journal={Bayesian inference and maximum entropy methods in science and engineering},
  volume={735},
  pages={395--405},
  year={2004}
}

@article{skilling2006nested,
  title={Nested sampling for general Bayesian computation},
  author={Skilling, John},
  journal="",
  year={2006}
}

@article{abbott2023gwtc3,
  title={GWTC-3: Compact binary coalescences observed by LIGO and Virgo during the second part of the third observing run},
  author={{Abbott, Richard} and others},
  journal={Physical Review X},
  volume={13},
  number={4},
  pages={041039},
  year={2023},
  publisher={APS}
}

@article{somiya2012kagra,
  title={Detector configuration of KAGRA--the Japanese cryogenic gravitational-wave detector},
  author={Somiya, Kentaro and KAGRA Collaboration},
  journal={Classical and Quantum Gravity},
  volume={29},
  number={12},
  pages={124007},
  year={2012},
  publisher={IOP Publishing}
}

@article{vitale2018nsbhH0,
  title={Measuring the Hubble constant with neutron star black hole mergers},
  author={Vitale, Salvatore and Chen, Hsin-Yu},
  journal={Physical review letters},
  volume={121},
  number={2},
  pages={021303},
  year={2018},
  publisher={APS}
}

@article{chen2022:190521H0,
  title={A standard siren cosmological measurement from the potential GW190521 electromagnetic counterpart ZTF19abanrhr},
  author={Chen, Hsin-Yu and Haster, Carl-Johan and Vitale, Salvatore and Farr, Will M and Isi, Maximiliano},
  journal={Monthly Notices of the Royal Astronomical Society},
  volume={513},
  number={2},
  pages={2152--2157},
  year={2022},
  publisher={Oxford University Press}
}

@article{chenTwoCentHubble2018,
  title = {A Two per Cent {{Hubble}} Constant Measurement from Standard Sirens within Five Years},
  author = {Chen, Hsin-Yu and Fishbach, Maya and Holz, Daniel E.},
  year = {2018},
  month = oct,
  journal = {Nature},
  volume = {562},
  number = {7728},
  pages = {545--547},
  issn = {0028-0836, 1476-4687},
  doi = {10.1038/s41586-018-0606-0},
  urldate = {2023-01-03},
  langid = {english}
}

@article{holzUsingGravitationalWave2005,
  title = {Using {{Gravitational}}-{{Wave Standard Sirens}}},
  author = {Holz, Daniel E. and Hughes, Scott A.},
  year = {2005},
  month = aug,
  journal = {The Astrophysical Journal},
  volume = {629},
  number = {1},
  pages = {15--22},
  issn = {0004-637X, 1538-4357},
  doi = {10.1086/431341},
  urldate = {2023-11-27},
  langid = {english}
}

@article{schutzDeterminingHubbleConstant1986,
  title = {Determining the {{Hubble}} Constant from Gravitational Wave Observations},
  author = {Schutz, Bernard F.},
  year = {1986},
  month = sep,
  journal = {Nature},
  volume = {323},
  number = {6086},
  pages = {310--311},
  issn = {0028-0836, 1476-4687},
  doi = {10.1038/323310a0},
  urldate = {2023-01-03},
  langid = {english}
}

@article{dalalShortGRBBinary2006,
  title = {Short {{GRB}} and Binary Black Hole Standard Sirens as a Probe of Dark Energy},
  author = {Dalal, Neal and Holz, Daniel E. and Hughes, Scott A. and Jain, Bhuvnesh},
  year = {2006},
  month = sep,
  journal = {Physical Review D},
  volume = {74},
  number = {6},
  pages = {063006},
  issn = {1550-7998, 1550-2368},
  doi = {10.1103/PhysRevD.74.063006},
  urldate = {2024-01-15},
  langid = {english}
}

@article{nissanke2013determining,
  title={Determining the Hubble constant from gravitational wave observations of merging compact binaries},
  author={Nissanke, Samaya and Holz, Daniel E and Dalal, Neal and Hughes, Scott A and Sievers, Jonathan L and Hirata, Christopher M},
  journal={arXiv preprint arXiv:1307.2638},
  year={2013}
}

@article{ezquiaga2022spectral,
  title={Spectral sirens: Cosmology from the full mass distribution of compact binaries},
  author={Ezquiaga, Jose Mar{\'\i}a and Holz, Daniel E},
  journal={Physical Review Letters},
  volume={129},
  number={6},
  pages={061102},
  year={2022},
  publisher={APS}
}

@article{fishbach2019gw170817H0,
  title={A standard siren measurement of the Hubble constant from GW170817 without the electromagnetic counterpart},
  author={Fishbach, Maya and Gray, R and Hernandez, I Maga{\~n}a and Qi, H and Sur, A and Acernese, F and Aiello, L and Allocca, A and Aloy, MA and Amato, A and others},
  journal={The Astrophysical Journal Letters},
  volume={871},
  number={1},
  pages={L13},
  year={2019},
  publisher={IOP Publishing}
}

@article{ligo2017gw170817H0,
  title={A gravitational-wave standard siren measurement of the Hubble constant},
  author={{Abbott, B. P.} and others},
  journal={Nature},
  volume={551},
  number={7678},
  pages={85--88},
  year={2017},
  publisher={Nature Publishing Group UK London}
}

@article{rosado2011agwb1,
  title={Gravitational wave background from binary systems},
  author={Rosado, Pablo A},
  journal={Physical Review D—Particles, Fields, Gravitation, and Cosmology},
  volume={84},
  number={8},
  pages={084004},
  year={2011},
  publisher={APS}
}

@article{zhu2011agwb2,
  title={Stochastic gravitational wave background from coalescing binary black holes},
  author={Zhu, Xing-Jiang and Howell, Eric and Regimbau, Tania and Blair, David and Zhu, Zong-Hong},
  journal={The Astrophysical Journal},
  volume={739},
  number={2},
  pages={86},
  year={2011},
  publisher={IOP Publishing}
}

@article{marassi2011agwb3,
  title={Imprint of the merger and ring-down on the gravitational wave background from black hole binaries coalescence},
  author={Marassi, Stefania and Schneider, Raffaella and Corvino, Giovanni and Ferrari, Valeria and Zwart, S Portegies},
  journal={Physical Review D—Particles, Fields, Gravitation, and Cosmology},
  volume={84},
  number={12},
  pages={124037},
  year={2011},
  publisher={APS}
}

@article{starobinskii1979earlygwb1,
  title={Spectrum of relict gravitational radiation and the early state of the universe},
  author={Starobinskii, AA},
  journal={JETP Letters},
  volume={30},
  number={11},
  pages={682--685},
  year={1979}
}

@article{kibble1976stringsearlygwb2,
  title={Topology of cosmic domains and strings},
  author={Kibble, Thomas WB},
  journal={Journal of Physics A: Mathematical and General},
  volume={9},
  number={8},
  pages={1387},
  year={1976},
  publisher={IOP Publishing}
}

@article{easther2006earlygwb3,
  title={Stochastic gravitational wave production after inflation},
  author={Easther, Richard and Lim, Eugene A},
  journal={Journal of Cosmology and Astroparticle Physics},
  volume={2006},
  number={04},
  pages={010},
  year={2006},
  publisher={IOP Publishing}
}

@article{rinaldi2024evidence,
  title={Evidence of evolution of the black hole mass function with redshift},
  author={Rinaldi, Stefano and Del Pozzo, Walter and Mapelli, Michela and Lorenzo-Medina, Ana and Dent, Thomas},
  journal={Astronomy \& Astrophysics},
  volume={684},
  pages={A204},
  year={2024},
  publisher={EDP Sciences}
}

@article{renzini2024popstock,
  title={Projections of the uncertainty on the compact binary population background using popstock},
  author={Renzini, Arianna I and Golomb, Jacob},
  journal={arXiv preprint arXiv:2407.03742},
  year={2024}
}

@article{KAGRA:2021gwtc3population,
    author = "{Abbott, R.} and others",
    collaboration = "LIGO, Virgo, and KAGRA Collaboration",
    title = "{Population of Merging Compact Binaries Inferred Using Gravitational Waves through GWTC-3}",
    eprint = "2111.03634",
    archivePrefix = "arXiv",
    primaryClass = "astro-ph.HE",
    reportNumber = "LIGO-P2100239 ; Data release: https://zenodo.org/record/5655785, LIGO-P2100239",
    doi = "10.1103/PhysRevX.13.011048",
    journal = "Phys. Rev. X",
    volume = "13",
    number = "1",
    pages = "011048",
    year = "2023"
}

@article{madau2014cosmic,
  title={Cosmic star-formation history},
  author={Madau, Piero and Dickinson, Mark},
  journal={Annual Review of Astronomy and Astrophysics},
  number={1},
  year = 2014,
  month = aug,
  volume = {52},
  pages = {415-486},
  doi = {10.1146/annurev-astro-081811-125615},
  archivePrefix = {arXiv},
  eprint = {1403.0007},
  primaryClass = {astro-ph.CO},
  adsurl = {https://ui.adsabs.harvard.edu/abs/2014ARA&A..52..415M},
  publisher={Annual Reviews}
}

@article{torniamenti2024hierarchical,
  title={Hierarchical binary black hole mergers in globular clusters: Mass function and evolution with redshift},
  author={Torniamenti, Stefano and Mapelli, Michela and P{\'e}rigois, Carole and Sedda, Manuel Arca and Artale, Maria Celeste and Dall’Amico, Marco and Vaccaro, Maria Paola},
  journal={Astronomy \& Astrophysics},
  volume={688},
  pages={A148},
  year={2024},
  publisher={EDP Sciences}
}

@article{agazie2023nanograv,
  title={The NANOGrav 15 yr data set: Evidence for a gravitational-wave background},
  author={Agazie, Gabriella and Anumarlapudi, Akash and Archibald, Anne M and Arzoumanian, Zaven and Baker, Paul T and B{\'e}csy, Bence and Blecha, Laura and Brazier, Adam and Brook, Paul R and Burke-Spolaor, Sarah and others},
  journal={The Astrophysical Journal Letters},
  volume={951},
  number={1},
  pages={L8},
  year={2023},
  publisher={IOP Publishing}
}

@misc{kamionkowskiHubbleTensionEarly2022,
  title = {The {{Hubble Tension}} and {{Early Dark Energy}}},
  author = {Kamionkowski, Marc and Riess, Adam G.},
  year = {2022},
  month = nov,
  number = {arXiv:2211.04492},
  eprint = {2211.04492},
  primaryclass = {astro-ph, physics:gr-qc, physics:hep-ph},
  publisher = {arXiv},
  urldate = {2023-01-19},
  archiveprefix = {arxiv},
  langid = {english}
}

@article{riess1998observational,
  title={Observational evidence from supernovae for an accelerating universe and a cosmological constant},
  author={{Riess, Adam G} and others},
  journal={The Astronomical Journal},
  volume={116},
  number={3},
  pages={1009},
  year={1998},
  publisher={IOP Publishing}
}

@article{aghanim2020planck,
  title={Planck 2018 results-VI. Cosmological parameters},
  author={{Aghanim, Nabila} and others},
  journal={Astronomy \& Astrophysics},
  volume={641},
  pages={A6},
  year={2020},
  publisher={EDP sciences}
}

@article{Ashton:2018jfp,
    author = "Ashton, Gregory and others",
    title = "{BILBY: A user-friendly Bayesian inference library for gravitational-wave astronomy}",
    eprint = "1811.02042",
    archivePrefix = "arXiv",
    primaryClass = "astro-ph.IM",
    doi = "10.3847/1538-4365/ab06fc",
    journal = "Astrophys. J. Suppl.",
    volume = "241",
    number = "2",
    pages = "27",
    year = "2019"
}

@article{mandic2012spectralparameterGWB,
  title={Parameter estimation in searches for the stochastic gravitational-wave background},
  author={Mandic, Vuk and Thrane, Eric and Giampanis, Stefanos and Regimbau, Tania},
  journal={Physical review letters},
  volume={109},
  number={17},
  pages={171102},
  year={2012},
  publisher={APS}
}

@article{aasi2015advancedligo,
  title={Advanced ligo},
  author={{Aasi, Junaid} and others},
  journal={Classical and quantum gravity},
  volume={32},
  number={7},
  pages={074001},
  year={2015},
  publisher={IOP Publishing}
}

@article{acernese2014advancedvirgo,
  title={Advanced Virgo: a second-generation interferometric gravitational wave detector},
  author={{Acernese, Fausto} and others},
  journal={Classical and Quantum Gravity},
  volume={32},
  number={2},
  pages={024001},
  year={2014},
  publisher={IOP Publishing}
}

@article{reitze2019cosmic,
  title={Cosmic explorer: the US contribution to gravitational-wave astronomy beyond LIGO},
  author={{Reitze, David} and others},
  journal={arXiv preprint arXiv:1907.04833},
  doi={10.48550/arXiv.1907.04833},
  year={2019}
}

@article{renzini2023pygwb,
  title={pygwb: A python-based library for gravitational-wave background searches},
  author={{Renzini, Arianna I} and others},
  journal={The Astrophysical Journal},
  volume={952},
  number={1},
  pages={25},
  year={2023},
  publisher={IOP Publishing}
}

@article{talbot2024gwpopulation,
  title={GWPopulation: Hardware agnostic population inference for compact binaries and beyond},
  author={Talbot, Colm and Farah, Amanda and Galaudage, Shanika and Golomb, Jacob and Tong, Hui},
  journal={arXiv preprint arXiv:2409.14143},
  year={2024}
}

@article{ligoGWBimplications:2017,
    author = "{Abbott, Benjamin P.} and others",
    collaboration = "LIGO Scientific, Virgo",
    title = "{GW170817: Implications for the Stochastic Gravitational-Wave Background from Compact Binary Coalescences}",
    eprint = "1710.05837",
    archivePrefix = "arXiv",
    primaryClass = "gr-qc",
    reportNumber = "LIGO-P1700272",
    doi = "10.1103/PhysRevLett.120.091101",
    journal = "Phys. Rev. Lett.",
    volume = "120",
    number = "9",
    pages = "091101",
    year = "2018"
}

@article{ChristensenmeasuringGWB1992,
    author = "Christensen, N.",
    title = "{Measuring the stochastic gravitational radiation background with laser interferometric antennas}",
    doi = "10.1103/PhysRevD.46.5250",
    journal = "Phys. Rev. D",
    volume = "46",
    pages = "5250--5266",
    year = "1992"
}

@article{FlanagansensitivityligoGWB1993,
    author = "Flanagan, Eanna E.",
    title = "{The Sensitivity of the laser interferometer gravitational wave observatory (LIGO) to a stochastic background, and its dependence on the detector orientations}",
    eprint = "astro-ph/9305029",
    archivePrefix = "arXiv",
    reportNumber = "GRP-347",
    doi = "10.1103/PhysRevD.48.2389",
    journal = "Phys. Rev. D",
    volume = "48",
    pages = "2389--2407",
    year = "1993"
}

@article{Riess:2021jrx,
    author = "{Riess, Adam G.} and others",
    title = "{A Comprehensive Measurement of the Local Value of the Hubble Constant with 1 km s$^{-1}$ Mpc$^{-1}$ Uncertainty from the Hubble Space Telescope and the SH0ES Team}",
    eprint = "2112.04510",
    archivePrefix = "arXiv",
    primaryClass = "astro-ph.CO",
    doi = "10.3847/2041-8213/ac5c5b",
    journal = "Astrophys. J. Lett.",
    volume = "934",
    number = "1",
    pages = "L7",
    year = "2022"
}

@article{Speagle:2019ivv,
    author = "Speagle, Joshua S.",
    title = "{dynesty: a dynamic nested sampling package for estimating Bayesian posteriors and evidences}",
    eprint = "1904.02180",
    archivePrefix = "arXiv",
    primaryClass = "astro-ph.IM",
    doi = "10.1093/mnras/staa278",
    journal = "Mon. Not. Roy. Astron. Soc.",
    volume = "493",
    number = "3",
    pages = "3132--3158",
    year = "2020"
}

@article{taylorminingGWcatalogs2018,
    author = "Taylor, Stephen R. and Gerosa, Davide",
    title = "{Mining Gravitational-wave Catalogs To Understand Binary Stellar Evolution: A New Hierarchical Bayesian Framework}",
    eprint = "1806.08365",
    archivePrefix = "arXiv",
    primaryClass = "astro-ph.HE",
    doi = "10.1103/PhysRevD.98.083017",
    journal = "Phys. Rev. D",
    volume = "98",
    number = "8",
    pages = "083017",
    year = "2018"
}

@article{mandelextractingdistributionparameters2018,
    author = "Mandel, Ilya and Farr, Will M. and Gair, Jonathan R.",
    title = "{Extracting distribution parameters from multiple uncertain observations with selection biases}",
    eprint = "1809.02063",
    archivePrefix = "arXiv",
    primaryClass = "physics.data-an",
    doi = "10.1093/mnras/stz896",
    journal = "Mon. Not. Roy. Astron. Soc.",
    volume = "486",
    number = "1",
    pages = "1086--1093",
    year = "2019"
}

@article{Vitaleinferringpopulation2022,
  title={Inferring the properties of a population of compact binaries in presence of selection effects},
  author={Vitale, Salvatore and Gerosa, Davide and Farr, Will M and Taylor, Stephen R},
  booktitle={Handbook of Gravitational Wave Astronomy},
  pages={1--60},
  year={2022},
  publisher={Springer},
  eprint = "2007.05579",
  archivePrefix = "arXiv",
  primaryClass = "astro-ph.IM",
  doi = "10.48550/arXiv.2007.05579",
  journal = ""
}

@article{LallemanBBHdistributionredshift2025,
    author = "Lalleman, Max and Turbang, Kevin and Callister, Thomas and van Remortel, Nick",
    title = "{No evidence that the binary black hole mass distribution evolves with redshift}",
    eprint = "2501.10295",
    archivePrefix = "arXiv",
    primaryClass = "astro-ph.HE",
    month = "1",
    journal = "",
    year = "2025"
}

@article{LIGOScientific:2016fpe,
    author = "{Abbott, B. P.} and others",
    collaboration = "LIGO Scientific, Virgo",
    title = "{GW150914: Implications for the stochastic gravitational wave background from binary black holes}",
    eprint = "1602.03847",
    archivePrefix = "arXiv",
    primaryClass = "gr-qc",
    reportNumber = "LIGO-P1500222",
    doi = "10.1103/PhysRevLett.116.131102",
    journal = "Phys. Rev. Lett.",
    volume = "116",
    number = "13",
    pages = "131102",
    year = "2016"
}

@article{LIGOScientific:2017zlf,
    author = "{Abbott, Benjamin P.} and others",
    collaboration = "LIGO Scientific, Virgo",
    title = "{GW170817: Implications for the Stochastic Gravitational-Wave Background from Compact Binary Coalescences}",
    eprint = "1710.05837",
    archivePrefix = "arXiv",
    primaryClass = "gr-qc",
    reportNumber = "LIGO-P1700272",
    doi = "10.1103/PhysRevLett.120.091101",
    journal = "Phys. Rev. Lett.",
    volume = "120",
    number = "9",
    pages = "091101",
    year = "2018"
}

@article{Evans:2023euw,
    author = "{Evans, Matthew} and others",
    title = "{Cosmic Explorer: A Submission to the NSF MPSAC ngGW Subcommittee}",
    eprint = "2306.13745",
    archivePrefix = "arXiv",
    primaryClass = "astro-ph.IM",
    month = "6",
    year = "2023",
    journal = ""
}

@article{Branchesi:2023mws,
    author = "{Branchesi, Marica} and others",
    title = "{Science with the Einstein Telescope: a comparison of different designs}",
    eprint = "2303.15923",
    archivePrefix = "arXiv",
    primaryClass = "gr-qc",
    reportNumber = "ET-0084A-23",
    doi = "10.1088/1475-7516/2023/07/068",
    journal = "JCAP",
    volume = "07",
    pages = "068",
    year = "2023"
}

@article{Buscicchio:2022raf,
    author = "Buscicchio, Riccardo and Ain, Anirban and Ballelli, Matteo and Cella, Giancarlo and Patricelli, Barbara",
    title = "{Detecting non-Gaussian gravitational wave backgrounds: A unified framework}",
    eprint = "2209.01400",
    archivePrefix = "arXiv",
    primaryClass = "gr-qc",
    doi = "10.1103/PhysRevD.107.063027",
    journal = "Phys. Rev. D",
    volume = "107",
    number = "6",
    pages = "063027",
    year = "2023"
}

@article{di2021realm,
  title={In the realm of the Hubble tension—a review of solutions},
  author={Di Valentino, Eleonora and Mena, Olga and Pan, Supriya and Visinelli, Luca and Yang, Weiqiang and Melchiorri, Alessandro and Mota, David F and Riess, Adam G and Silk, Joseph},
  journal={Classical and Quantum Gravity},
  volume={38},
  number={15},
  pages={153001},
  year={2021},
  publisher={IOP Publishing}
}

@article{Ferrari:1998ut,
    author = "Ferrari, Valeria and Matarrese, Sabino and Schneider, Raffaella",
    title = "{Gravitational wave background from a cosmological population of core collapse supernovae}",
    eprint = "astro-ph/9804259",
    archivePrefix = "arXiv",
    doi = "10.1046/j.1365-8711.1999.02194.x",
    journal = "Mon. Not. Roy. Astron. Soc.",
    volume = "303",
    pages = "247",
    year = "1999"
}

@article{Chowdhury:2024fdr,
    author = "Chowdhury, Sourav Roy and Khlopov, Maxim",
    title = "{Stochastic gravitational wave background due to core collapse resulting in neutron stars}",
    eprint = "2409.01542",
    archivePrefix = "arXiv",
    primaryClass = "gr-qc",
    doi = "10.1103/PhysRevD.110.063037",
    journal = "Phys. Rev. D",
    volume = "110",
    number = "6",
    pages = "063037",
    year = "2024"
}

@article{regimbau2008astrophysical,
  title={Astrophysical sources of a stochastic gravitational-wave background},
  author={Regimbau, Tania and Mandic, Vuk},
  journal={Classical and Quantum Gravity},
  volume={25},
  number={18},
  pages={184018},
  year={2008},
  publisher={IOP Publishing}
}

@article{flanagan1998measuring,
  title={Measuring gravitational waves from binary black hole coalescences. I. Signal to noise for inspiral, merger, and ringdown},
  author={Flanagan, {\'E}anna {\'E} and Hughes, Scott A},
  journal={Physical Review D},
  volume={57},
  number={8},
  pages={4535},
  year={1998},
  publisher={APS}
}

@article{wu2012accessibility,
  title={Accessibility of the gravitational-wave background due to binary coalescences to second and third generation gravitational-wave detectors},
  author={Wu, Chengjiang and Mandic, Vuk and Regimbau, Tania},
  journal={Physical Review D—Particles, Fields, Gravitation, and Cosmology},
  volume={85},
  number={10},
  pages={104024},
  year={2012},
  publisher={APS}
}

@article{alamCompletedSDSSIVExtended2021,
  title = {Completed {{SDSS-IV}} Extended {{Baryon Oscillation Spectroscopic Survey}}: {{Cosmological}} Implications from Two Decades of Spectroscopic Surveys at the {{Apache Point Observatory}}},
  shorttitle = {Completed {{SDSS-IV}} Extended {{Baryon Oscillation Spectroscopic Survey}}},
  author = {{Alam, Shadab} and others},
  year = {2021},
  month = apr,
  journal = {Physical Review D},
  volume = {103},
  number = {8},
  pages = {083533},
  issn = {2470-0010, 2470-0029},
  doi = {10.1103/PhysRevD.103.083533},
  urldate = {2025-01-25},
  langid = {english}
}

@article{balkenholMeasurementCMBTemperature2023,
  title = {A {{Measurement}} of the {{CMB Temperature Power Spectrum}} and {{Constraints}} on {{Cosmology}} from the {{SPT-3G}} 2018 {{TT}}/{{TE}}/{{EE Data Set}}},
  author = {{Balkenhol, L.} and others},
  year = {2023},
  month = jul,
  journal = {Physical Review D},
  volume = {108},
  number = {2},
  eprint = {2212.05642},
  primaryclass = {astro-ph},
  pages = {023510},
  issn = {2470-0010, 2470-0029},
  doi = {10.1103/PhysRevD.108.023510},
  urldate = {2025-01-25},
  archiveprefix = {arXiv},
  langid = {english}
}

@article{broutPantheonAnalysisCosmological2022,
  title = {The {{Pantheon}}+ {{Analysis}}: {{Cosmological Constraints}}},
  shorttitle = {The {{Pantheon}}+ {{Analysis}}},
  author = {{Brout, Dillon} and others},
  year = {2022},
  month = oct,
  journal = {The Astrophysical Journal},
  volume = {938},
  number = {2},
  pages = {110},
  issn = {0004-637X, 1538-4357},
  doi = {10.3847/1538-4357/ac8e04},
  urldate = {2025-01-25},
  langid = {english}
}

@misc{collaborationDESI2024VI2024,
  title = {{{DESI}} 2024 {{VI}}: {{Cosmological Constraints}} from the {{Measurements}} of {{Baryon Acoustic Oscillations}}},
  shorttitle = {{{DESI}} 2024 {{VI}}},
  collaboration = {DESI},
  author = {{Adame, A. G.} and others (The DESI Collaboration)},
  year = {2024},
  month = nov,
  number = {arXiv:2404.03002},
  eprint = {2404.03002},
  primaryclass = {astro-ph},
  publisher = {arXiv},
  doi = {10.48550/arXiv.2404.03002},
  urldate = {2025-01-25},
  archiveprefix = {arXiv},
  langid = {english}
}

@article{freedmanMeasurementsHubbleConstant2021,
  title = {Measurements of the {{Hubble Constant}}: {{Tensions}} in {{Perspective}}*},
  shorttitle = {Measurements of the {{Hubble Constant}}},
  author = {Freedman, Wendy L.},
  year = {2021},
  month = sep,
  journal = {The Astrophysical Journal},
  volume = {919},
  number = {1},
  pages = {16},
  issn = {0004-637X, 1538-4357},
  doi = {10.3847/1538-4357/ac0e95},
  urldate = {2025-01-25},
  langid = {english}
}

@article{jimenezLocalDistantUniverse2019,
  title = {The Local and Distant {{Universe}}: Stellar Ages and \${{H}}\_0\$},
  shorttitle = {The Local and Distant {{Universe}}},
  author = {Jimenez, Raul and Cimatti, Andrea and Verde, Licia and Moresco, Michele and Wandelt, Benjamin},
  year = {2019},
  month = mar,
  journal = {Journal of Cosmology and Astroparticle Physics},
  volume = {2019},
  number = {03},
  eprint = {1902.07081},
  primaryclass = {astro-ph},
  pages = {043--043},
  issn = {1475-7516},
  doi = {10.1088/1475-7516/2019/03/043},
  urldate = {2025-02-09},
  archiveprefix = {arXiv},
  langid = {english}
}

@article{macaulayFirstCosmologicalResults2019,
  title = {First Cosmological Results Using {{Type Ia}} Supernovae from the {{Dark Energy Survey}}: Measurement of the {{Hubble}} Constant},
  shorttitle = {First Cosmological Results Using {{Type Ia}} Supernovae from the {{Dark Energy Survey}}},
  author = {{Macaulay, E.} and others (The DES Collaboration)},
  year = {2019},
  month = jun,
  journal = {Monthly Notices of the Royal Astronomical Society},
  volume = {486},
  number = {2},
  pages = {2184--2196},
  issn = {0035-8711, 1365-2966},
  doi = {10.1093/mnras/stz978},
  urldate = {2025-01-25},
  copyright = {https://academic.oup.com/journals/pages/open\_access/funder\_policies/chorus/standard\_publication\_model},
  langid = {english}
}

@article{wongH0LiCOWXIII242020,
  title = {{{H0LiCOW}} -- {{XIII}}. {{A}} 2.4 per Cent Measurement of {{H0}} from Lensed Quasars: 5.3{$\sigma$} Tension between Early- and Late-{{Universe}} Probes},
  shorttitle = {{{H0LiCOW}} -- {{XIII}}. {{A}} 2.4 per Cent Measurement of {{H0}} from Lensed Quasars},
  author = {{Wong, Kenneth C} and others},
  year = {2020},
  month = oct,
  journal = {Monthly Notices of the Royal Astronomical Society},
  volume = {498},
  number = {1},
  pages = {1420--1439},
  issn = {0035-8711, 1365-2966},
  doi = {10.1093/mnras/stz3094},
  urldate = {2025-02-09},
  copyright = {https://academic.oup.com/journals/pages/open\_access/funder\_policies/chorus/standard\_publication\_model},
  langid = {english}
}

@article{riessLargeMagellanicCloud2019a,
  title = {Large {{Magellanic Cloud Cepheid Standards Provide}} a 1\% {{Foundation}} for the {{Determination}} of the {{Hubble Constant}} and {{Stronger Evidence}} for {{Physics}} beyond {{$\Lambda$CDM}}},
  author = {Riess, Adam G. and Casertano, Stefano and Yuan, Wenlong and Macri, Lucas M. and Scolnic, Dan},
  year = {2019},
  month = may,
  journal = {The Astrophysical Journal},
  volume = {876},
  number = {1},
  pages = {85},
  issn = {0004-637X, 1538-4357},
  doi = {10.3847/1538-4357/ab1422},
  urldate = {2025-02-10},
  langid = {english}
}

@article{suyu2010dissecting,
  title={Dissecting the gravitational lens B1608+ 656. II. Precision measurements of the Hubble constant, spatial curvature, and the dark energy equation of state},
  author={Suyu, SH and Marshall, PJ and Auger, MW and Hilbert, S and Blandford, RD and Koopmans, LVE and Fassnacht, CD and Treu, T},
  journal={The Astrophysical Journal},
  volume={711},
  number={1},
  pages={201},
  year={2010},
  publisher={IOP Publishing}
}

@article{hogg1999distance,
  title={Distance measures in cosmology},
  author={Hogg, David W},
  journal={arXiv preprint astro-ph/9905116},
  year={1999}
}

@book{dodelson2003modern,
  title={Modern cosmology},
  author={Dodelson, Scott and Schmidt, Fabian},
  year={2003},
  publisher={Elsevier}
}

@article{thorne1987gravitational,
  title={Gravitational radiation.},
  author={Thorne, Kip S},
  journal={Three hundred years of gravitation},
  pages={330--458},
  year={1987}
}

@article{chernoff1993gravitational,
  title={Gravitational radiation, inspiraling binaries, and cosmology},
  author={Chernoff, David F and Finn, Lee Samuel},
  journal={arXiv preprint gr-qc/9304020},
  year={1993}
}

@article{taylor2012cosmology1,
  title={Cosmology using advanced gravitational-wave detectors alone},
  author={Taylor, Stephen R and Gair, Jonathan R and Mandel, Ilya},
  journal={Physical Review D—Particles, Fields, Gravitation, and Cosmology},
  volume={85},
  number={2},
  pages={023535},
  year={2012},
  publisher={APS}
}

@article{taylor2012cosmology2,
  title={Cosmology with the lights off: Standard sirens in the Einstein Telescope era},
  author={Taylor, Stephen R and Gair, Jonathan R},
  journal={Physical Review D—Particles, Fields, Gravitation, and Cosmology},
  volume={86},
  number={2},
  pages={023502},
  year={2012},
  publisher={APS}
}

@article{you2021standard,
  title={Standard-siren cosmology using gravitational waves from binary black holes},
  author={You, Zhi-Qiang and Zhu, Xing-Jiang and Ashton, Gregory and Thrane, Eric and Zhu, Zong-Hong},
  journal={The Astrophysical Journal},
  volume={908},
  number={2},
  pages={215},
  year={2021},
  publisher={IOP Publishing}
}

@article{ezquiaga2021jumping,
  title={Jumping the gap: searching for LIGO’s biggest black holes},
  author={Ezquiaga, Jose Mar{\'\i}a and Holz, Daniel E},
  journal={The Astrophysical Journal Letters},
  volume={909},
  number={2},
  pages={L23},
  year={2021},
  publisher={IOP Publishing}
}

@article{farr2019future,
  title={A future percent-level measurement of the Hubble expansion at redshift 0.8 with advanced LIGO},
  author={Farr, Will M and Fishbach, Maya and Ye, Jiani and Holz, Daniel E},
  journal={The Astrophysical Journal Letters},
  volume={883},
  number={2},
  pages={L42},
  year={2019},
  publisher={IOP Publishing}
}

@article{messenger2012measuring,
  title={Measuring a Cosmological Distance-Redshift Relationship Using Only Gravitational Wave Observations of Binary Neutron Star Coalescences},
  author={Messenger, Chris and Read, Jocelyn},
  journal={Physical review letters},
  volume={108},
  number={9},
  pages={091101},
  year={2012},
  publisher={APS}
}

@article{chatterjee2021cosmology,
  title={Cosmology with Love: Measuring the Hubble constant using neutron star universal relations},
  author={Chatterjee, Deep and Hegade KR, Abhishek and Holder, Gilbert and Holz, Daniel E and Perkins, Scott and Yagi, Kent and Yunes, Nicol{\'a}s},
  journal={Physical Review D},
  volume={104},
  number={8},
  pages={083528},
  year={2021},
  publisher={APS}
}

@article{macleod2008precision,
  title={Precision of Hubble constant derived using black hole binary absolute distances and statistical redshift information},
  author={MacLeod, Chelsea L and Hogan, Craig J},
  journal={Physical Review D—Particles, Fields, Gravitation, and Cosmology},
  volume={77},
  number={4},
  pages={043512},
  year={2008},
  publisher={APS}
}

@article{delpozzo2012inference,
  title={Inference of cosmological parameters from gravitational waves: Applications to second generation interferometers},
  author={Del Pozzo, Walter},
  journal={Physical Review D—Particles, Fields, Gravitation, and Cosmology},
  volume={86},
  number={4},
  pages={043011},
  year={2012},
  publisher={APS}
}

@article{nair2018measuring,
  title={Measuring the Hubble constant: Gravitational wave observations meet galaxy clustering},
  author={Nair, Remya and Bose, Sukanta and Saini, Tarun Deep},
  journal={Physical Review D},
  volume={98},
  number={2},
  pages={023502},
  year={2018},
  publisher={APS}
}

@article{nishizawa2017measurement,
  title={Measurement of Hubble constant with stellar-mass binary black holes},
  author={Nishizawa, Atsushi},
  journal={Physical Review D},
  volume={96},
  number={10},
  pages={101303},
  year={2017},
  publisher={APS}
}

@article{DES:2019ccw,
    author = "{Soares-Santos, M.} and others",
    collaboration = "DES, LIGO Scientific, Virgo",
    title = "{First Measurement of the Hubble Constant from a Dark Standard Siren using the Dark Energy Survey Galaxies and the LIGO/Virgo Binary\textendash{}Black-hole Merger GW170814}",
    eprint = "1901.01540",
    archivePrefix = "arXiv",
    primaryClass = "astro-ph.CO",
    reportNumber = "FERMILAB-PUB-18-629-AE",
    doi = "10.3847/2041-8213/ab14f1",
    journal = "The Astrophysical Journal Letters",
    volume = "876",
    number = "1",
    pages = "L7",
    year = "2019"
}

@article{Mali:2024wpq,
    author = "Mali, Utkarsh and Essick, Reed",
    title = "{Striking a Chord with Spectral Sirens: Multiple Features in the Compact Binary Population Correlate with H$_{0}$}",
    eprint = "2410.07416",
    archivePrefix = "arXiv",
    primaryClass = "astro-ph.HE",
    doi = "10.3847/1538-4357/ad9de7",
    journal = "Astrophys. J.",
    volume = "980",
    number = "1",
    pages = "85",
    year = "2025"
}

@article{gray2020cosmological,
  title={Cosmological inference using gravitational wave standard sirens: A mock data analysis},
  author={Gray, Rachel and Hernandez, Ignacio Maga{\~n}a and Qi, Hong and Sur, Ankan and Brady, Patrick R and Chen, Hsin-Yu and Farr, Will M and Fishbach, Maya and Gair, Jonathan R and Ghosh, Archisman and others},
  journal={Physical Review D},
  volume={101},
  number={12},
  pages={122001},
  year={2020},
  publisher={APS}
}

@article{yu2020hunting,
  title={Hunting for the host galaxy groups of binary black holes and the application in constraining Hubble constant},
  author={Yu, Jiming and Wang, Yu and Zhao, Wen and Lu, Youjun},
  journal={Monthly Notices of the Royal Astronomical Society},
  volume={498},
  number={2},
  pages={1786--1800},
  year={2020},
  publisher={Oxford University Press}
}

@article{palmese2020statistical,
  title={A statistical standard siren measurement of the Hubble constant from the LIGO/Virgo gravitational wave compact object merger GW190814 and Dark Energy Survey galaxies},
  author={{Palmese, Antonella} and others},
  journal={The Astrophysical Journal Letters},
  volume={900},
  number={2},
  pages={L33},
  year={2020},
  publisher={IOP Publishing}
}

@article{borhanian2020dark,
  title={Dark sirens to resolve the Hubble--Lema{\^\i}tre tension},
  author={Borhanian, Ssohrab and Dhani, Arnab and Gupta, Anuradha and Arun, KG and Sathyaprakash, BS},
  journal={The Astrophysical Journal Letters},
  volume={905},
  number={2},
  pages={L28},
  year={2020},
  publisher={IOP Publishing}
}

@article{finke2021cosmology,
  title={Cosmology with LIGO/Virgo dark sirens: Hubble parameter and modified gravitational wave propagation},
  author={Finke, Andreas and Foffa, Stefano and Iacovelli, Francesco and Maggiore, Michele and Mancarella, Michele},
  journal={Journal of Cosmology and Astroparticle Physics},
  volume={2021},
  number={08},
  pages={026},
  year={2021},
  publisher={IOP Publishing}
}

@article{gray2023joint,
  title={Joint cosmological and gravitational-wave population inference using dark sirens and galaxy catalogues},
  author={{Gray, Rachel} and others},
  journal={Journal of Cosmology and Astroparticle Physics},
  volume={2023},
  number={12},
  pages={023},
  year={2023},
  publisher={IOP Publishing}
}

@article{oguri2016measuring,
  title={Measuring the distance-redshift relation with the cross-correlation of gravitational wave standard sirens and galaxies},
  author={Oguri, Masamune},
  journal={Physical Review D},
  volume={93},
  number={8},
  pages={083511},
  year={2016},
  publisher={APS}
}

@article{bera2020incompleteness,
  title={Incompleteness matters not: inference of H0 from binary black hole--galaxy cross-correlations},
  author={Bera, Sayantani and Rana, Divya and More, Surhud and Bose, Sukanta},
  journal={The Astrophysical Journal},
  volume={902},
  number={1},
  pages={79},
  year={2020},
  publisher={IOP Publishing}
}

@article{mukherjee2021accurate,
  title={Accurate precision cosmology with redshift unknown gravitational wave sources},
  author={Mukherjee, Suvodip and Wandelt, Benjamin D and Nissanke, Samaya M and Silvestri, Alessandra},
  journal={Physical Review D},
  volume={103},
  number={4},
  pages={043520},
  year={2021},
  publisher={APS}
}

@article{punturoEinsteinTelescopeThirdgeneration2010,
  title = {The {{Einstein Telescope}}: A Third-Generation Gravitational Wave Observatory},
  shorttitle = {The {{Einstein Telescope}}},
  author = {{Punturo, M} and others},
  date = {2010-10-07},
  year = {2010},
  journal = {Classical and Quantum Gravity},
  shortjournal = {Class. Quantum Grav.},
  volume = {27},
  number = {19},
  pages = {194002},
  issn = {0264-9381, 1361-6382},
  doi = {10.1088/0264-9381/27/19/194002},
  url = {https://iopscience.iop.org/article/10.1088/0264-9381/27/19/194002},
  urldate = {2023-11-22},
  langid = {english}
}

@article{maggioreScienceCaseEinstein2020,
  title = {Science {{Case}} for the {{Einstein Telescope}}},
  author = {{Maggiore, Michele} and others},
  date = {2020-03-01},
  journal = {Journal of Cosmology and Astroparticle Physics},
  shortjournal = {J. Cosmol. Astropart. Phys.},
  volume = {2020},
  year = {2020},
  number = {03},
  eprint = {1912.02622},
  eprinttype = {arxiv},
  eprintclass = {astro-ph, physics:gr-qc},
  pages = {050--050},
  issn = {1475-7516},
  doi = {10.1088/1475-7516/2020/03/050},
  url = {http://arxiv.org/abs/1912.02622},
  urldate = {2023-11-22},
  langid = {english}
}

@article{evansHorizonStudyCosmic2021,
  title = {A {{Horizon Study}} for {{Cosmic Explorer}}: {{Science}}, {{Observatories}}, and {{Community}}},
  shorttitle = {A {{Horizon Study}} for {{Cosmic Explorer}}},
  author = {{Evans, Matthew} and others},
  date = {2021-10-06},
  year = {2021},
  eprint = {2109.09882},
  eprinttype = {arxiv},
  eprintclass = {astro-ph, physics:gr-qc},
  doi = {10.48550/arXiv.2109.09882},
  url = {http://arxiv.org/abs/2109.09882},
  urldate = {2023-04-10},
  langid = {english},
  pubstate = {preprint},
  journal = ""
}

@misc{Fischbach:stochmon,
    author = {Makenzi Fischbach and Davis, Derek and Renzini, Arianna},
    title = {Investigating data quality metrics for stochastic gravitational-wave detection},
    journal = {Caltech LIGO SURF program 2021},
    year = {2021},
    url = {https://dcc-backup.ligo.org/public/0176/T2100203/001/Fischbach_First_Interim_Report.pdf}
}

@article{belczynski2010effect,
  title={The effect of metallicity on the detection prospects for gravitational waves},
  author={Belczynski, Krzysztof and Dominik, Michal and Bulik, Tomasz and O’Shaughnessy, Richard and Fryer, Chris and Holz, Daniel E},
  journal={The Astrophysical Journal Letters},
  volume={715},
  number={2},
  pages={L138},
  year={2010},
  publisher={IOP Publishing}
}

@article{kushnir2016gw150914,
  title={GW150914: spin-based constraints on the merger time of the progenitor system},
  author={Kushnir, Doron and Zaldarriaga, Matias and Kollmeier, Juna A and Waldman, Roni},
  journal={Monthly Notices of The Royal Astronomical Society},
  volume={462},
  number={1},
  pages={844--849},
  year={2016},
  publisher={The Royal Astronomical Society}
}

@article{gallegos2021binary,
  title={Binary Black Hole Formation with Detailed Modeling: Stable Mass Transfer Leads to Lower Merger Rates},
  author={Gallegos-Garcia, Monica and Berry, Christopher PL and Marchant, Pablo and Kalogera, Vicky},
  journal={The Astrophysical Journal},
  volume={922},
  number={2},
  pages={110},
  year={2021},
  publisher={IOP Publishing}
}

@article{van2022redshift,
  title={The redshift evolution of the binary black hole merger rate: A weighty matter},
  author={Van Son, LAC and De Mink, SE and Callister, T and Justham, S and Renzo, M and Wagg, T and Broekgaarden, FS and Kummer, F and Pakmor, R and Mandel, I},
  journal={The Astrophysical Journal},
  volume={931},
  number={1},
  pages={17},
  year={2022},
  publisher={IOP Publishing}
}

@ARTICLE{drasco2003DetectionMethods,
       author = {{Drasco}, Steve and {Flanagan}, {\'E}anna {\'E}.},
        title = "{Detection methods for non-Gaussian gravitational wave stochastic backgrounds}",
      journal = {\prd},
         year = 2003,
        month = apr,
       volume = {67},
       number = {8},
          eid = {082003},
        pages = {082003},
          doi = {10.1103/PhysRevD.67.082003},
archivePrefix = {arXiv},
       eprint = {gr-qc/0210032},
 primaryClass = {gr-qc},
       adsurl = {https://ui.adsabs.harvard.edu/abs/2003PhRvD..67h2003D}
}

@ARTICLE{smith2018OptimalSearch,
       author = {{Smith}, Rory and {Thrane}, Eric},
        title = "{Optimal Search for an Astrophysical Gravitational-Wave Background}",
      journal = {Physical Review X},
         year = 2018,
        month = apr,
       volume = {8},
       number = {2},
          eid = {021019},
        pages = {021019},
          doi = {10.1103/PhysRevX.8.021019},
archivePrefix = {arXiv},
       eprint = {1712.00688},
 primaryClass = {gr-qc},
       adsurl = {https://ui.adsabs.harvard.edu/abs/2018PhRvX...8b1019S}
}

@article{caprini2018cosmological,
  title={Cosmological backgrounds of gravitational waves},
  author={Caprini, Chiara and Figueroa, Daniel G},
  journal={Classical and Quantum Gravity},
  volume={35},
  number={16},
  pages={163001},
  year={2018},
  publisher={IOP Publishing}
}

@article{turbang2024metallicity,
  title={The metallicity dependence and evolutionary times of merging binary black holes: Combined constraints from individual gravitational-wave detections and the stochastic background},
  author={Turbang, Kevin and Lalleman, Max and Callister, Thomas A and van Remortel, Nick},
  journal={The Astrophysical Journal},
  volume={967},
  number={2},
  pages={142},
  year={2024},
  publisher={IOP Publishing}
}

@article{ferraiuolo2025inferring,
    author = "Ferraiuolo, S. and Mastrogiovanni, S. and Escoffier, S. and
Kajfasz, E.",
    title = "{Inferring astrophysics and cosmology with individual compact
binary coalescences and their gravitational-wave stochastic background}",
    eprint = "2503.14686",
    archivePrefix = "arXiv",
    primaryClass = "astro-ph.CO",
    doi = "10.1051/0004-6361/202555124",
    journal = "Astron. Astrophys.",
    volume = "701",
    pages = "A36",
    year = "2025"
}

@article{mukherjee2020probing,
  title={Probing the theory of gravity with gravitational lensing of gravitational waves and galaxy surveys},
  author={Mukherjee, Suvodip and Wandelt, Benjamin D and Silk, Joseph},
  journal={Monthly Notices of the Royal Astronomical Society},
  volume={494},
  number={2},
  pages={1956--1970},
  year={2020},
  publisher={Oxford University Press}
}

@article{mukherjee2024cross,
  title={Cross-correlating Dark Sirens and Galaxies: Constraints on H 0 from GWTC-3 of LIGO--Virgo--KAGRA},
  author={Mukherjee, Suvodip and Krolewski, Alex and Wandelt, Benjamin D and Silk, Joseph},
  journal={The Astrophysical Journal},
  volume={975},
  number={2},
  pages={189},
  year={2024},
  publisher={IOP Publishing}
}

@article{VoyagerAdhikariCryogenicSiliconInterferometer2020,
  title = {A Cryogenic Silicon Interferometer for Gravitational-Wave Detection},
  author = {{Adhikari, R X} and others},
  date = {2020-08-20},
  year = {2020},
  journal = {Classical and Quantum Gravity},
  shortjournal = {Class. Quantum Grav.},
  volume = {37},
  number = {16},
  pages = {165003},
  issn = {0264-9381, 1361-6382},
  doi = {10.1088/1361-6382/ab9143},
  url = {https://iopscience.iop.org/article/10.1088/1361-6382/ab9143},
  langid = {english}
}

@misc{fritschel2022asharp,
  title = {Report from the LSC Post-O5 study group Technical Report T2200287 (LIGO)},
  author = {Fritschel, P and Kuns, K and Driggers, J and Effler, A and Lantz, B and Ottaway, D and Ballmer, S and Dooley, K and Adhikari, R and Evans, M and Farr, B and Gonzalez, G and Schmidt, P and Raja, S},
  year = {2022},
  url = {https://dcc.ligo.org/LIGO-T2200287/public}
}

@article{ade2016planck,
  title={Planck 2015 results-xiii. cosmological parameters},
  author={{Ade, Peter AR} and others},
  journal={Astronomy \& Astrophysics},
  volume={594},
  pages={A13},
  year={2016},
  publisher={EDP sciences}
}

@article{gupta2023using,
  title={Using grey sirens to resolve the Hubble--Lema{\^\i}tre tension},
  author={Gupta, Ish},
  journal={Monthly Notices of the Royal Astronomical Society},
  volume={524},
  number={3},
  pages={3537--3558},
  year={2023},
  publisher={Oxford University Press}
}

@article{bom2024dark,
  title={A dark standard siren measurement of the Hubble constant following LIGO/Virgo/KAGRA O4a and previous runs},
  author={Bom, C\_R and Alfradique, V and Palmese, A and Teixeira, G and Santana-Silva, L and Santos, A and Darc, P},
  journal={Monthly Notices of the Royal Astronomical Society},
  volume={535},
  number={1},
  pages={961--975},
  year={2024},
  publisher={Oxford University Press}
}
